\documentclass[12pt,aps,prd,preprint,tightenlines,superscriptaddress,
amsmath,amssymb,nofootinbib]{revtex4-1} 

\input{header}
\begin{document}

\preprint{CERN-LHCC-2019-017, LHCC-P-015, UCI-TR-2019-25}

\title{
TECHNICAL PROPOSAL \\
\PRE{\vspace*{0.15in}} 
{\Large \FASERnu \\} 
\PRE{\vspace*{0.12in}}
FASER Collaboration
\PRE{\vspace*{0.00in}}
}

\begin{figure*}[h]
\PRE{\vspace*{-0.3in}}
\centering
\includegraphics[width=0.45\textwidth]{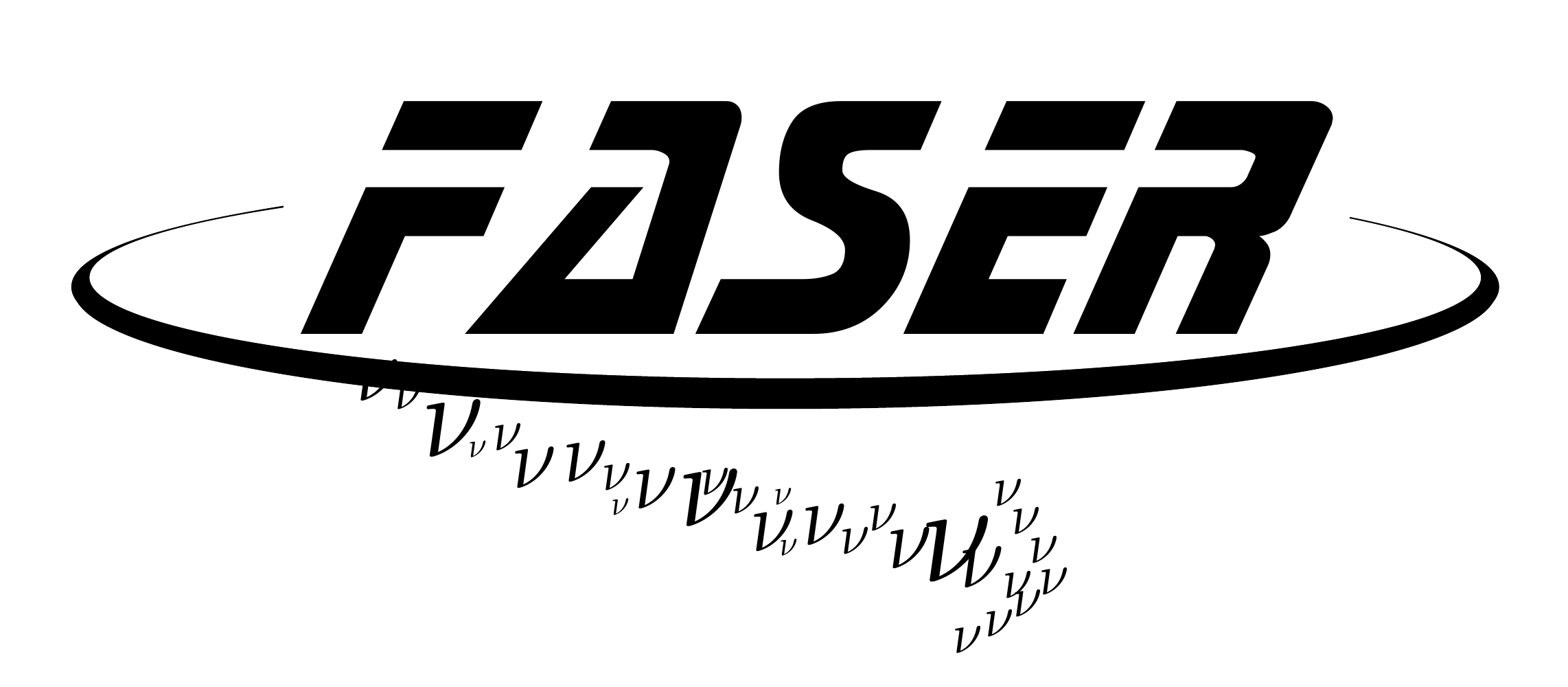}
\end{figure*}

\author{Henso~Abreu}
\affiliation{Department of Physics and Astronomy, Technion---Israel Institute of Technology, Haifa 32000, Israel}

\author{Marco~Andreini}
\email[FASER Associate.]{\strut}
\affiliation{CERN, CH-1211 Geneva 23, Switzerland}

\author{Claire~Antel}
\affiliation{D\'epartement de Physique Nucl\'eaire et Corpusculaire, 
University of Geneva, CH-1211 Geneva 4, Switzerland}

\author{Akitaka~Ariga}
\email[Contact emails: ]{akitaka.ariga@lhep.unibe.ch, tomoko.ariga@cern.ch, felixk@slac.stanford.edu}
\affiliation{Albert Einstein Center for Fundamental Physics, Laboratory for High Energy Physics, Universit\"at Bern, Sidlerstrasse 5, CH-3012 Bern, Switzerland}

\author{Tomoko~Ariga}
\email[Contact emails: ]{akitaka.ariga@lhep.unibe.ch, tomoko.ariga@cern.ch, felixk@slac.stanford.edu}
\affiliation{Albert Einstein Center for Fundamental Physics, Laboratory for High Energy Physics, Universit\"at Bern, Sidlerstrasse 5, CH-3012 Bern, Switzerland}
\affiliation{Kyushu University, Nishi-ku, 819-0395 Fukuoka, Japan}

\author{Caterina Bertone}
\email[FASER Associate.]{\strut}
\affiliation{CERN, CH-1211 Geneva 23, Switzerland}

\author{Jamie~Boyd}
\affiliation{CERN, CH-1211 Geneva 23, Switzerland}

\author{Andy~Buckley}
\email[FASER Associate.]{\strut}
\affiliation{School of Physics and Astronomy, University of Glasgow, Glasgow, G12~8QQ, United Kingdom} 

\author{Franck~Cadoux}
\affiliation{D\'epartement de Physique Nucl\'eaire et Corpusculaire, 
University of Geneva, CH-1211 Geneva 4, Switzerland}

\author{David~W.~Casper}
\affiliation{Department of Physics and Astronomy, 
University of California, Irvine, CA 92697-4575, USA}

\author{Francesco~Cerutti}
\email[FASER Associate.]{\strut}
\affiliation{CERN, CH-1211 Geneva 23, Switzerland}

\author{Xin~Chen}
\affiliation{Physics Department,Tsinghua University, Beijing, China}
 
\author{Andrea~Coccaro}
\affiliation{INFN Sezione di Genova, Via Dodecaneso, 33--16146, Genova, Italy}

\author{Salvatore~Danzeca} 
\email[FASER Associate.]{\strut} 
\affiliation{CERN, CH-1211 Geneva 23, Switzerland}

\author{Liam~Dougherty} 
\email[FASER Associate.]{\strut} 
\affiliation{CERN, CH-1211 Geneva 23, Switzerland} 
 
\author{Candan~Dozen}
\affiliation{Physics Department,Tsinghua University, Beijing, China}

\author{Peter~B.~Denton} 
\email[FASER Associate.]{\strut} 
\affiliation{Department of Physics, Brookhaven National Laboratory, Upton, NY 11973, USA} 

\author{Yannick~Favre}
\affiliation{D\'epartement de Physique Nucl\'eaire et Corpusculaire, 
University of Geneva, CH-1211 Geneva 4, Switzerland}

\author{Deion~Fellers}
\affiliation{University of Oregon, Eugene, OR 97403, USA}

\author{Jonathan~L.~Feng}
\affiliation{Department of Physics and Astronomy, 
University of California, Irvine, CA 92697-4575, USA}

\author{Didier~Ferrere}
\affiliation{D\'epartement de Physique Nucl\'eaire et Corpusculaire, 
University of Geneva, CH-1211 Geneva 4, Switzerland}

\author{Jonathan~Gall}
\email[FASER Associate.]{\strut} 
\affiliation{CERN, CH-1211 Geneva 23, Switzerland}

\author{Iftah~Galon}
\affiliation{New High Energy Theory Center, Rutgers, The State University of New Jersey, Piscataway, New Jersey 08854-8019, USA}

\author{Stephen~Gibson}
\affiliation{Royal Holloway, University of London, Egham, TW20 0EX, UK}

\author{Sergio~Gonzalez-Sevilla}
\affiliation{D\'epartement de Physique Nucl\'eaire et Corpusculaire, 
University of Geneva, CH-1211 Geneva 4, Switzerland}

\author{Shih-Chieh~Hsu}
\affiliation{Department of Physics, University of Washington, PO Box 351560, Seattle, WA 98195-1560, USA}

\author{Zhen~Hu}
\affiliation{Physics Department, Tsinghua University, Beijing, China}

\author{Giuseppe~Iacobucci}
\affiliation{D\'epartement de Physique Nucl\'eaire et Corpusculaire, 
University of Geneva, CH-1211 Geneva 4, Switzerland}

\author{Sune~Jakobsen}
\affiliation{CERN, CH-1211 Geneva 23, Switzerland}

\author{Roland~Jansky}
\affiliation{D\'epartement de Physique Nucl\'eaire et Corpusculaire, 
University of Geneva, CH-1211 Geneva 4, Switzerland}

\author{Enrique~Kajomovitz}
\affiliation{Department of Physics and Astronomy, 
Technion---Israel Institute of Technology, Haifa 32000, Israel}

\author{Felix~Kling}
\email[Contact emails: ]{akitaka.ariga@lhep.unibe.ch, tomoko.ariga@cern.ch, felixk@slac.stanford.edu}
\affiliation{SLAC National Accelerator Laboratory, 2575 Sand Hill Road, Menlo Park, CA 94025, USA}

\author{Umut Kose}
\affiliation{CERN, CH-1211 Geneva 23, Switzerland}

\author{Susanne~Kuehn}
\affiliation{CERN, CH-1211 Geneva 23, Switzerland}

\author{Mike~Lamont}
\email[FASER Associate.]{\strut} 
\affiliation{CERN, CH-1211 Geneva 23, Switzerland} 

\author{Helena~Lefebvre} 
\affiliation{Royal Holloway, University of London, Egham, TW20 0EX, UK}

\author{Lorne~Levinson}
\affiliation{Department of Particle Physics and Astrophysics, Weizmann Institute of Science, Rehovot 76100, Israel}

\author{Ke~Li} 
\affiliation{Department of Physics, University of Washington, PO Box 351560, Seattle, WA 98195-1560, USA}

\author{Josh~McFayden}
\affiliation{CERN, CH-1211 Geneva 23, Switzerland}

\author{Sam~Meehan}
\affiliation{CERN, CH-1211 Geneva 23, Switzerland}

\author{Dimitar~Mladenov}
\affiliation{CERN, CH-1211 Geneva 23, Switzerland}

\author{Mitsuhiro~Nakamura}
\affiliation{Nagoya University, Furo-cho, Chikusa-ku, Nagoya 464-8602, Japan}

\author{Toshiyuki~Nakano}
\affiliation{Nagoya University, Furo-cho, Chikusa-ku, Nagoya 464-8602, Japan}

\author{Marzio~Nessi}
\affiliation{CERN, CH-1211 Geneva 23, Switzerland}

\author{Friedemann~Neuhaus}
\affiliation{Institut f\"ur Physik, Universität Mainz, Mainz, Germany}

\author{John~Osborne}
\email[FASER Associate.]{\strut}  
\affiliation{CERN, CH-1211 Geneva 23, Switzerland}

\author{Hidetoshi~Otono}
\affiliation{Kyushu University, Nishi-ku, 819-0395 Fukuoka, Japan}

\author{Serge Pelletier}
\email[FASER Associate.]{\strut}  
\affiliation{CERN, CH-1211 Geneva 23, Switzerland}

\author{Brian~Petersen}
\affiliation{CERN, CH-1211 Geneva 23, Switzerland}

\author{Francesco~Pietropaolo}
\affiliation{CERN, CH-1211 Geneva 23, Switzerland}

\author{Michaela~Queitsch-Maitland}
\affiliation{CERN, CH-1211 Geneva 23, Switzerland}

\author{Filippo Resnati}
\affiliation{CERN, CH-1211 Geneva 23, Switzerland}

\author{Marta~Sabat\'e-Gilarte}
\email[FASER Associate.]{\strut} 
\affiliation{CERN, CH-1211 Geneva 23, Switzerland}
\affiliation{University of Seville, Seville, Spain}

\author{Jakob~Salfeld-Nebgen}
\affiliation{CERN, CH-1211 Geneva 23, Switzerland}

\author{Francisco~Sanchez~Galan} 
\email[FASER Associate.]{\strut} 
\affiliation{CERN, CH-1211 Geneva 23, Switzerland}

\author{Pablo~Santos~Diaz}
\email[FASER Associate.]{\strut} 
\affiliation{CERN, CH-1211 Geneva 23, Switzerland}

\author{Osamu~Sato}
\affiliation{Nagoya University, Furo-cho, Chikusa-ku, Nagoya 464-8602, Japan}

\author{Paola~Scampoli}
\affiliation{Albert Einstein Center for Fundamental Physics, Laboratory for High Energy Physics, Universit\"at Bern, Sidlerstrasse 5, CH-3012 Bern, Switzerland}
\affiliation{Dipartimento di Fisica ``Ettore Pancini'', Universit\`a di Napoli Federico II, Complesso Universitario di Monte S. Angelo, I-80126 Napoli, Italy }

\author{Kristof~Schmieden}
\affiliation{CERN, CH-1211 Geneva 23, Switzerland}

\author{Matthias~Schott}
\affiliation{Institut f\"ur Physik, Universität Mainz, Mainz, Germany}

\author{Holger~Schulz}
\email[FASER Associate.]{\strut} 
\affiliation{Department of Physics, University of Cincinnati, Cincinnati, OH 45219, USA} 

\author{Anna~Sfyrla}
\affiliation{D\'epartement de Physique Nucl\'eaire et Corpusculaire, 
University of Geneva, CH-1211 Geneva 4, Switzerland}

\author{Savannah~Shively}
\affiliation{Department of Physics and Astronomy, 
University of California, Irvine, CA 92697-4575, USA}

\author{Jordan~Smolinsky}
\affiliation{Department of Physics, University of Florida, Gainesville, FL 32611, USA}

\author{Aaron~M.~Soffa}
\affiliation{Department of Physics and Astronomy, 
University of California, Irvine, CA 92697-4575, USA}

\author{Yosuke~Takubo}
\affiliation{Institute of Particle and Nuclear Study, 
KEK, Oho 1-1, Tsukuba, Ibaraki 305-0801, Japan}

\author{Eric~Torrence}
\affiliation{University of Oregon, Eugene, OR 97403, USA}

\author{Sebastian~Trojanowski}
\affiliation{Consortium for Fundamental Physics, School of Mathematics and  Statistics, University of Sheffield, Hounsfield Road, Sheffield, S3 7RH, UK \PRE{\vspace*{0.2in}}}

\author{Serhan~Tufanli}
\affiliation{CERN, CH-1211 Geneva 23, Switzerland}

\author{Dengfeng~Zhang}
\affiliation{Physics Department, Tsinghua University, Beijing, China}

\author{Gang~Zhang
\PRE{\vspace*{.2in}}}
\affiliation{Physics Department, Tsinghua University, Beijing, China}

\begin{abstract}
\PRE{\vspace*{0.1in}}
\FASERnu is a proposed small and inexpensive emulsion detector designed to detect collider neutrinos for the first time and study their properties.  \FASERnu will be located directly in front of FASER, 480 m from the ATLAS interaction point along the beam collision axis in the unused service tunnel TI12.  From 2021-23 during Run 3 of the 14 TeV LHC, roughly 1,300 electron neutrinos, 20,000 muon neutrinos, and 20 tau neutrinos will interact in \FASERnu with TeV-scale energies.  With the ability to observe these interactions, reconstruct their energies, and distinguish flavors, \FASERnu will probe the production, propagation, and interactions of neutrinos at the highest human-made energies ever recorded.   

The \FASERnu detector will be composed of 1000 emulsion layers interleaved with tungsten plates. The total volume of the emulsion and tungsten is $25\,\cm \times 25\,\cm \times 1.35\,\m$, and the tungsten target mass is 1.2 tonnes.  From 2021-23, 7 sets of emulsion layers will be installed, with replacement roughly every $20-50~\ifb$ in planned Technical Stops.  In this document, we summarize \FASERnu's physics goals and discuss the estimates of neutrino flux and interaction rates. We then describe the \FASERnu detector in detail, including plans for assembly, transport, installation, and emulsion replacement, and procedures for emulsion readout and analyzing the data.  We close with cost estimates for the detector components and infrastructure work and a timeline for the experiment.
\end{abstract}

%\pacs{}

%\pagenumbering{roman}
\maketitle

\renewcommand{\baselinestretch}{0.9}\normalsize
\tableofcontents
\renewcommand{\baselinestretch}{1.0}\normalsize

\clearpage
%%%%%%%%%%%%%%%%%%%%%%%%%%%%%%%%%%%%%%%%%%%%%%%%%%%%%%
%%%%%%%%%%%%%%%%%%%%%%%%%%%%%%%%%%%%%%%%%%%%%%%%%%%%%%
\section{Introduction and Overview}
\label{sec:introduction}
%%%%%%%%%%%%%%%%%%%%%%%%%%%%%%%%%%%%%%%%%%%%%%%%%%%%%%
%%%%%%%%%%%%%%%%%%%%%%%%%%%%%%%%%%%%%%%%%%%%%%%%%%%%%%

\FASERnu is a proposed emulsion detector designed to detect and study the interactions of neutrinos produced at the LHC~\cite{Abreu:2019yak}. \FASERnu will be located along the beam collision axis, 480 m from the ATLAS interaction point (IP) in the unused tunnel TI12, and directly in front of the Forward Search Experiment (FASER) spectrometer~\cite{Feng:2017uoz, Ariga:2018zuc, Ariga:2018pin, Ariga:2018uku}. At this special location, \FASERnu will be able to record the interactions of $\sim 10,000$ neutrinos at the TeV energy scale, including neutrinos and anti-neutrinos of all flavors.  These interactions will be the highest energy neutrino-nucleus interactions ever recorded for electron and tau neutrinos, and they will allow a precise measurement of muon neutrino interaction rates in an energy range that has never been directly constrained.  Such measurements will shed light on neutrino properties and will also constrain the forward production of heavy mesons, with important implications for other accelerator, collider, and astroparticle experiments.

The \FASERnu detector will be composed of 1000 emulsion layers~\cite{emulsion} interleaved with tungsten plates, with a total tungsten target mass of 1.2 tonnes.  The total volume of the emulsion layers and tungsten plates is $25\,\cm \times 25\,\cm \times 1.35\,\m$.  We propose that \FASERnu be installed in TI12 in time to collect data during Run 3 of the 14 TeV LHC. From 2021-23, 7 emulsion detectors will be installed, with replacement roughly every $20-50~\ifb$ in planned Technical Stops.  Based on data from pilot emulsion detectors installed in TI12 in 2018, this replacement rate will result in an acceptably low track density to allow for event reconstruction.  The \FASERnu plans benefit significantly from these pilot data, as well as from the infrastructure work in TI12 that is already underway to make the area ready for FASER. The XSEN Collaboration has also submitted a Letter of Intent~\cite{Beni:2019pyp} to construct a complementary experiment in the tunnel TI18, which is located at a symmetric position on the other side of ATLAS. 

The physics motivations and detector concept for \FASERnu have been discussed previously in the \FASERnu Letter of Intent (LOI)~\cite{Abreu:2019yak}.  In this document, we describe the technical aspects of the experiment in more detail.  We begin in \secref{physics} with a brief summary of the physics goals of the experiment.  We then give an overview of the detector location and environment in \secref{environment}, and we discuss our estimates of the neutrino flux and interaction rates in \secref{flux}. In \secref{tungsten_emulsion_detector}, we describe the detector in detail, including our plans for assembling and transporting the detector and replacing the emulsion films during Technical Stops, as well as safety-related matters. In \secref{interface_detector}, we also discuss the possibility of adding an interface detector, which would integrate \FASERnu with FASER, allowing neutrinos and anti-neutrinos to be distinguished and improving measurements of signal and background.  \secref{offline_analysis} summarizes plans for off-line analysis. Finally, we conclude with our estimates of cost and schedule in \secref{cost_and_schedule}.

%%%%%%%%%%%%%%%%%%%%%%%%%%%%%%%%%%%%%%%%%%%%%%%%%%%%%%
%%%%%%%%%%%%%%%%%%%%%%%%%%%%%%%%%%%%%%%%%%%%%%%%%%%%%%
\section{Physics Goals}
\label{sec:physics}
%%%%%%%%%%%%%%%%%%%%%%%%%%%%%%%%%%%%%%%%%%%%%%%%%%%%%%
%%%%%%%%%%%%%%%%%%%%%%%%%%%%%%%%%%%%%%%%%%%%%%%%%%%%%%

The LHC is the highest energy particle collider built so far, and it is therefore also the source of the most energetic human-made neutrinos created in a controlled laboratory environment. Proton-proton collisions typically lead to a large number of hadrons produced along the beam collision axis, which can inherit an $\mathcal{O}(1)$ fraction of the protons' momenta. The decay of those hadrons then leads to a large flux of high-energy neutrinos, which are highly collimated around the beam collision axis. We have estimated that in Run 3 of the 14 TeV LHC from 2021-23, roughly $10^{11}$ electron neutrinos, $10^{12}$ muon neutrinos, and $10^9$ tau neutrinos will be produced in the far-forward region of the ATLAS IP~\cite{Abreu:2019yak}. However, despite the fact that neutrinos are copiously produced at the LHC, no collider neutrino has been detected so far.

The \FASERnu detector, which will be placed along the beam collision axis, $480~\m$ downstream from the ATLAS IP, will take advantage of this neutrino beam to detect neutrinos from the LHC for the first time. During Run 3 of the LHC, assuming an integrated luminosity of $150~\ifb$, about 1300 electron neutrinos, 20,000 muon neutrinos, and 20 tau neutrinos are expected to interact with the \FASERnu detector. This will open a new window to study neutrino interactions at high energies and therefore extend the LHC’s physics program in a new direction.

In \figref{physics-nuxs} we show existing measurements of neutrino-nucleon charged current (CC) scattering cross sections for $\nu_e$ (left panel), $\nu_\mu$ (center panel), and $\nu_\tau$ (right panel). At low energies $E_\nu < 360~\gev$, the neutrino cross section for all three flavors has been constrained by neutrino experiments utilizing the CERN SPS (400 GeV proton) and Fermilab Tevatron (800 GeV proton) accelerators~\cite{Baltay:1988au,Kodama:2007aa,Tanabashi:2018oca}. At very high energies, $E_\nu >6.3~\tev$, IceCube has constrained the muon neutrino cross section using atmospheric neutrinos, albeit with relatively large uncertainties~\cite{Aartsen:2017kpd, Bustamante:2017xuy}. For a detailed discussion of these constraints, see Ref.~\cite{Abreu:2019yak}. In \figref{physics-nuxs} we additionally show the energy spectra of neutrinos that interact in \FASERnu, as obtained in Ref.~\cite{Abreu:2019yak}. We can see that the neutrino spectra are broad band and span over more than one order of magnitude in energy, indicating \FASERnu's potential to measure neutrino cross sections in currently unprobed energy ranges for all three neutrino flavors. 

\begin{figure}[t]
\centering
\includegraphics[width=\textwidth]{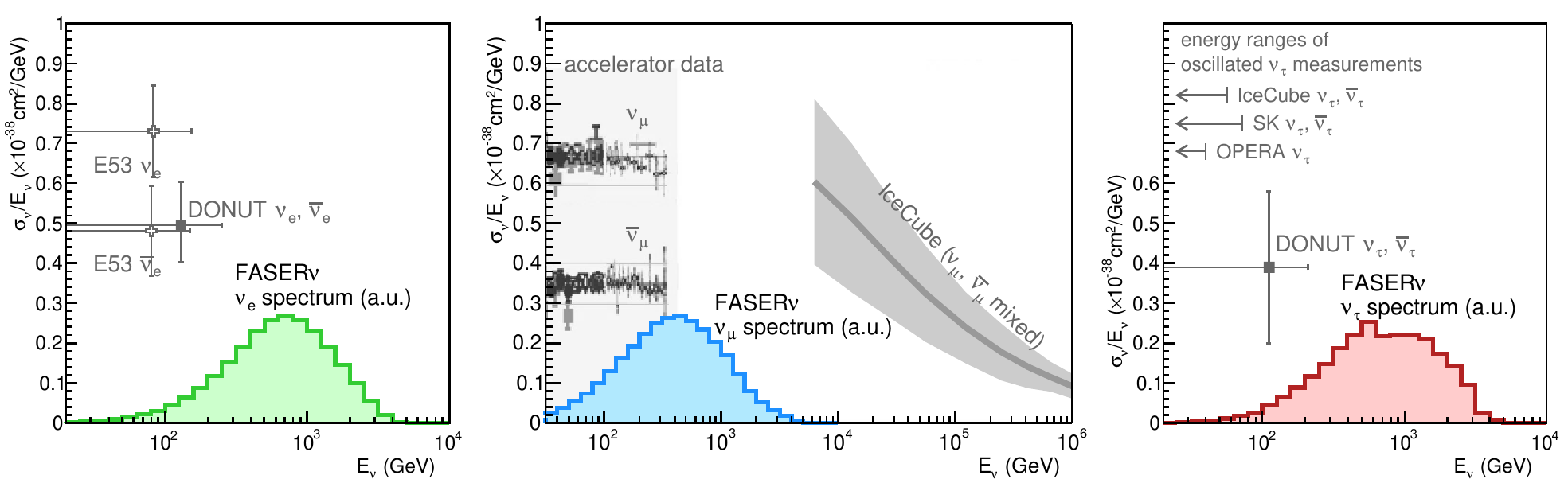}
\caption{Existing constraints on the CC neutrino scattering cross sections on an isoscalar target for electron neutrinos (left), muon neutrinos (center), and tau neutrinos (right) from previous accelerator experiments at low energies and IceCube at high energies. The colored histograms show the expected energy spectra of neutrinos that interact in \FASERnu.  For all three flavors, the \FASERnu energy spectra are peaked at energies that are currently unconstrained. From Ref.~\cite{Abreu:2019yak}.
}
\label{fig:physics-nuxs}
\end{figure}

This potential for measuring the neutrino interaction cross section has been studied in Ref.~\cite{Abreu:2019yak}. The expected sensitivity for \FASERnu to constrain neutrino CC cross sections is shown in \figref{physics-xsprojection}. The black dashed curve is the theoretical prediction for the average CC cross section per nucleon in tungsten, $\sigma = (\sigma_{\nu W} + \sigma_{\bar \nu W})/2$. The solid error bars show the sensitivity considering only statistical uncertainties. The shaded bands show the uncertainties from the range of neutrino production rates predicted by different MC generators, which serve as a rough estimate for the expected size of systematic uncertainties related to the neutrino flux. Our efforts to reduce these uncertainties are discussed in \secref{tuning}. The combination of statistical and production rate uncertainties, added in quadrature, is shown as the dashed error bars. These sensitivity estimates take into account the geometrical acceptance, vertex detection efficiency, and lepton identification efficiency, and  assume that the measurement is background free. We can see that \FASERnu significantly extends the neutrino cross section measurements to higher energies for both electron and tau neutrinos, while for muon neutrinos, \FASERnu will fill the gap between the existing measurements from accelerator experiments and IceCube. An additional interface detector between \FASERnu and the FASER spectrometer will further be able to distinguish $\nu_\mu$ and $\bar{\nu}_\mu$ events, as discussed in \secref{interface_detector}. 

\begin{figure}[t]
\centering
\includegraphics[width=0.32\textwidth]{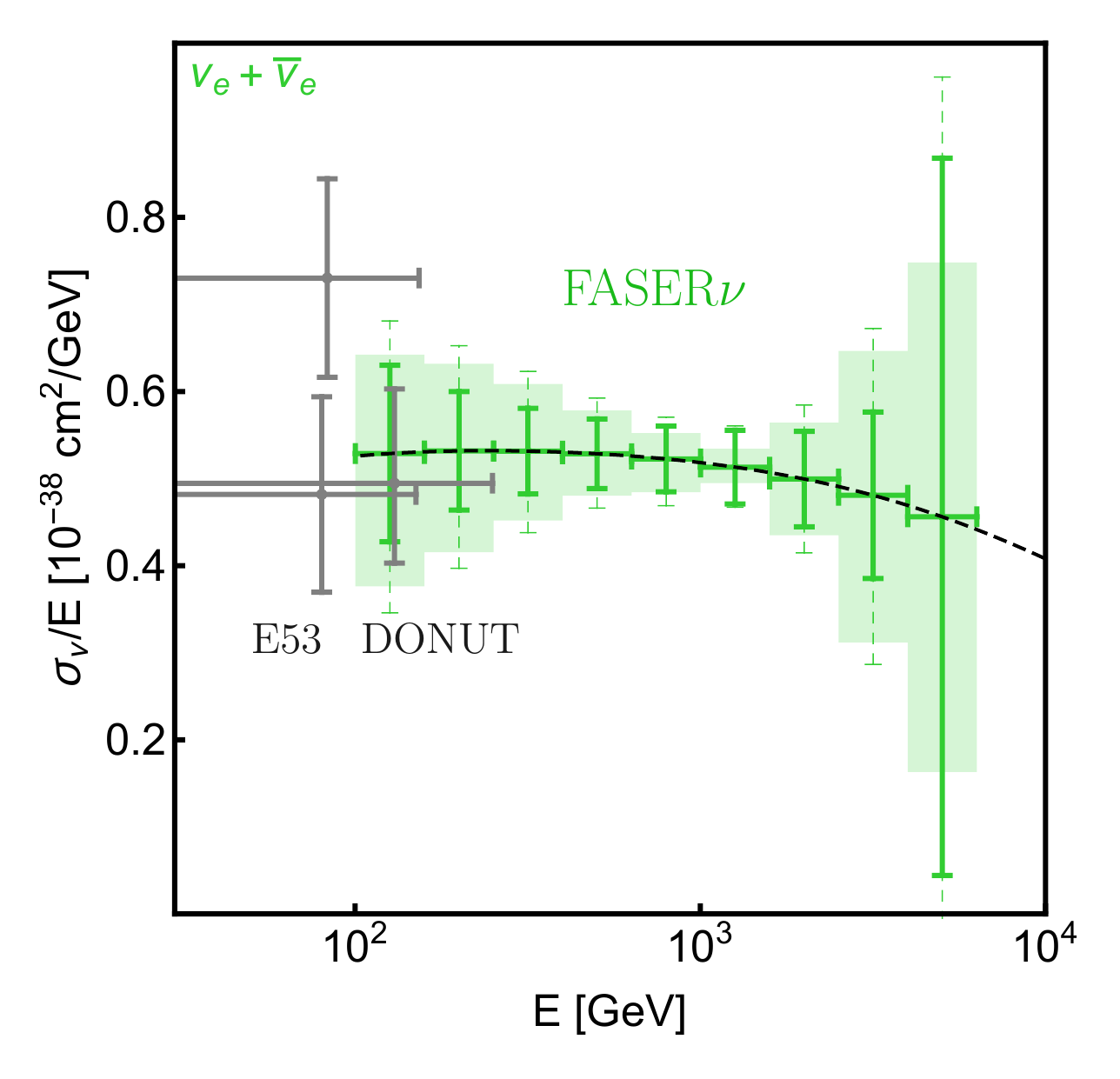}
\includegraphics[width=0.32\textwidth]{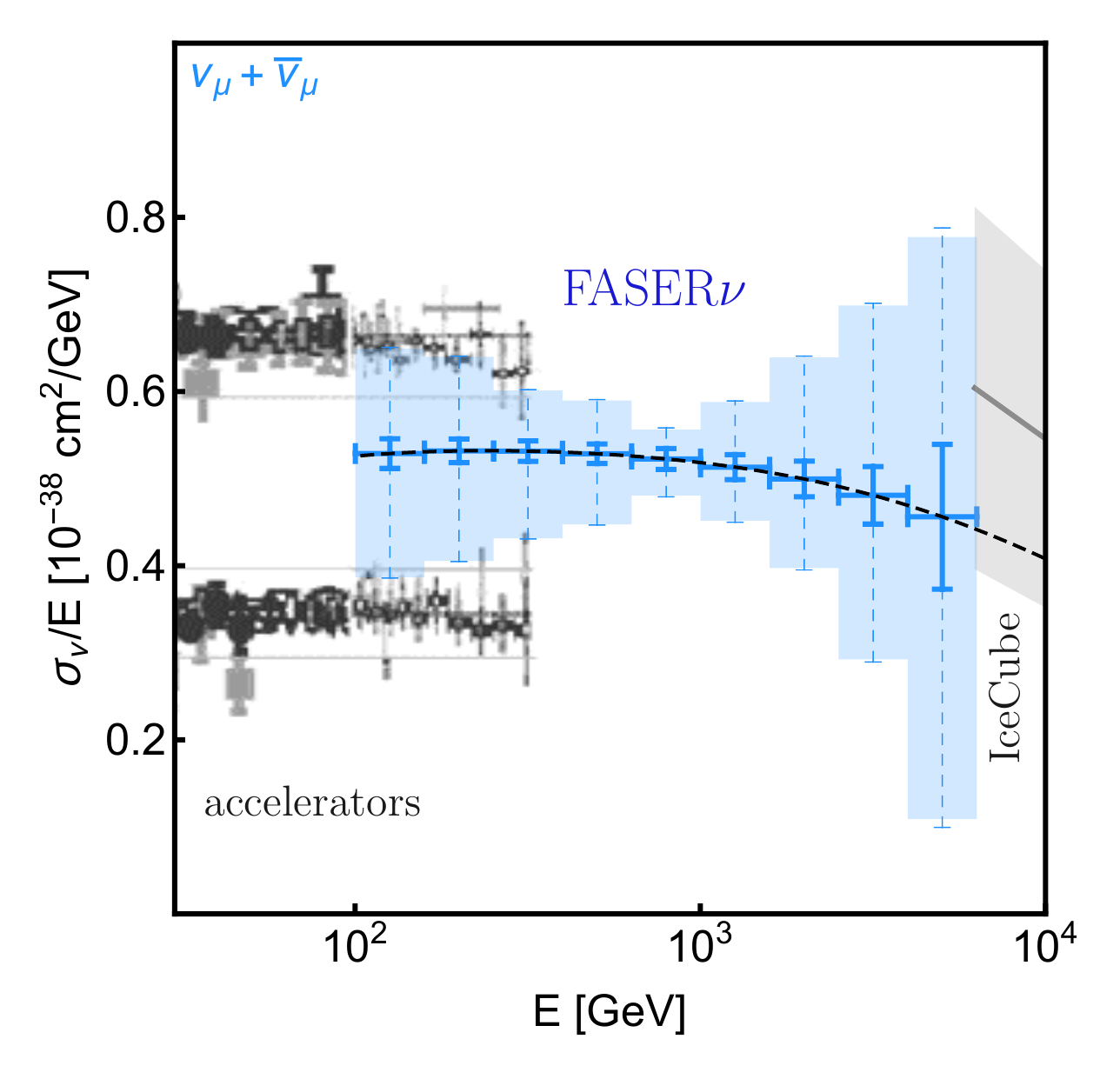}
\includegraphics[width=0.32\textwidth]{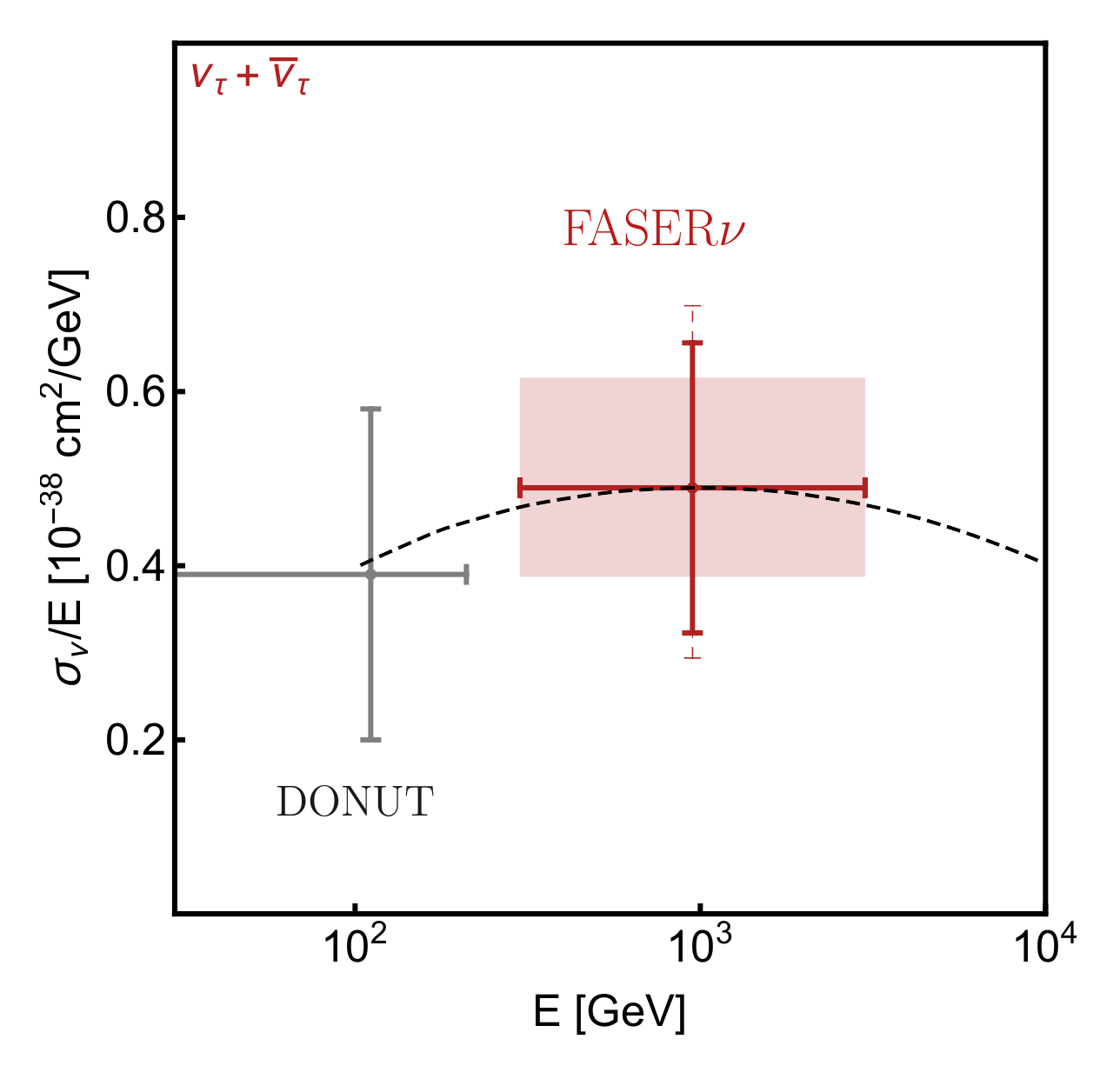}
\caption{\FASERnu's estimated $\nu$-nucleon CC cross section sensitivity for $\nu_e$ (left), $\nu_{\mu}$ (center), and $\nu_\tau$ (right) at Run 3 of the 14 TeV LHC with an integrated luminosity of $150~\ifb$ collected from 2021-23. Existing constraints are shown in gray. The black dashed curves are the theoretical predictions for the average deep inelastic scattering (DIS) cross section per tungsten-weighted nucleon. The solid error bars correspond to statistical uncertainties, the shaded regions show uncertainties from neutrino production rate corresponding to the range of predictions obtained from different MC generators, and the dashed error bars show their combination. 
}
\label{fig:physics-xsprojection}
\end{figure}

In addition to detecting collider neutrinos and anti-neutrinos of all three flavors and measuring their cross sections at higher energies than observed from any previous human-made source, \FASERnu can explore several other topics related to the physics of neutrino production, propagation, and interaction at the energy frontier: 
\begin{description}[leftmargin=0.16in]

\item [Tau Neutrino Detection] Of the seventeen particles in the standard model of particle physics, the tau neutrino is the least well measured. The DONuT and OPERA experiments have each observed about $10~\nu_\tau$ events~\cite{Kodama:2007aa, Agafonova:2018auq}, and these data sets provide the primary information about tau neutrinos at present. Additionally, SuperKamiokande and IceCube have recently reported higher statistics $\nu_{\tau}$ appearance in atmospheric oscillations~\cite{Li:2017dbe,Aartsen:2019tjl}, although with considerably larger uncertainties, resulting in a measurement with precision comparable to DONuT and OPERA. During LHC Run 3, \FASERnu will accumulate about $20~\nu_\tau$ CC interactions, of which about $13~\nu_\tau$ events are expected to be identified. This will significantly increase the world’s supply of reconstructed $\nu_\tau$ neutrinos and will allow them to be studied at much higher energies $E_\nu\sim\tev$.

\item [Event Shapes and Kinematics] Due to its high spatial resolution, the \FASERnu detector will be able to resolve the shape of each neutrino event, including, for example, the multiplicity and momentum distributions of charged particles. These event shapes will provide valuable input to tune MC tools used to simulate high-energy neutrino events, such as \textsc{Genie}.

\item [Heavy Flavor Associated Processes] In addition to the inclusive CC cross section, \FASERnu can also study specific exclusive neutrino interaction processes. One example is charm-associated neutrino interactions $\nu N \to \ell X_c +X$, which can be directly identified in \FASERnu due to the presence of the secondary charm decay vertex. Such measurements have previously been used to probe the strangeness content of the nucleon, the CKM matrix element $V_{cd}$, and charm fragmentation fractions~\cite{Goncharov:2001qe, Rabinowitz:1993xx, KayisTopaksu:2011mx}. Additionally, bottom-associated neutrino interactions $\nu N \to \ell X_b +X$, which are strongly CKM suppressed in the standard model (SM), might be sensitive to physics beyond the standard model (BSM), such as $W'$ bosons, charged Higgs boson, and leptoquarks at the TeV scale. 

\item [Neutrino Production and Hadronic Interaction Models] Aside from probing neutrino interactions, the neutrino measurements at \FASERnu can also be used to constrain neutrino production rates. Although the existing LHC detectors have great coverage of the central region, the production of particles in the very forward direction along the beam pipe is only poorly constrained. In this regime, the measurement of the neutrino flux and spectrum at \FASERnu will provide complementary constraints on neutrino production, which could help to validate and improve the underlying hadronic interactions models. Those models are used to simulate multi-parton interactions and underlying events at the LHC, and they are also used to simulate cosmic ray events. 
   
The measurement of forward neutrino production will also be a key input for high-energy neutrino measurements by large-scale Cherenkov observatories, such as IceCube~\cite{Aartsen:2016nxy}, ANTARES~\cite{Collaboration:2011nsa}, Baikal-GVD~\cite{Avrorin:2013uyc}, and KM3NeT/ARCA~\cite{Adrian-Martinez:2016fdl}. One of the main aims of these experiments is to search for high-energy astrophysical neutrinos. This is subject to atmospheric neutrino background with an important prompt component from the decays of heavy mesons. Such a component is expected to become dominant at the highest energies, but it has not yet been identified in the IceCube data~\cite{Aartsen:2016xlq}. A direct measurement of the currently poorly-constrained prompt flux by \FASERnu would provide important data, not only for IceCube, but also for all current and future high-energy neutrino telescopes.

\item [Sterile Neutrino Oscillations] Given the high neutrino energy $E_\nu \sim \tev$ and short baseline $L = 480~\m$, SM neutrino oscillation effects are expected to be negligible at \FASERnu. However, the presence of an additional sterile neutrino with a mass splitting of the order of $\Delta m^2 \sim (40~\ev)^2$ could lead to observable sterile neutrino oscillations in the \FASERnu neutrino spectrum. By searching for either appearance of extra neutrinos above the expected rate or disappearance below the expected rate, \FASERnu could put constraints on such sterile neutrino models. 

\end{description}

%%%%%%%%%%%%%%%%%%%%%%%%%%%%%%%%%%%%%%%%%%%%%%%%%%%%%%
%%%%%%%%%%%%%%%%%%%%%%%%%%%%%%%%%%%%%%%%%%%%%%%%%%%%%%
\section{Detector Location and Environment}
\label{sec:environment}
%%%%%%%%%%%%%%%%%%%%%%%%%%%%%%%%%%%%%%%%%%%%%%%%%%%%%%
%%%%%%%%%%%%%%%%%%%%%%%%%%%%%%%%%%%%%%%%%%%%%%%%%%%%%%

%%******************************************
\subsection{Detector Location}
%%******************************************

The \FASERnu detector will be placed in tunnel TI12 along the beam collision axis or line of sight (LOS) directly in front of the FASER detector. The CERN survey team has performed detailed measurements and mapped out the LOS in TI12, assuming no crossing angle between the beams at the ATLAS IP. In reality the LHC will operate with a small half-crossing angle of about $150~\murad$, which we will discuss below. 

The tunnel TI12 connects the LHC to the much shallower SPS, and therefore slopes steeply upwards as it leaves the LHC tunnel. Because of this geometry, the LOS is below the current tunnel floor as it enters the tunnel, and then emerges from the floor. To maximize the length of the FASER detector that can be centered on the LOS, the floor of TI12 will be lowered.  From the beginning~\cite{Ariga:2018pin}, the excavation plans have included a space at the front of the FASER spectrometer (toward the ATLAS IP) to accommodate \FASERnu; the current trench shape is shown in \figref{environment-trench}. Note that the front part of the trench, where \FASERnu will be located, has been widened and deepened relative to previous designs~\cite{Abreu:2019yak,Ariga:2018pin}, as shown in \figref{trench}. This enlargement was dictated by civil engineering considerations that required the redirection of a drainage pipe, but the additional space has the added benefit of providing more room for \FASERnu installation and emulsion replacement. It has been checked that the concrete in TI12 is strong enough to hold the \FASERnu detector. The minimum distance from the nominal LOS to the side wall is $150~\mm$, which defines the maximum width of \FASERnu (including the mechanical structure) to be $300~\mm$ if \FASERnu is centered around the LOS. The length of the trench in front of the FASER spectrometer is 1350 mm.  The trench excavation is scheduled to be completed by the end of March 2020. 

\begin{figure}
    \centering
    \includegraphics[width=0.46\textwidth]{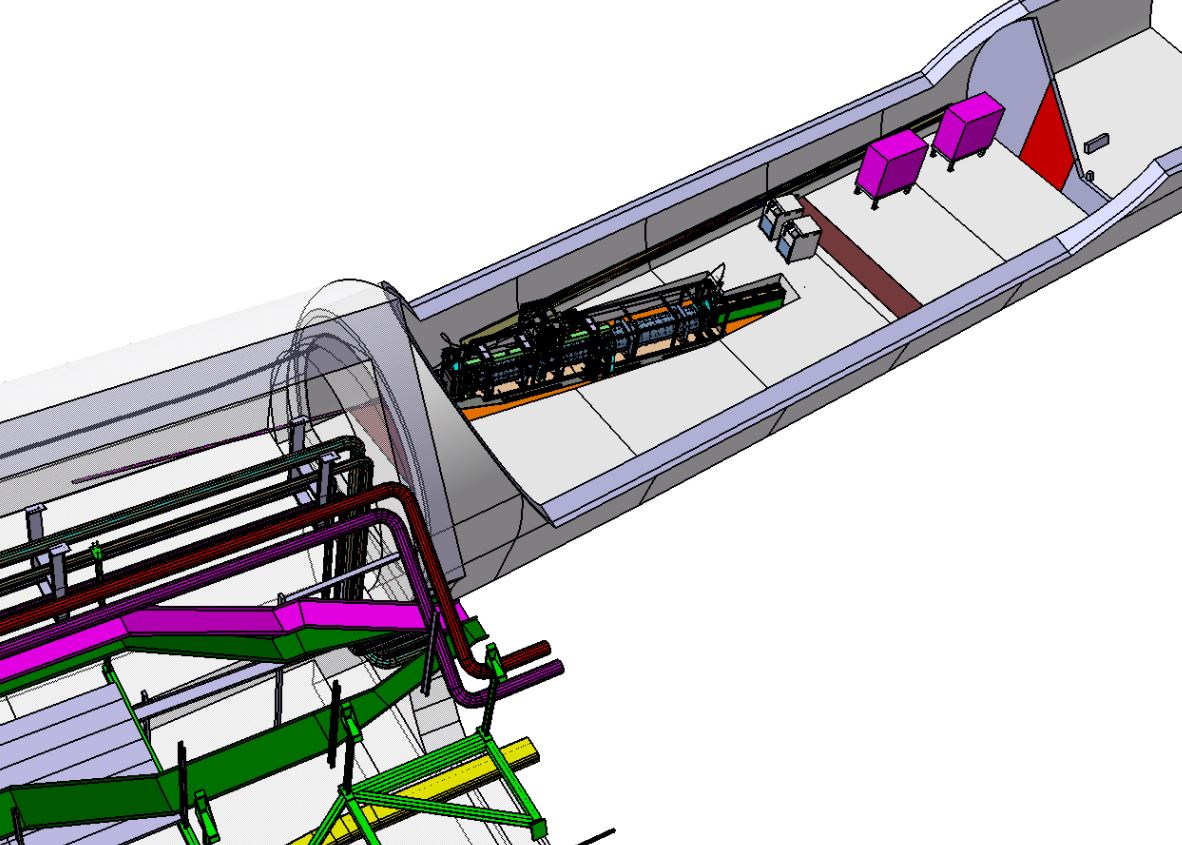}
    \includegraphics[width=0.53\textwidth]{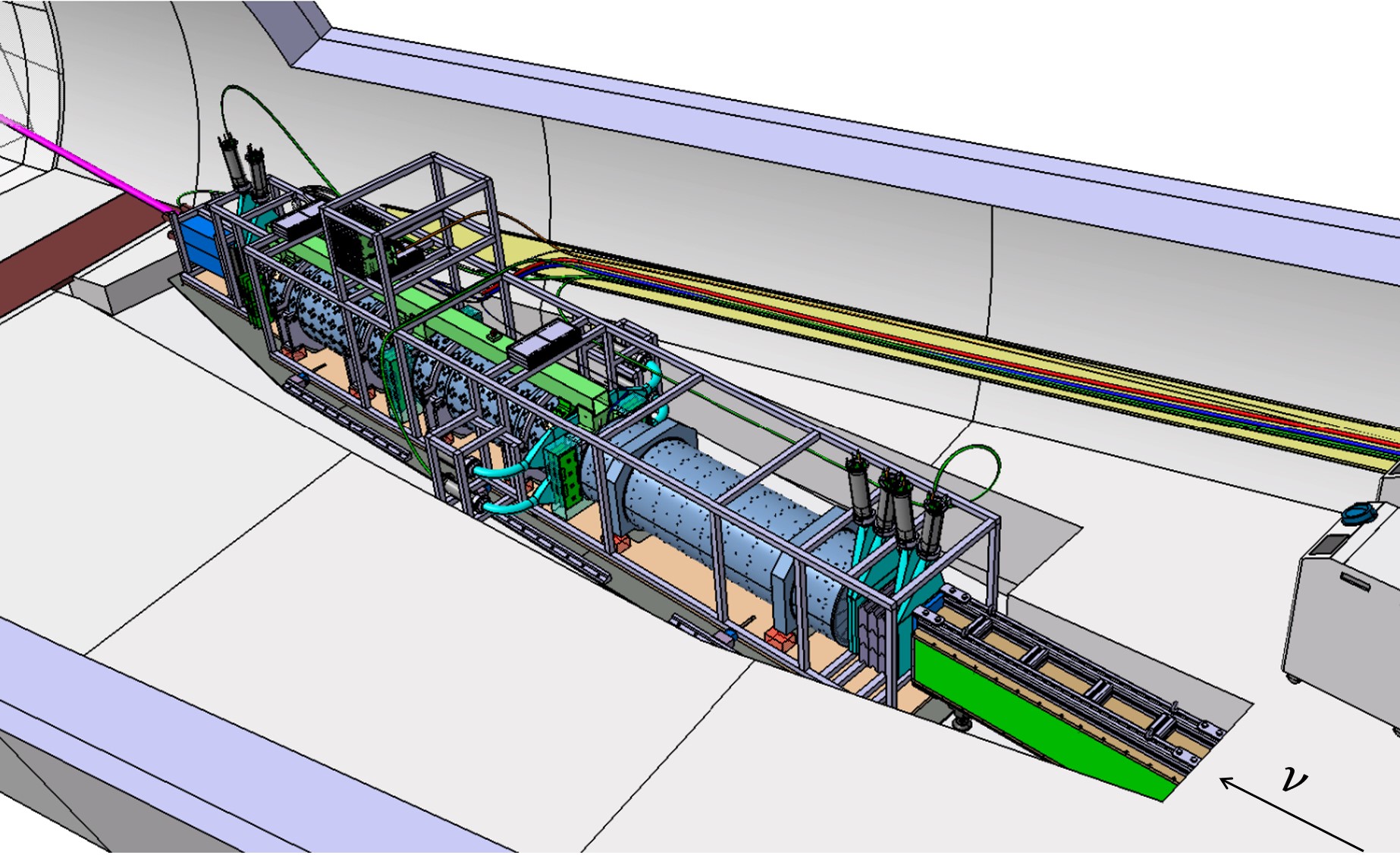}
    \caption{\textbf{Left}: Location of FASER and \FASERnu in the UJ12/TI12  region, 480 m downstream from the ATLAS IP. Particles from the ATLAS IP arrive from the right. \textbf{Right}: A view of FASER and \FASERnu in the trench being excavated in TI12 to allow them to be located along the line of sight (LOS). \FASERnu, shown in green, is located at the front of FASER (toward the ATLAS IP) and consists of emulsion and tungsten layers and their support structure, which together occupy a volume of $30\,\cm \times 30\,\cm \times 1.35\,\m$.   
    }
    \label{fig:environment-trench}
\end{figure}

\begin{figure}
    \centering
    \includegraphics[height=5cm]{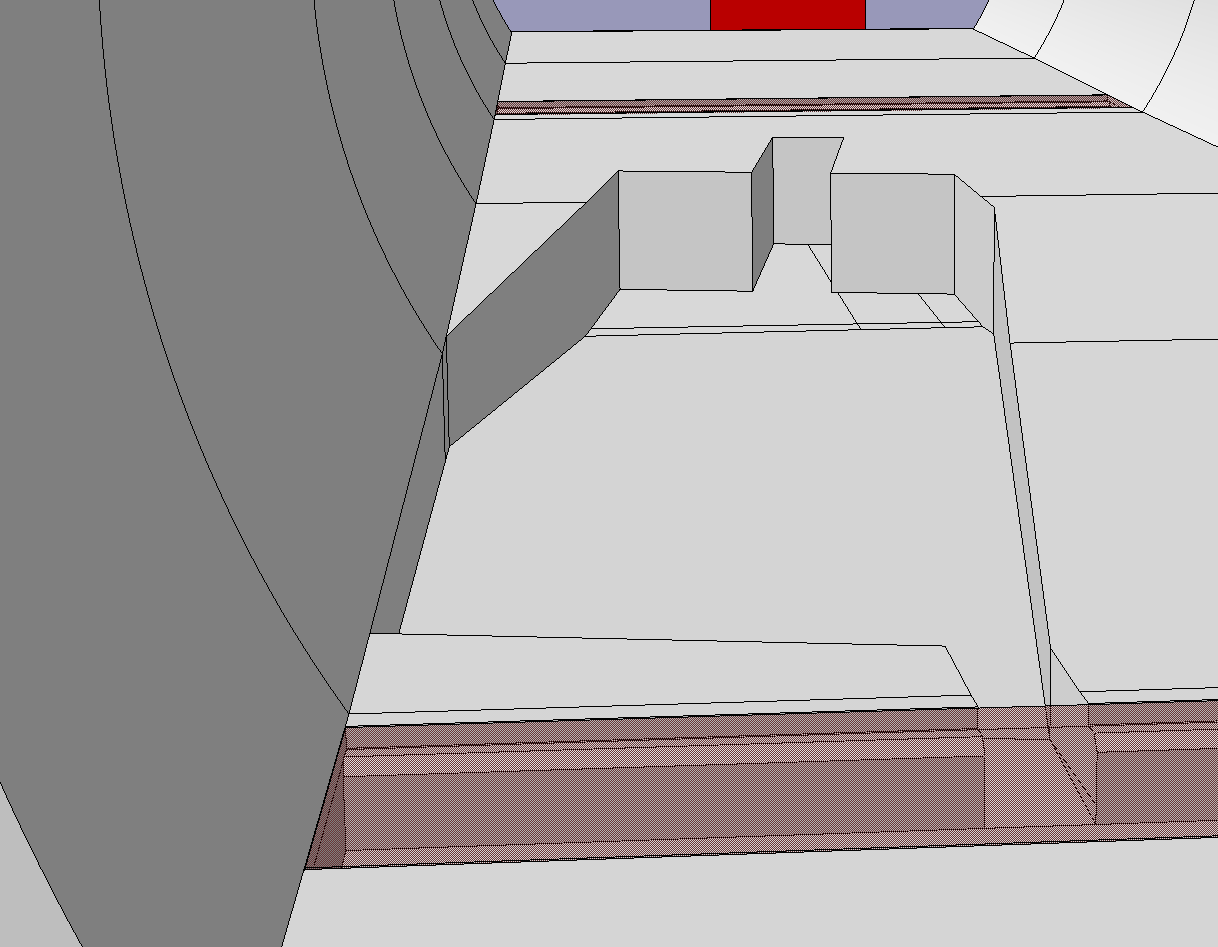} \hspace*{.4in}
    \includegraphics[height=5cm]{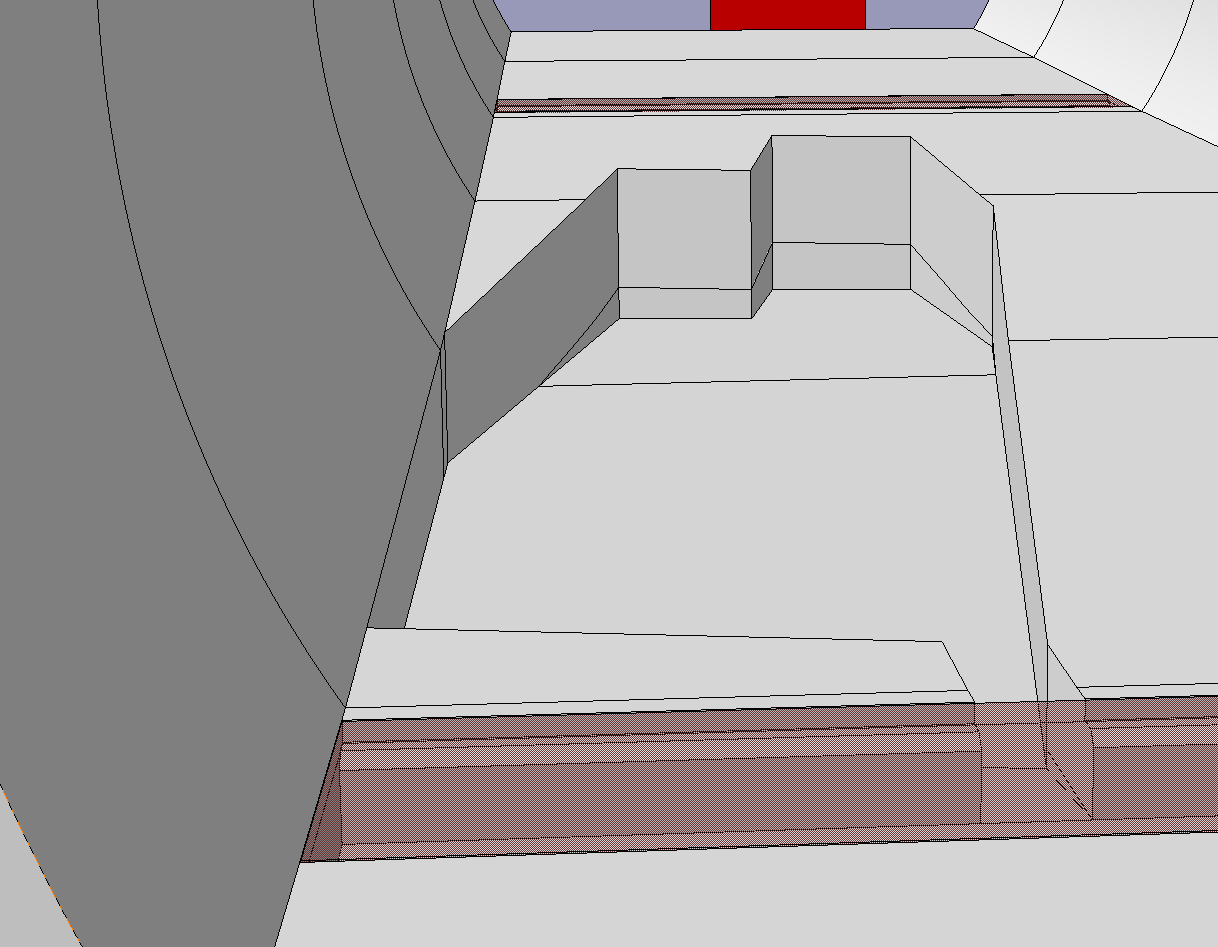} \vspace*{.2in} \\
    \includegraphics[height=4.5cm]{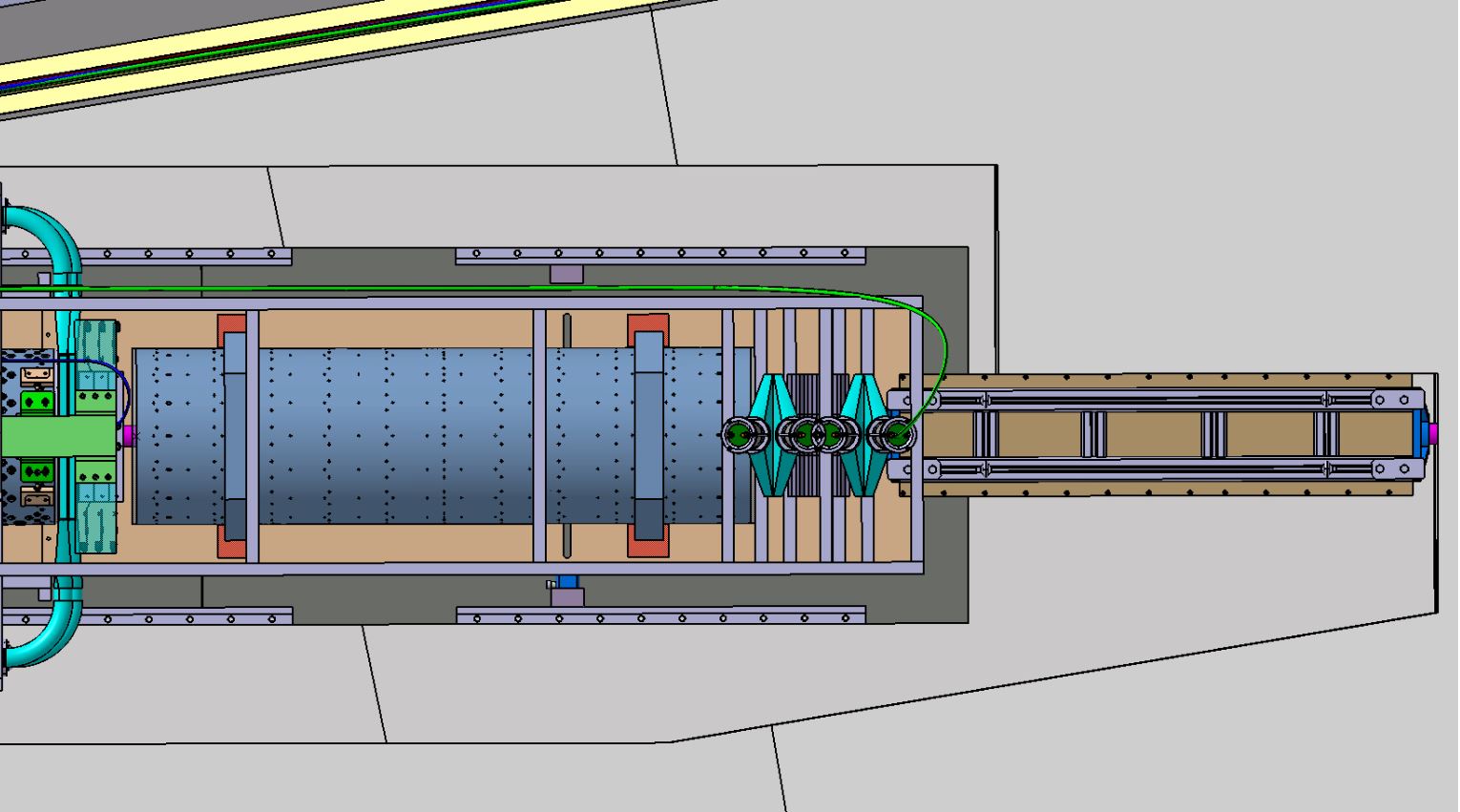}
    \includegraphics[height=4.5cm]{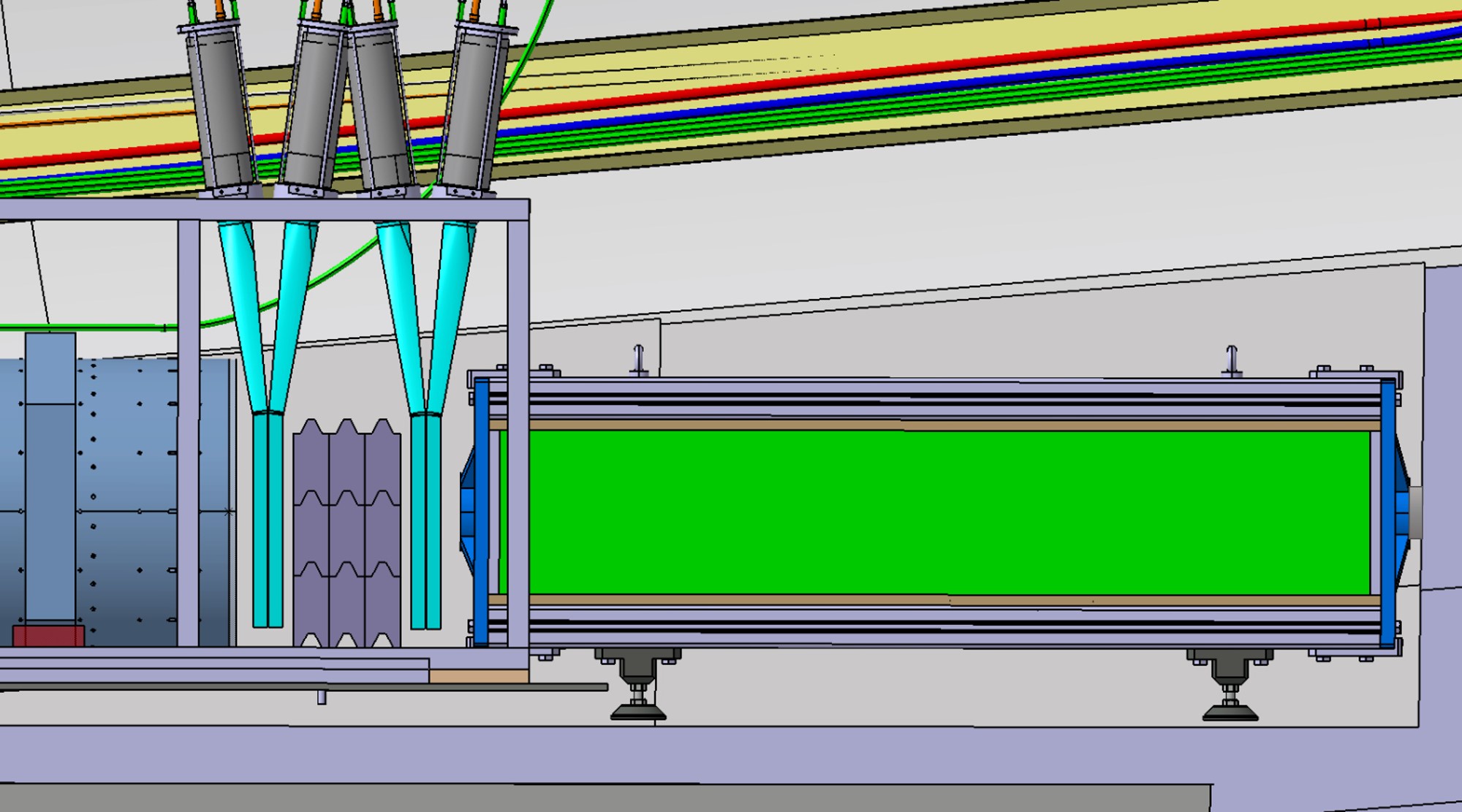}
    \caption{\textbf{Top left}: The original plans for the FASER trench.   \textbf{Top right}: The current shape of the FASER trench.  Civil engineering considerations have required the drainage pipe to be redirected, and as a consequence, the front of the trench has been widened and deepened, providing more room for \FASERnu. The trench width around the FASER spectrometer is unchanged. \textbf{Bottom}: Top and side views of the front of the trench.  The box on the right indicates the location of \FASERnu, which contains emulsion and tungsten layers and occupies a volume of 30 cm $\times$ 30 cm $\times$ 1.35 m. }
    \label{fig:trench}
\end{figure}

The FASER location is currently being prepared with lighting and power, and a passarelle (stairs) and support structures are being put in place to safely transport detector components over the LHC.  \FASERnu will benefit from all of the infrastructure plans already underway to prepare TI12 for the FASER spectrometer.

%%******************************************
\subsection{Run 3 Operation and Beam Configuration}
\label{sec:operation}
%%******************************************

As mentioned above, the LHC runs with a beam half-crossing angle of about $150~\murad$ at the ATLAS IP to avoid long range beam-beam effects and parasitic collisions inside the common beam pipe.  At the location of \FASERnu, a half-crossing angle of $150~\murad$ corresponds to a shift of the beam collision axis of $7.2~\cm$ relative to the nominal LOS, which assumes no crossing angle. Given that \FASERnu's cross sectional area is $25~\cm \times 25~\cm$, if \FASERnu is centered on the nominal LOS, the actual LOS will pass through \FASERnu for half-crossing angles of $150~\murad$ in any direction.  Our simulations have shown that, when keeping \FASERnu centered around the nominal LOS, such shifts of the LOS will reduce the neutrino flux, and hence the interaction rate, by not more than 10\% for muon neutrinos and less than that for the other neutrino flavors. 

 During LHC Run 2, it was decided to flip the crossing angle direction periodically (e.g., once per year).  With the current trench design, \FASERnu can be shifted to track crossing angles of this size in either vertical direction. A change to a horizontal crossing angle would always point the LOS away from the LHC, and in this case the detector would not be able to be centred fully on the LOS.

The beam crossing angle plans for Run 3 have not yet been finalized, but the expected luminosity and crossing plane (direction) for the different years of running in Run 3 are shown in Table~\ref{table:operation}.  The table also shows how many emulsion detectors will be needed for each year to keep the track multiplicity in the detector at a manageable level. We stress that changes to the crossing plane (horizontal or vertical), the direction of the crossing angle (e.g., up or down), and the size of the crossing angle are all possible. However, in all cases under consideration, the crossing angle values will be similar to or smaller than those used in Run 2.  We will therefore align each emulsion detector with the actual LOS whenever possible. 

\begin{table}[tbp]
\centering
\begin{tabular}{|c|c|c|c|}
\hline
\hline
    \multirow{2}{*}{{\bf Year}}& 
    \multirow{2}{*}{{\bf Crossing Plane}}&     
    \multirow{2}{*}{{\bf Luminosity}}&     
    {\bf Sets of } \\
    & & &{\bf Emul.~Detectors}  \\
\hline
\hline
    2021   & vertical (down) & $10 - 20~\ifb$ & 1\\
    2022   & vertical (down) & $80 - 100~\ifb$ & 3\\
    2023   & \, vertical / horizontal (TBD)\, & $80 - 100~\ifb$ & 3\\
    {[}2024{]}   & {[}horizontal{]} & {[}$80 - 100~\ifb${]} & {[}3{]} \\
    \hline
    Total  & & $170 - 220~\ifb$ & 7\\
     \, {[}Total incl. 2024{]} \,  & & \, {[}$250 - 320~\ifb${]} \, &{[}10{]} \\
\hline
\hline
\end{tabular}
\caption{Expected beam operating parameters during Run 3. In the case of a horizontal cross angle, this will always shift the LOS away from the LHC. The current schedule includes running from 2021-23, but there are ongoing discussions to run also in 2024. The last column shows how many sets of emulsion layers are needed each year to keep the track multiplicity in the detector at a manageable level.
}
\label{table:operation}
\end{table}

Additional effects related to beam divergence and the LHC filling scheme have been discussed in detail in the FASER Technical Proposal~\cite{Ariga:2018pin} and are either small or negligible. 

%%******************************************
\subsection{Temperature and Humidity}
%%******************************************

\begin{figure}
\centering
\includegraphics[width=0.49\textwidth]{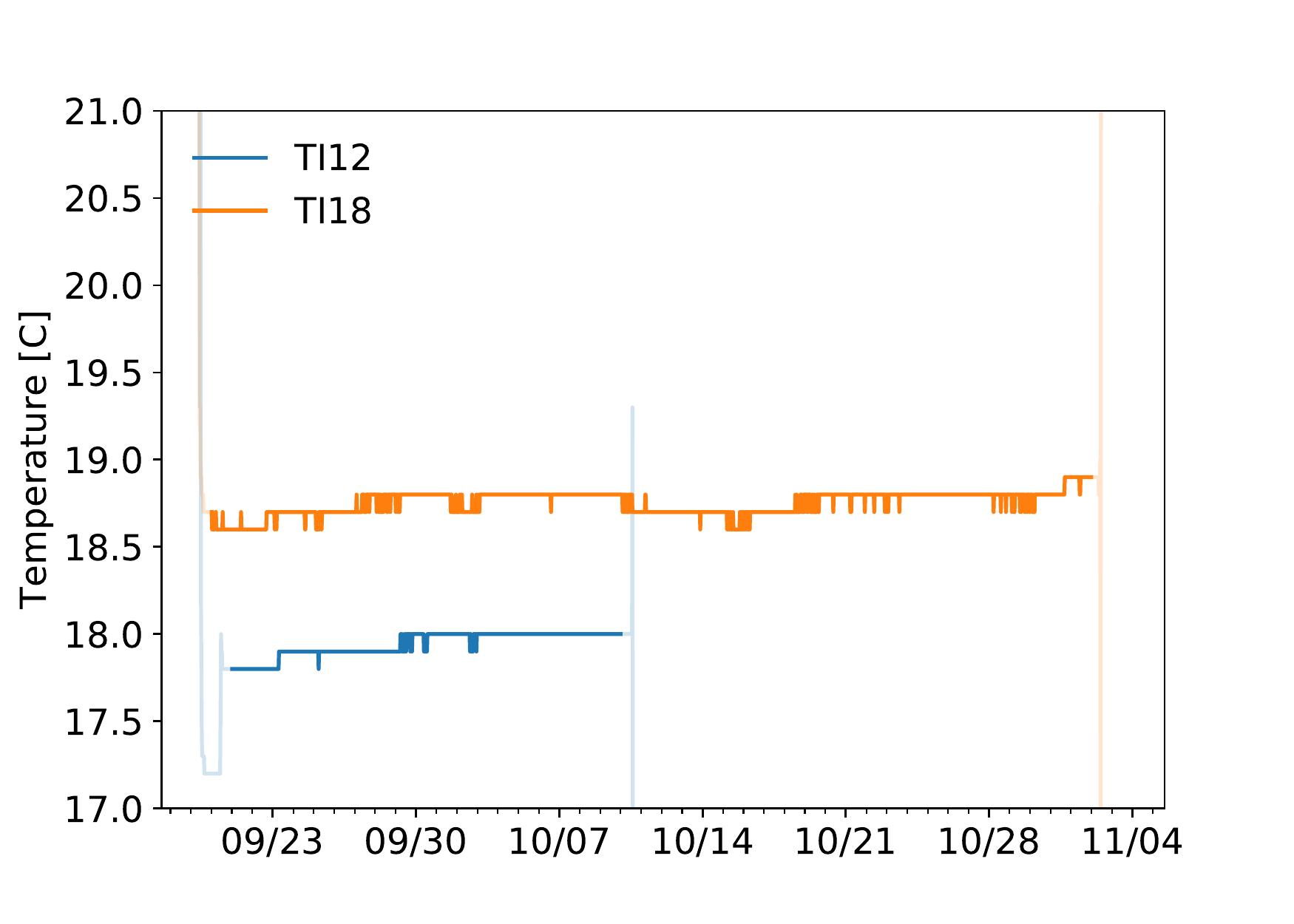}
\includegraphics[width=0.49\textwidth]{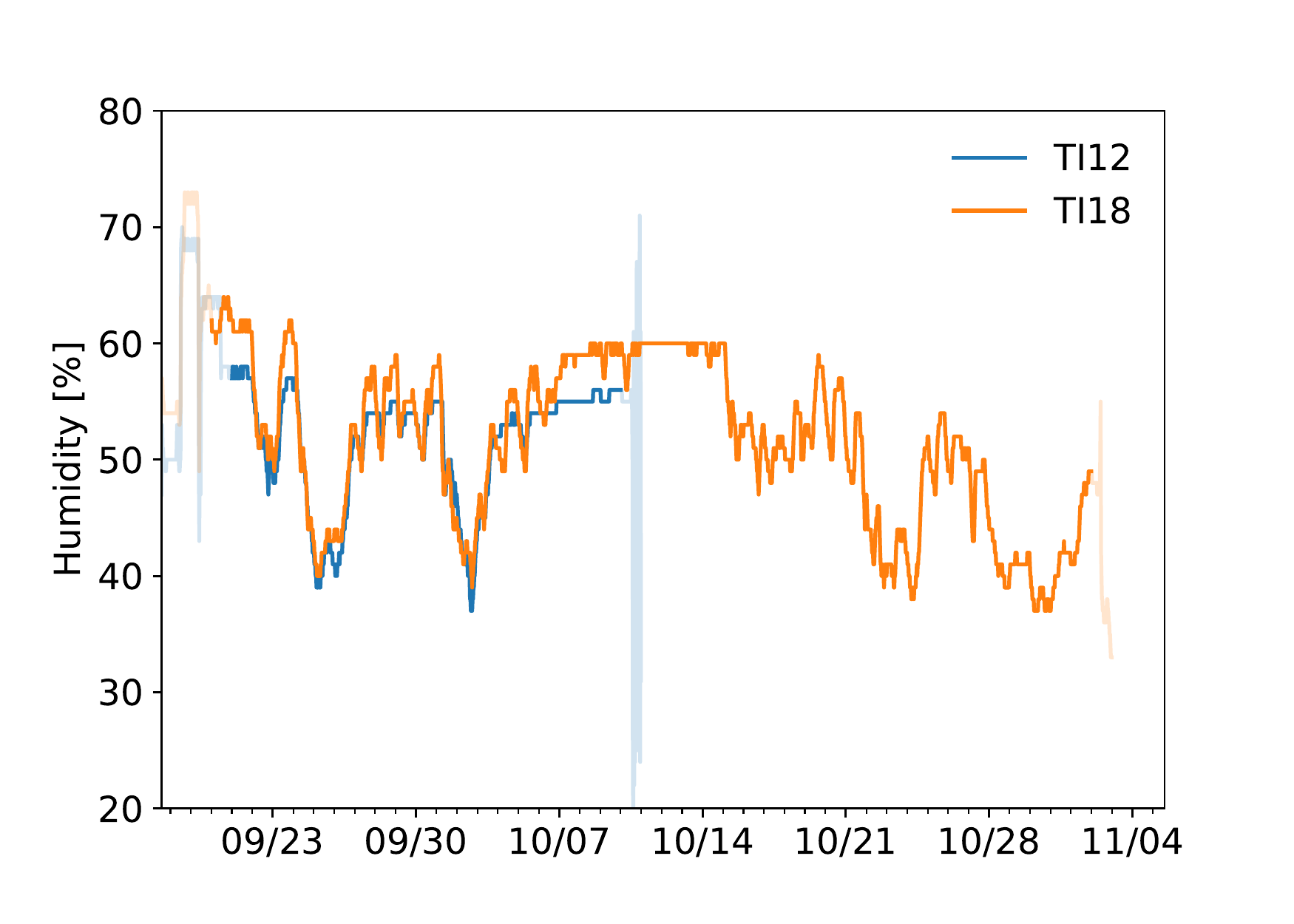}
\caption{Measured temperature (left) and humidity (right) during the period when the emulsion detector was installed on the LOS in the tunnels TI12 and TI18. The lighter parts of the curves correspond to the time before placing the detector in (after removing the detector from) the tunnel. }
\label{fig:environment-temp}
\end{figure}

\begin{table}[]
\centering
\begin{tabular}{|c|l|c|c|}
    \hline
    \hline
    Location & Measurement Time & Average [\si{\degree C}]& Deviation (rms) [\si{\degree C}]\\
    \hline
    TI12 & 2018/9/20 - 2018/10/10 & 17.94 & 0.08 \\
    \hline
    TI18 & 2018/9/20 - 2018/11/2  & 18.75 & 0.07 \\
    \hline
    \hline
\end{tabular}
\caption{Temperature measurements in tunnels TI12 and TI18 during the pilot emulsion detector runs in 2018.}
\label{table:environment-temp}
\end{table}

\begin{figure}
\centering
\includegraphics[width=\textwidth]{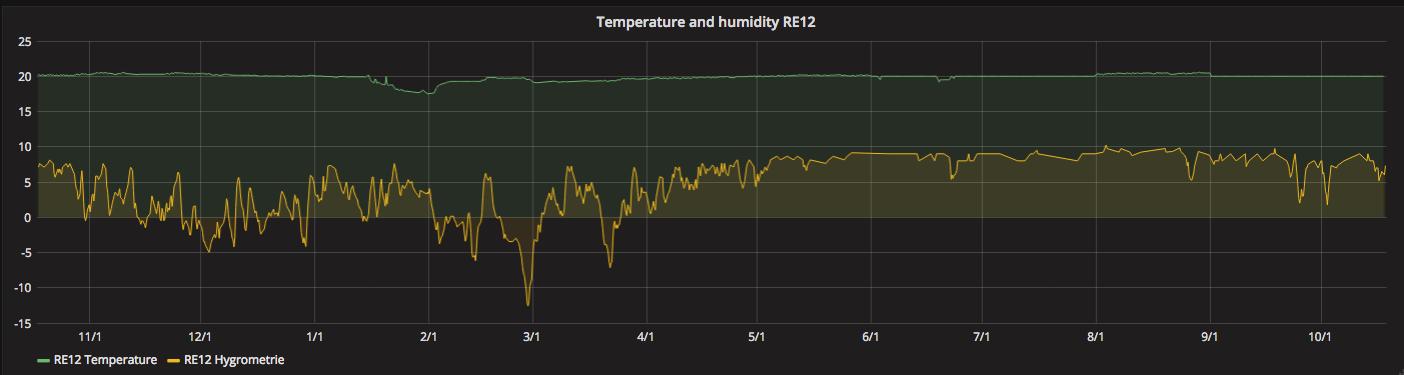}
\caption{Year-long temperature (green) and dew point (yellow) measurements in 2018 in the LHC tunnel close to TI12 (in the UJ12 region) from LHC monitoring.}
\label{fig:environment-temp-2018}
\end{figure}

As part of the {\it in situ} emulsion detector measurements performed in 2018 to measure the charged particle flux, temperature and humidity sensors of type T\&D TR-72wf were installed in the TI12 and TI18 tunnels on the LOS. (TI18 is the tunnel that is also 480 m from the ATLAS IP, but on the other side of ATLAS from TI12.) The measured temperature and humidity as functions of time are shown in \figref{environment-temp}. During this period the temperature in TI12 was constant at about 18$^\circ$C with temperature variations of about $0.1^\circ$C; see \tableref{environment-temp}. The humidity varied between 40\% and 60\%, with a value around 55\% for most of the time. \Figref{environment-temp-2018} shows the variation over a longer timescale of about a year in 2018, but using the LHC environmental monitoring system sensors that are closest to TI12. The temperature in the LHC tunnel during LHC operation is also very stable over the longer time scale. 

Additional heat could originate from operating the FASER detector. The effect of FASER on the tunnel temperature is not known yet, but will be monitored by four temperature sensors around the \FASERnu detector. To reduce the impact of possible heat from the chillers, it was decided to move the chillers further away from \FASERnu. 

%%******************************************
\subsection{Radiation Levels}
%%******************************************

The radiation level has been simulated by FLUKA and measured by battery-operated radiation monitoring devices (BatMons) in the TI12 and TI18 tunnels, with fully consistent results. As discussed in the FASER Technical Proposal~\cite{Ariga:2018pin}, non-radiation-hard electronics can be used at the location of FASER/\FASERnu with an estimated dose less than $5\times 10^{-3}$~Gy per year and a $1~\mev$ neutron equivalent fluence of less than $5\times 10^7$ per year. For thermal neutrons, the flux of $3\times 10^6~\cm^{-2}$ from the simulations agrees well with the measured one of $4\times 10^6~\cm^{-2}$, and this flux is at a low level that will not affect emulsion detectors as demonstrated by pilot measurements. 
In fact, the \textit{in situ} measurements with emulsion detectors in 2018 showed that the dominant component of charged particles was not low-energy particles due to neutrons.

%%******************************************
\subsection{Particle Fluxes and Backgrounds}
\label{sec:bg}
%%******************************************

The expected particle fluxes passing through FASER and \FASERnu have been estimated with dedicated extensive FLUKA simulations~\cite{Ferrari:2005zk,Battistoni:2015epi} performed by the CERN STI group~\cite{FLUKAstudy}. These studies include high-energy particles produced at the ATLAS IP, $480$~m away from the detector, and also particles produced in beam-gas collisions and proton-loss-induced showers in the dispersion suppressor region closer to \FASERnu. The results of these studies have been summarized in the FASER LOI~\cite{Ariga:2018zuc} and Technical Proposal~\cite{Ariga:2018pin}. In addition, further FLUKA simulations have been carried out for \FASERnu to determine the flux of particles produced by high-energy muons interacting in the rock in front of the detector, as well as in the detector itself. These have been discussed in the \FASERnu LOI~\cite{Abreu:2019yak}.

As noted above, to validate these numerical simulations, measurements have also made during LHC Run 2 with pilot emulsion detectors installed during Technical Stops in 2018 in both tunnels TI12 and TI18. The heterogeneous structure of these detectors, which employed emulsion films interleaved with tungsten layers in their downstream sections, made it possible to separately measure both the flux of all particles with energies above $50$~MeV and the flux of all particles with higher energies $E\gtrsim 1$~GeV. In addition, an active monitoring device (a TimePix3 Beam Loss Monitor~\cite{TimePix}) was installed to correlate the rate of detected particles with beam conditions, showing that as expected the rate of high energy particles is directly correlated with the instantaneous luminosity at the ATLAS IP.

The results of these simulations and measurements show that the flux of high-energy particles passing through FASER/\FASERnu is dominated by particles coming from the ATLAS IP that are correlated with the corresponding instantaneous luminosity. This particle flux has been discussed extensively in the aforementioned references. Here we briefly summarize the most important findings that are relevant for \FASERnu.

%%******************************************
\subsubsection{Muons and the Related Electromagnetic Component} 
%%******************************************

Other than neutrinos, by far the dominant flux of high-energy particles passing through \FASERnu are muons produced at the ATLAS IP or further downstream. According to simulations, muons and the related electromagnetic component correspond to more than $99.999\%$ of particles with $E>100~\gev$. The flux of muons with $E_\mu>10~\gev$ predicted in simulations, $\Phi\simeq 2\times 10^4~\fb/\cm^2$, agrees remarkably well with the one measured within $10~\mrad$ around the collision axis, $\Phi\simeq (1.9\pm 0.2)\times 10^4~\fb/\cm^2$. This corresponds to about $N_\mu\simeq 2\times 10^9$ muons crossing \FASERnu for the LHC Run 3 integrated luminosity of $L=150~\ifb$, with roughly equal numbers of positive and negative muons, as predicted by simulations. On the other hand, because of the complicated impact of the LHC optics on muon trajectories on their way to \FASERnu, the high-energy part of the muon spectrum is dominated by $\mu^-$. For the same reason, the flux of muons passing through TI12 is not uniformly distributed in the transverse plane. In particular, it is much larger for off-axis positions, while \FASERnu, placed along the LOS, is near a local minimum of the flux. This is illustrated in the left panel of \figref{muonflux}, which shows the flux of negative muons obtained in the FLUKA simulations performed by the CERN STI group~\cite{FLUKAstudy}.

\begin{figure}[tpb]
\centering
\includegraphics[clip,trim = 1cm 0 3cm 0 , width=0.49\textwidth]{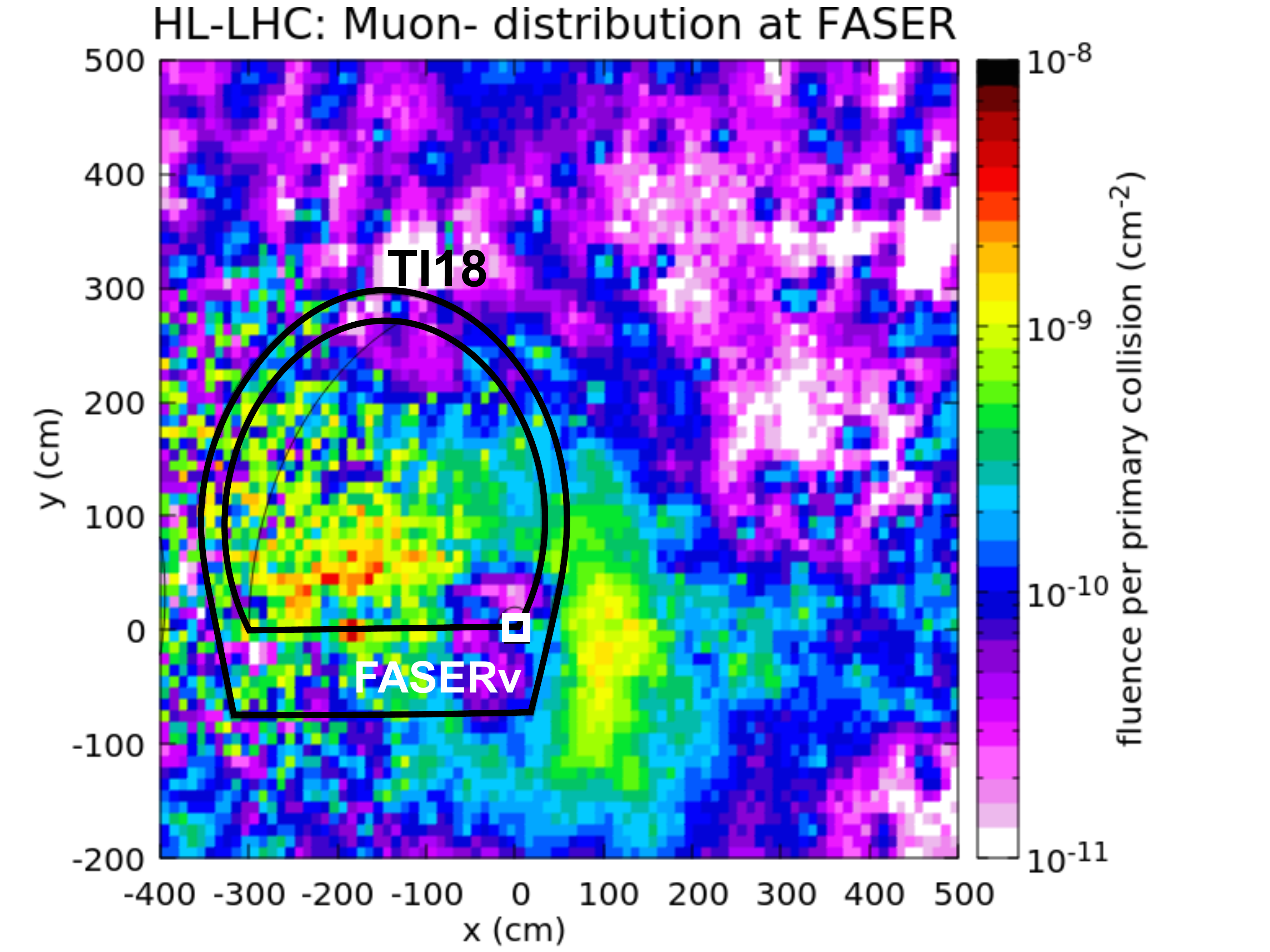}
\includegraphics[width=0.49\textwidth]{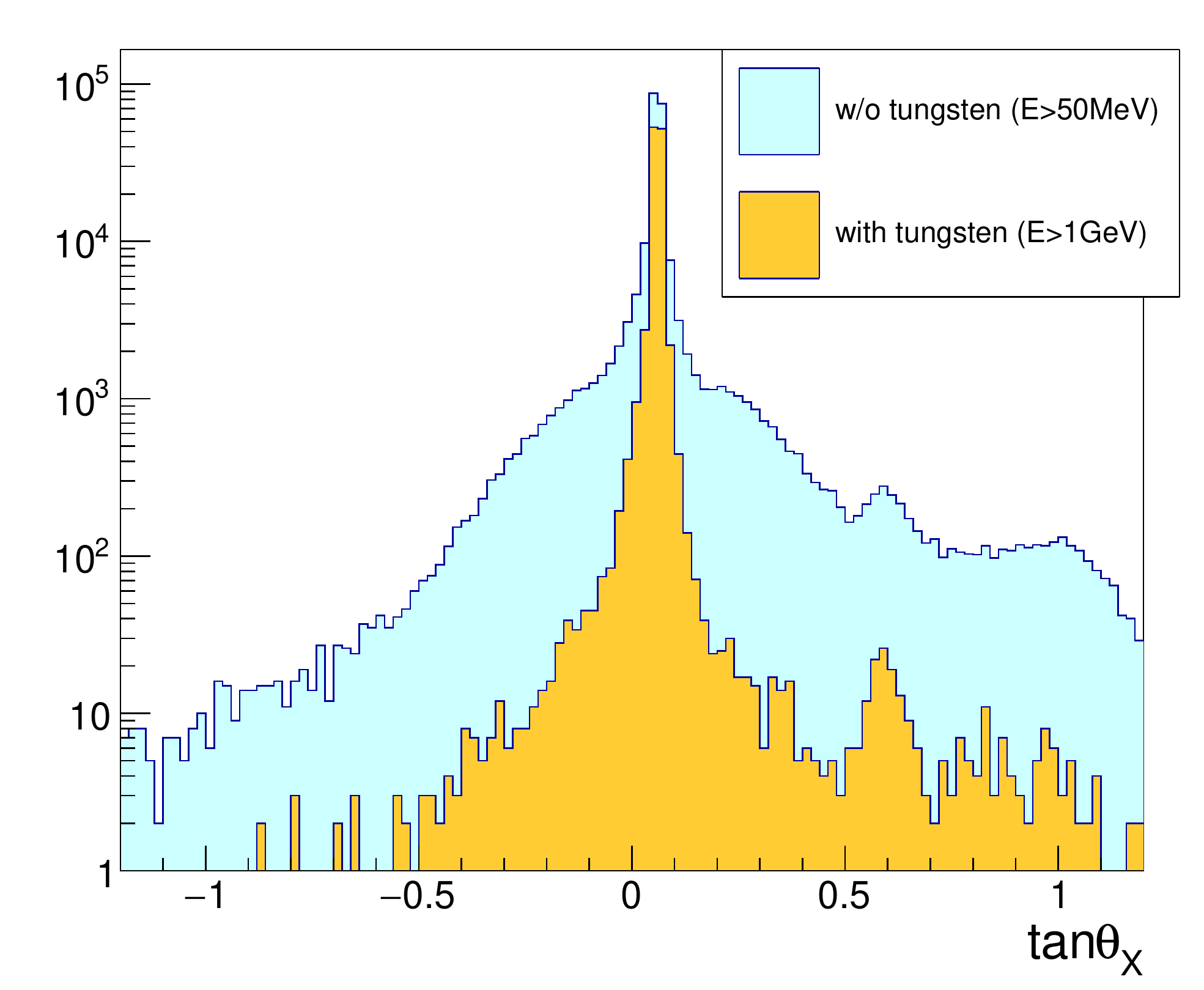}
\caption{\textbf{Left}: The distribution of negative muons crossing the tunnel TI18, which is in a symmetric position on the opposite side of the ATLAS IP with respect to \FASERnu. The equivalent position of \FASERnu along the LOS is at the origin of the coordinate system and is indicated by a white square. \textbf{Right}: The angular distributions of charged particles measured by the emulsion films with and without tungsten plates, corresponding to energy cutoffs of about 1 GeV and 50 MeV due to multiple Coulomb scattering, respectively. From Ref.~\cite{Ariga:2018pin}.}
\label{fig:muonflux}
\end{figure}

The total measured flux of charged particles with energies above $50~\mev$ is $\Phi\simeq (3\pm 0.3)\times 10^4~\fb/\cm^2$. The angular distribution of these charged particles, as measured by the pilot emulsion detectors, is shown in the right panel of \figref{muonflux}.  The pilot detectors had a cross sectional area of about $10~\cm\times 25~\cm$, centered on the LOS. As can be seen, most of the charged particles are in a narrow peak coming from the direction of the ATLAS IP, and the peak is especially narrow if one considers only energies above 1 GeV.  

Importantly, single high-energy through-going muons are not a background for neutrino searches, as their mis-identification rate is very low, dropping to a level below $10^{-10}$ after the first $1~\cm$ of the detector. They are, however, useful for obtaining precise alignment between the emulsion films in different layers, and, of course, the muon track density governs the frequency with which the emulsion layers must be replaced. 

The muons also produce other particles.  In particular, muon-induced photons, produced both in the rock in front of the detector and in the detector volume itself, are the second largest particle flux in \FASERnu, as shown in \tableref{BGtable}. However, these photons initiated EM showers, either in the front layers of the detector or in close vicinity to the parent muon inside the detector volume, and so they can be distinguished from neutrino-induced vertices based on signal characteristics discussed below in \secref{offline_analysis}.

\begin{table}[t]
\centering
\begin{tabular}{|c|c|c|c|c|c|}
\hline
\hline
\multirow{2}{*}{Particle} & \multicolumn{4}{c|}{Expected number of particles passing through \FASERnu} \\
\cline{2-5}
 & $E>10$~GeV &  $E>100$~GeV &  $E>300$~GeV & $E>1$~TeV \\ 
\hline
Neutrons $n$ & $27.8$k / $138$k & $1.5$k / $11.5$k & $150$ / $1.1$k & $2.2$ / $42$ \\
\hline
Anti-neutrons $\bar{n}$ & $15.5$k / $98$k & $900$ / $9$k & $110$ / $1.5$k & $2.8$ / $46$ \\
\hline
$\Lambda$ & $5.3$k / $36$k & $390$ / $4.1$k & $39$ /  $800$ & $0.9$ / $58$\\
\hline
Anti-$\Lambda$ & $3.4$k / $31$k & $290$ / $3.5$k & $31$ / $200$ & $0.6$ / $14$ \\
\hline
$K^0_S$ & $1.3$k / $30$k & $240$ / $6.8$k & $52$ / $390$ & $1.8$ / $6.2$ \\
\hline
$K^0_L$ & $1.6$k / $31$k & $270$ / $5.7$k & $55$ / $500$ & $1.2$ / $18$ \\
\hline
$\Xi^0$ & $240$ / $1.3$k & $13$ / $190$ & $2.3$ / $12$ & $0.1$ / $-$ \\
\hline
Anti-$\Xi^0$ & $150$ / $1$k & $10$ / $200$ & $1.4$ / $19$ & $-$\\
\hline
Photons $\gamma$ & $2.2$M / $62$M & $160$k / $16.3$M & $38.2$k / $6.3$M & $5.9$k / $1.1$M \\
\hline
\hline
$\nu_\mu+\bar\nu_\mu$ (signal int.) & 23.1k& 20.4k & 13.3k & 3.4k \\
\hline
\hline
\end{tabular}
\caption{The expected number of $\mu^-$-induced particles passing through \FASERnu in LHC Run 3 with an integrated luminosity of $150~\ifb$, as estimated by a dedicated FLUKA study.  In each entry, the first number is the number of particles emerging from the rock in front of \FASERnu, and the second is the number of particles produced in muon interactions in the tungsten plates in \FASERnu.  $2\times 10^9$ muons are expected to pass through \FASERnu in Run 3. Note that the statistical uncertainties of the numbers presented in this table can reach even factors of a few, especially for the less abundant neutral hadrons.
\label{table:BGtable}}
\end{table}

%%******************************************
\subsubsection{Muon-induced Neutral Hadrons} 
%%******************************************

Photon-nuclear interactions of muons in the rock in front of \FASERnu, as well as within the detector volume, can produce secondary neutrons and other neutral hadrons that go through the detector. Most of these hadrons will interact in the tungsten layers of \FASERnu with the hadronic interaction length $\lambda_{\textrm{int}}\sim 10~\cm$. These interactions, however, will typically take place away from the parent through-going muon track and can, therefore, more easily mimic neutrino interactions. The relevant numbers of expected particles are shown in \tableref{BGtable} for several energy ranges: $E>10, 100, 300, 1000~\gev$. As can be seen, we expect up to $\mathcal{O}(10^5)$ neutral hadrons with $E> 10~\gev$, but the number drops rapidly with increasing energy. For $E > 300~\gev$, the number of neutral hadrons drops below the total number of expected neutrino interactions, and the predicted signal to background ratio grows rapidly for higher energy cuts. This is also illustrated in the left panel of \figref{BGspectrum}, where we show the spectra of the neutral hadron background and the neutrino interaction signal obtained in the FLUKA simulations. In the right panel of \figref{BGspectrum}, we show the angular spectrum of neutrons emerging from the rock in front of \FASERnu.  A similar spectrum is expected for neutrons produced inside the detector, as well as for other neutral hadrons. As can be seen, most of the high-energy neutral hadrons with $E>100~\gev$ will come from directions consistent with the ATLAS IP, while those with lower energy often have different directionality. 

\begin{figure}[tpb]
\centering
\includegraphics[width=0.43\textwidth]{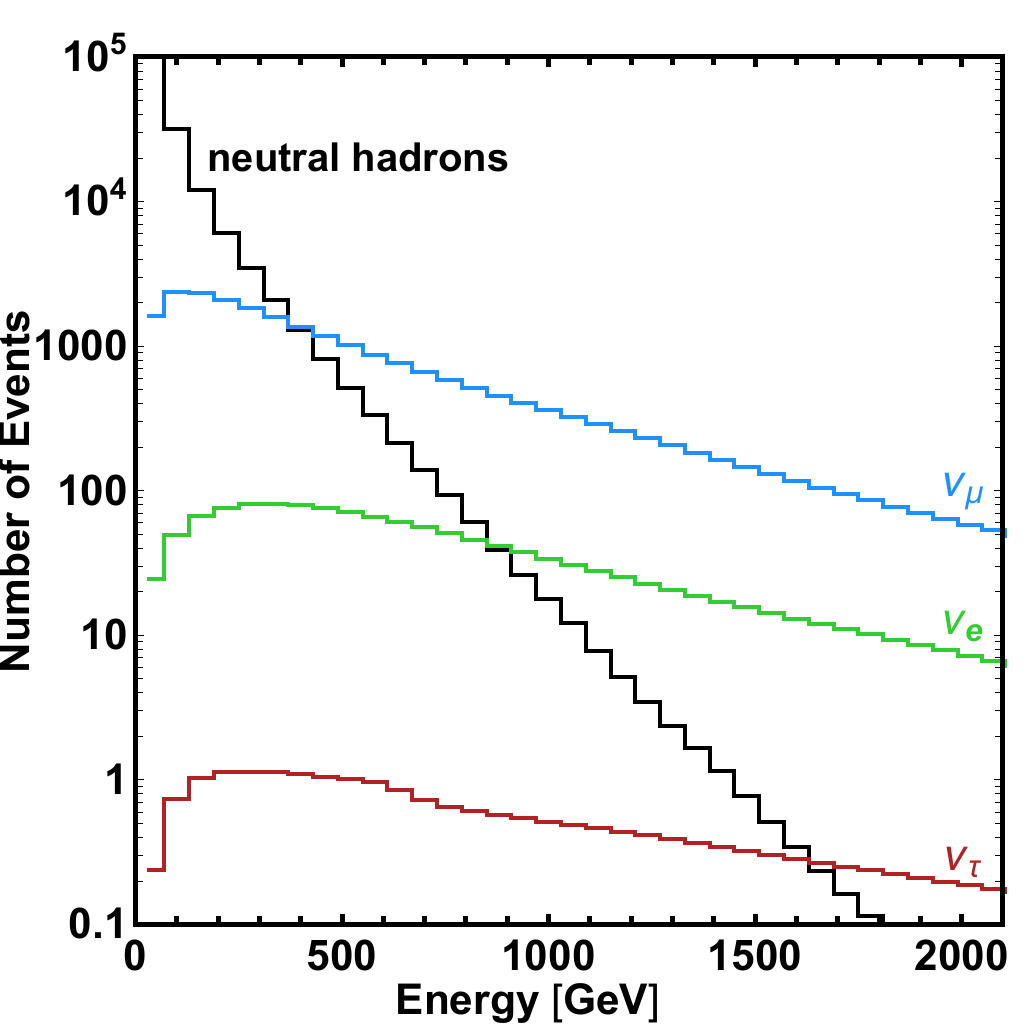}
\includegraphics[width=0.43\textwidth]{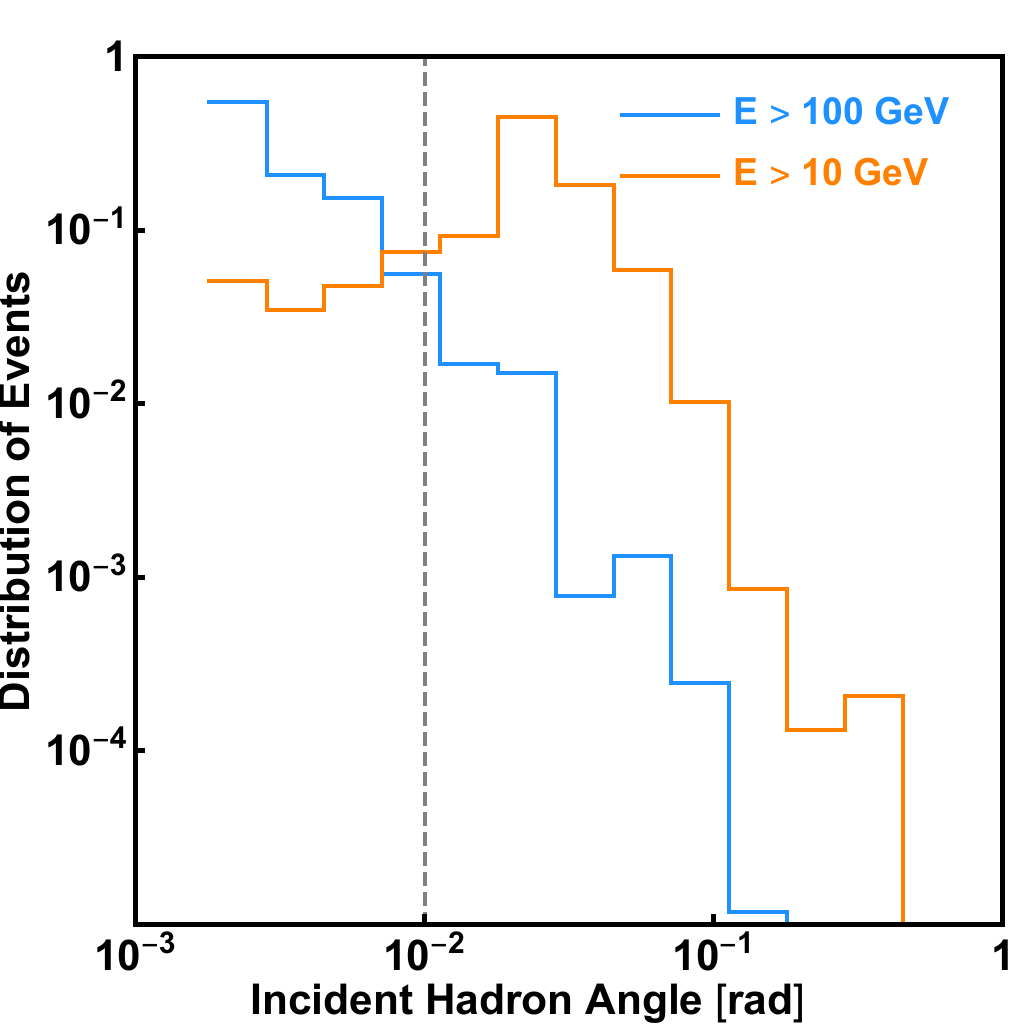}
\caption{The energy and angular distributions of neutral hadrons that are produced by negative muon interactions and pass through \FASERnu in Run 3.  \textbf{Left}: The energy spectrum of neutral hadrons produced either in the rock in front of FASER or within the detector itself (black), along with the energy spectra of neutrinos interacting in the detector: $\nu_e$ (red), $\nu_\mu$ (blue), and $\nu_\tau$ (green).  \textbf{Right}: The angular distribution of neutrons produced in the rock in front of \FASERnu and passing into the detector for energies $E>10~\gev$ (orange) and $E>100~\gev$ (blue). The angle is given with respect to the beam collision axis.  The estimated angular resolution of a $\sim 100~\gev$ hadron is about $10~\mrad$, as indicated by the vertical dashed line. From Ref.~\cite{Abreu:2019yak}.}
\label{fig:BGspectrum}
\end{figure}

%%******************************************
\subsubsection{Background for Neutrino Searches} 
%%******************************************

The interactions of high-energy muon-induced neutral hadron background can be distinguished from the neutrino-induced signal events at the level of analysis. This is especially important for the interactions of electron and tau neutrinos, which are far less abundant than muon neutrinos.  The background can be greatly reduced by requiring at least 5 charged tracks emerging from a single vertex. An additional improvement in background rejection will be achieved by identifying outgoing charged leptons at the neutrino interaction vertex. This will be achieved by the use of multivariate techniques employing, e.g., the kinematical features of the highest momentum particle (HMP) produced in the vertex, as well as by analyzing its interactions in the rest of the detector. The relevant signal features for such an analysis have been identified and discussed in Ref.~\cite{Abreu:2019yak} for the three neutrino flavors. More detailed study will be performed with a dedicated \textsc{Geant4} detector model, which is discussed below in \secref{offline_analysis}.

%%%%%%%%%%%%%%%%%%%%%%%%%%%%%%%%%%%%%%%%%%%%%%%%%%%%%%
%%%%%%%%%%%%%%%%%%%%%%%%%%%%%%%%%%%%%%%%%%%%%%%%%%%%%%
\section{Neutrino Flux and Uncertainty Estimates}
\label{sec:flux}
%%%%%%%%%%%%%%%%%%%%%%%%%%%%%%%%%%%%%%%%%%%%%%%%%%%%%%
%%%%%%%%%%%%%%%%%%%%%%%%%%%%%%%%%%%%%%%%%%%%%%%%%%%%%%

One of the main goals of \FASERnu is the measurement of CC neutrino-nucleus interaction cross sections at high energies. As can be seen in \figref{physics-xsprojection}, this measurement is limited not by statistics, but by the uncertainties associated with the incoming neutrino flux. This implies that neutrino flux estimates are key inputs to the neutrino cross section measurements. 

In \secref{tuning} we describe the simulation of forward hadron production at the LHC, how tuning uncertainties can be quantified, and how existing and future data can be used to reduce these uncertainties. In \secref{bdsim} we describe how we use BDSIM to simulate the propagation of hadrons through the forward LHC magnets and infrastructure before decaying to neutrinos. Finally, in \secref{nuc} we discuss nuclear effects that need to be taken into account when modeling neutrino interactions with the \FASERnu detector and their effect on both the interaction rate and the event kinematics. 

%%******************************************
\subsection{Hadronic Interaction Models and Tuning Uncertainties}
\label{sec:tuning}
%%******************************************

The neutrinos that can be observed at \FASERnu are produced in the decays of hadrons, mainly pions, kaons and $D$ mesons. In the forward region, the production of these hadrons can be simulated using hadronic interaction models, such as \textsc{Epos-Lhc}~\cite{Pierog:2013ria}, \textsc{Qgsjet-ii-04}~\cite{Ostapchenko:2010vb}, \textsc{Sibyll 2.3c}~\cite{Ahn:2009wx, Riehn:2015oba, Riehn:2017mfm, Fedynitch:2018cbl}, and \textsc{Pythia~8}~\cite{Sjostrand:2006za, Sjostrand:2007gs}, which have been designed to describe inelastic collisions at both particle colliders and cosmic ray experiments. Despite their sophisticated modeling of microscopic physics, these models contain a sizable number of phenomenological parameters, typically in kinematic regimes where perturbative methods do not apply.  It is therefore necessary to adjust or ``tune'' these parameters to obtain physics predictions that are able to describe observed experimental data as well as possible.  The procedure to do so is well understood and exercised routinely at LHC experiments with a mature set of tools. Simulated events for a given set of parameters are analysed with \textsc{Rivet}~\cite{Buckley:2010ar} to yield histograms directly comparable with published experimental data. We can then apply numerical optimization techniques, such as \textsc{Professor}~\cite{Buckley:2009bj}, on a suitable test statistic to yield a best-fit point. 

Currently, most of the hadronic interaction models, including \textsc{Epos-Lhc}, \textsc{Qgsjet-ii-04}, and \textsc{Sibyll 2.3c}, only provide a single tune corresponding to a central prediction. However, to describe the neutrino flux uncertainties at \FASERnu, it is crucial that the uncertainties related to the tuning of these models are also quantified.  

\begin{figure}[t]
\centering
\includegraphics[width=0.34\textwidth]{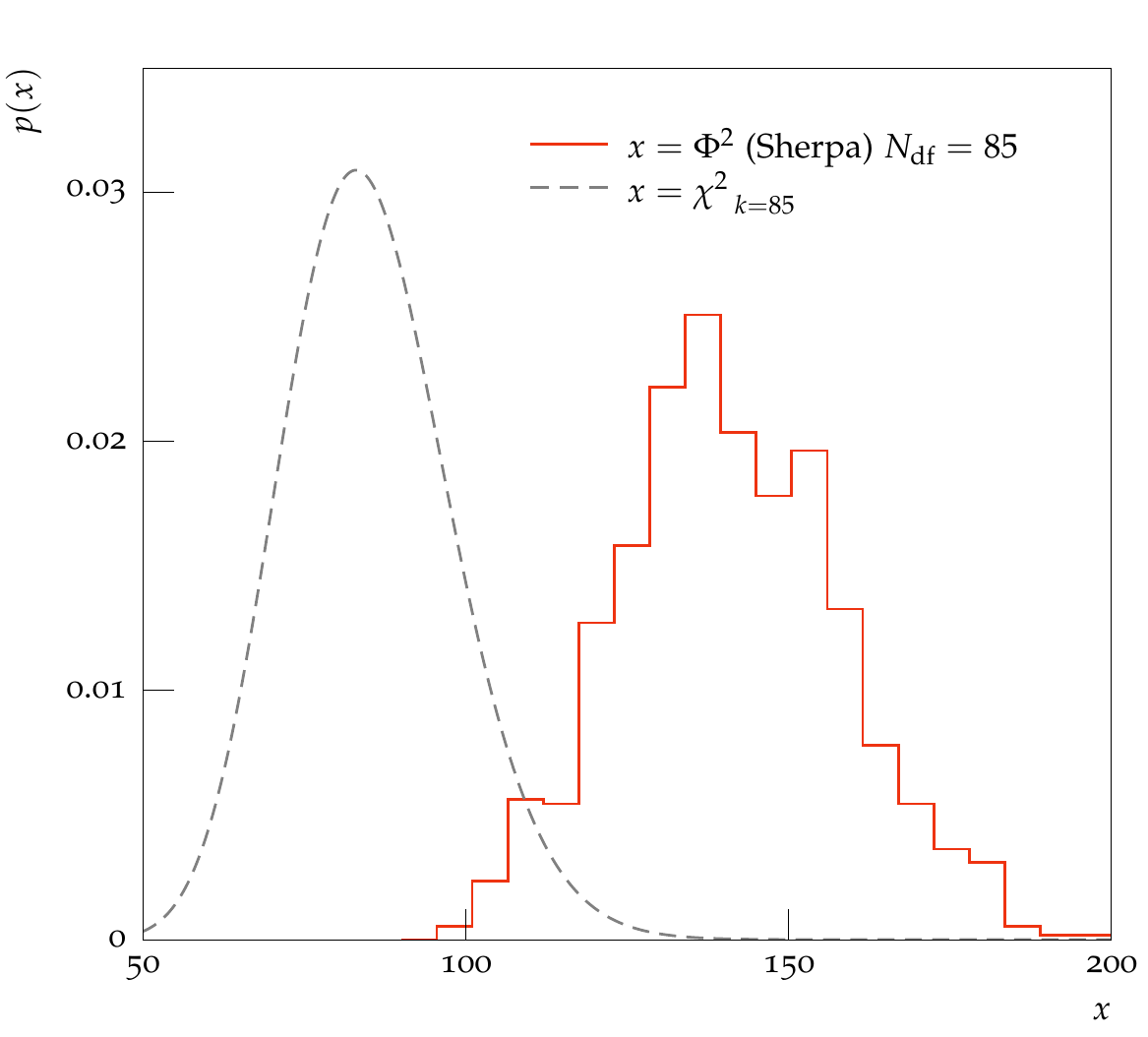}
\includegraphics[width=0.28\textwidth]{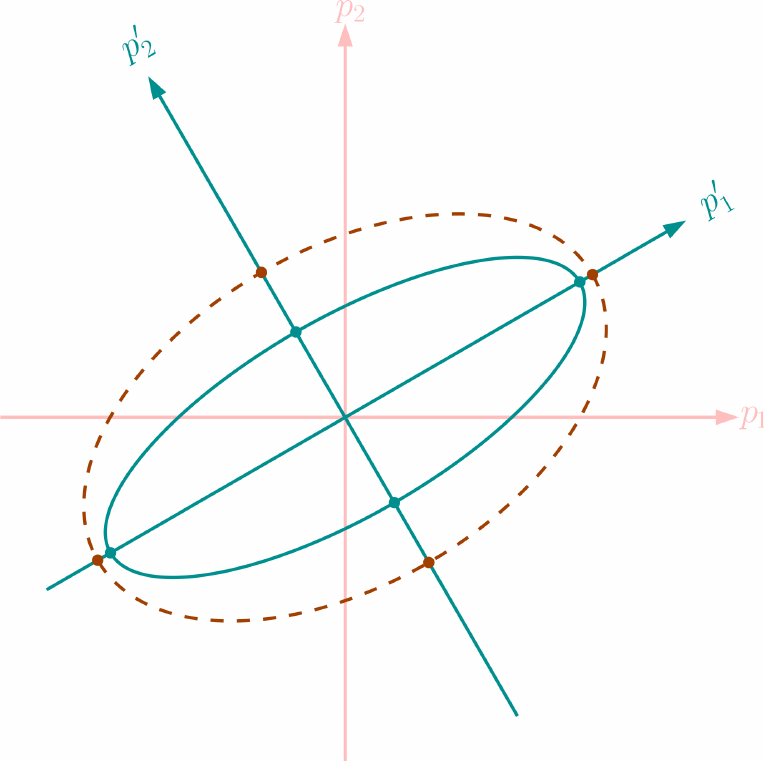}
\includegraphics[width=0.34\textwidth]{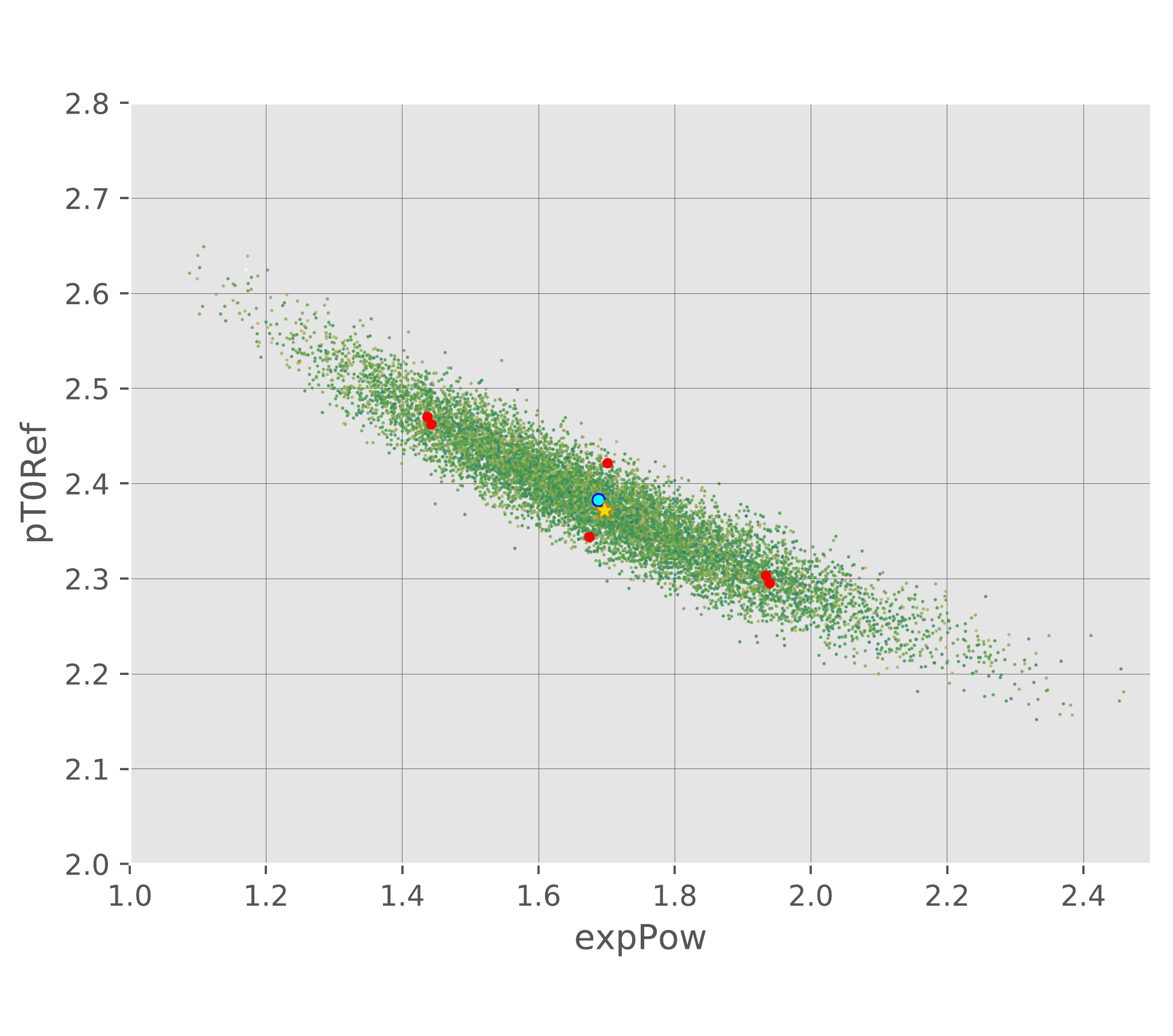}
\caption{
    \textbf{Left}: Comparison of a $\chi^2$ distribution (dashed gray) with 85 degrees of freedom with the \emph{actual}  distribution of our tuning goodness-of-fit measure $\Phi^2$ (solid red) with as many degrees of freedom. This illustrates that one cannot apply a standard error construction to find, say, the 68\% confidence volume of the parameter space around the best-fit point. If one were to do so, the uncertainties would be gravely underestimated in this example. Contributing to the breakdown of the $\chi^2$ error construction are shortcomings in the physics modeling, data tension, and unknown correlations in the tuning application.
    \textbf{Center}: Pictorial representation of a covariance matrix decomposition and determination of tuning uncertainties. We want to find representative points on the surface of the hyper-ellipsoid that correspond to a certain level of confidence around the best-fit point. A suitable decomposition lets us find the principal axes $p_1'$ and $p_2'$ along which to explore the goodness-of-fit measure. The blue ellipse illustrates the standard error construction (``$\Delta\chi^2 +1$''), which is invalid if the goodness-of-fit does not follow a $\chi^2$ distribution. Instead, we want to find the intercepts of the principal axes with the hyper-contour of the bootstrapped $\Phi^2$. They are shown as dots on the axes in the rotated system. \textbf{Right}: A two-dimensional projection of bootstrapped tuning results for a 4-dimensional problem. The best-fit point is marked with a star and the points representing the tuning uncertainties are marked in red.
}
\label{fig:chi2}
\end{figure}

The estimation and propagation of tuning uncertainties is based on an approach that is not too dissimilar from the way PDF uncertainties are evaluated by, e.g., the CTEQ Collaboration. We consider the best-fit point of the tuning to be the central value and are interested in finding points in the parameter space that deviate from the central value in such a way that they are representative of a given confidence level. There are two obstacles that need to be considered. First, the test statistic does not follow a $\chi^2$ distribution due to unknown (unpublished) correlations present in the data and the imperfections present in the physics modeling of, e.g., the underlying event and hadronization. An illustrative example is given in the left panel of \figref{chi2}. Second, there are, in principle, infinitely many points on the multidimensional manifold that correspond to the confidence volume around the central fit value. The first issue is overcome by a bootstrapping procedure of our test statistic ($\Phi^2$), which allows us to find the correct critical value $\Delta\Phi^2$ that corresponds to the chosen confidence level. The task of finding representative points on the confidence manifold is solved by applying a principal component analysis to the post-fit covariance matrix. This allows one to determine the intercepts of the principal axes with the manifold (see \figref{chi2}).

The currently available best tunes of MC generators are typically focused on precisely modeling the physics relevant to ATLAS and CMS data; i.e., they do not have an emphasis on forward physics. We therefore propose a tuning campaign that focuses more on data from forward detectors. In addition to the standard data set used for tuning, we will include the following experiments:
\begin{description}[leftmargin=0.16in]

 \item [LHCf] The LHCf detector is a forward calorimeter at the ATLAS interaction point, which covers the pseudorapidity range $\eta>8.81$.  It can measure the energy and transverse momentum spectrum of neutral particles, in particular neutrons, photons, and $\pi^0$. Additionally, upcoming analyses of Run 3 data are expected to measure both the $\pi^0$ and $\eta$ spectra. 

 \item [TOTEM] The TOTEM experiment consists of `Roman Pot' detectors sensitive to elastic collisions, as well as forward telescopes sensitive to inelastic collisions. Of particular interest for us are the two T2 telescopes, which detect charged particles in the pseudorapidity range $5.3<|\eta|<6.5$ on both sides of CMS. Additionally, a low-intensity run with displaced interaction point has provided constraints on the charged particle density up to $\eta \approx 7$.

 \item [CASTOR] The CASTOR calorimeter measures the inclusive energy spectra of inelastic collisions. It covers the forward pseudorapidity range $-6.6<\eta<-5.2$ and can distinguish between electromagnetic and hadronic components. 
\end{description}

\begin{figure}[t]
\centering
\includegraphics[width=0.49\textwidth]{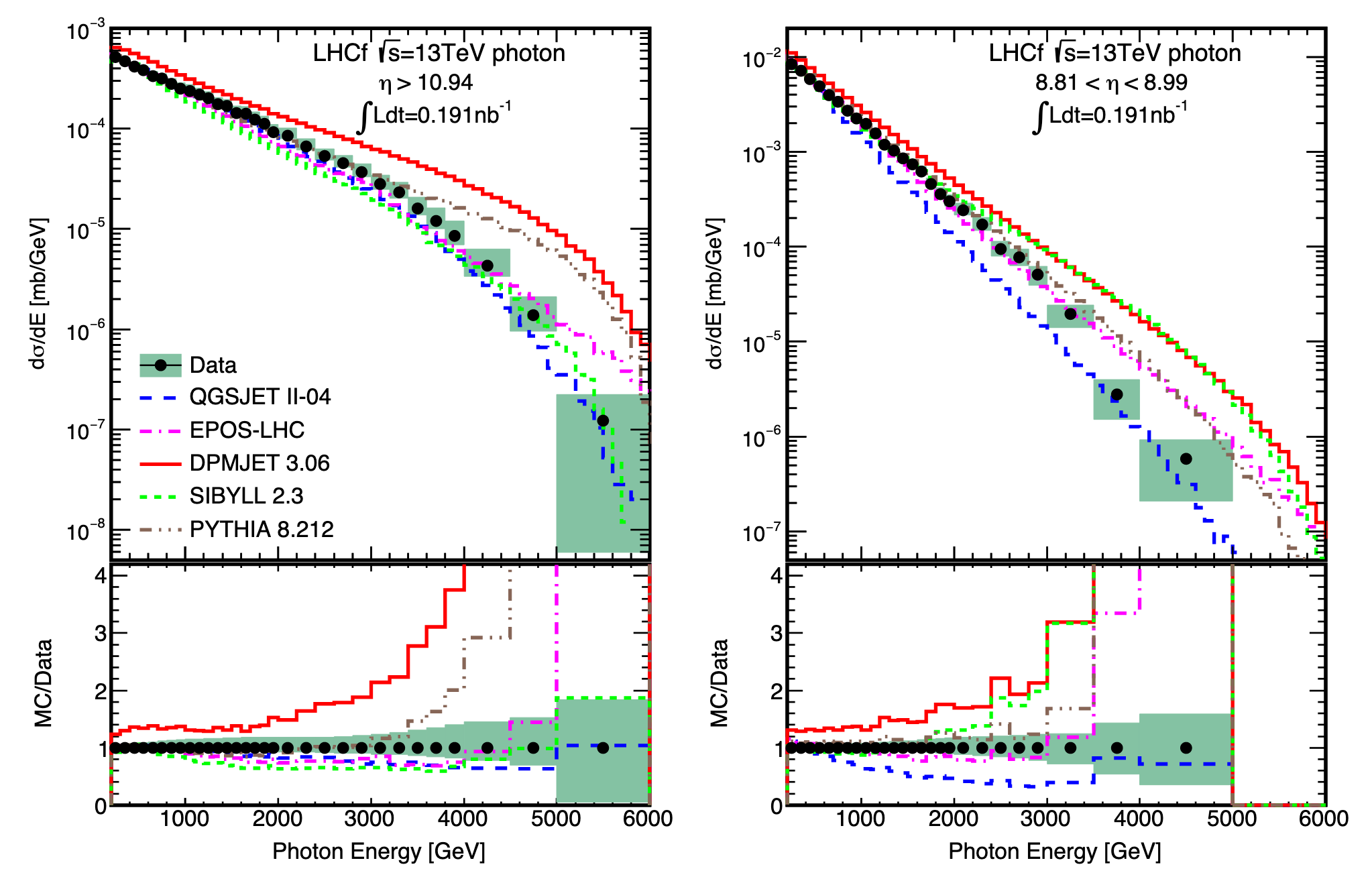}
\includegraphics[width=0.49\textwidth]{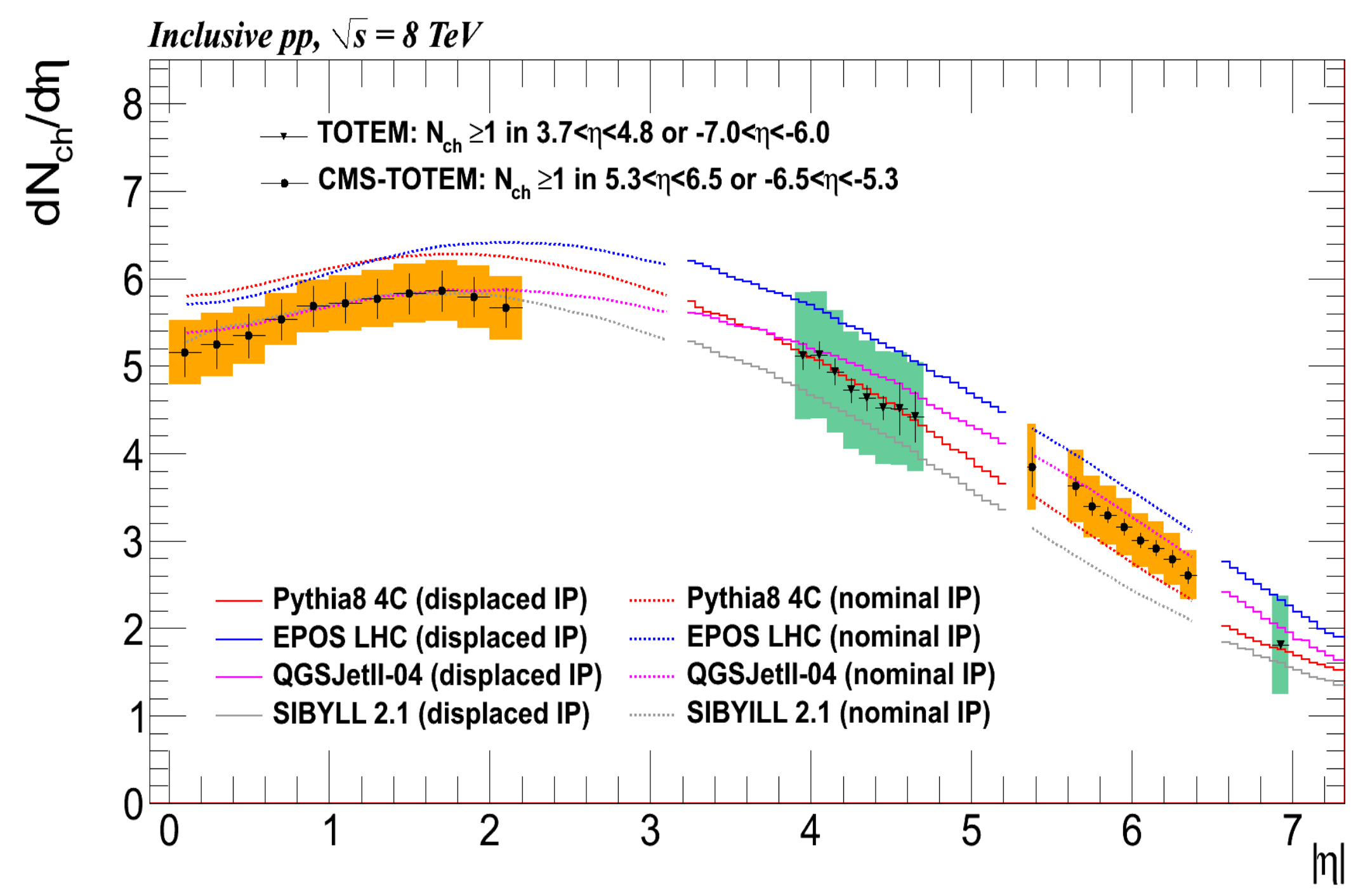}
\caption{Photon energy spectra in two pseudorapidity ranges measured at LHCf (left) and charged particle pseudorapidity density $dN_{\text{ch}}/d\eta$ measured at TOTEM (right). From Refs.~\cite{Adriani:2017jys} and~\cite{Aspell:2012ux}.
}
\label{fig:LHCf_Totem}
\end{figure}

An overview of forward data currently available in \textsc{HepData}, which we consider for tuning, is listed in \tableref{tuninganalyses}. Representative results from LHCf and TOTEM are shown in \figref{LHCf_Totem}. Additionally, measurements of heavy meson spectra at LHCb will be the primary input to constrain forward charm production. Note that most of the data is already available in \textsc{Rivet}. As a starting point, we plan to create a dedicated forward physics tune, including tuning uncertainties, for \textsc{Pythia~8}. However, we note that the same procedure can be applied to other simulators, such as \textsc{Epos} and \textsc{Sibyll}, as well.  

\begin{table}[t]
  \centering
  \begin{tabular}{|c|c|c|c|c|}
  \hline \hline
	{\bf Experiment} & {\bf Analysis}    &{\sc HepData}& {\sc Rivet}  &  {\bf  Refs. }  \\ 
	\hline 
	\hline
    LHCf& 
    photon energy spectrum at $13~\tev$ &
    $\surd$ &$\surd$ & \cite{Adriani:2017jys} \\
    \cline{2-5}
    ($\eta>8.81$)& diffractive photon energy spectrum at $13~\tev$
    & --- & --- & \cite{ATLAS:2017rme} \\
    \cline{2-5}
    &neutron energy spectrum  at $13~\tev$ &
    $\surd$ &$\surd$ & \cite{Adriani:2018ess} \\
    \cline{2-5}
    & neutron energy spectrum at $7~\tev$ &
    $\surd$ &$\surd$ & \cite{Adriani:2015nwa} \\
    \cline{2-5}
    & $\pi^0$ energy and $p_T$ spectrum at 2.76 and $7~\tev$ &
    $\surd$ &$\surd$ & \cite{Adriani:2015iwv,Adriani:2012ap} \\
    \hline
    TOTEM& 
    $dN_{\text{ch}}/d\eta$ at $8~\tev$ for
    $3.9\!<\!\eta\!<\!4.7$ and $\eta\approx6.9$ &
    $\surd$ &$\surd$ & \cite{Antchev:2014lez} \\
    \cline{2-5}
    ($5.3\!<\!\eta\!<\!6.5$)& 
    $dN_{\text{ch}}/d\eta$ at $8~\tev$ &
    $\surd$ &$\surd$ & \cite{Chatrchyan:2014qka} \\
    \cline{2-5}
    &$dN_{\text{ch}}/d\eta$ at $7~\tev$   &
    $\surd$ &$\surd$ & \cite{Aspell:2012ux} \\
    \hline
    CASTOR& 
    inclusive energy spectrum at $13~\tev$ &
    $\surd$ &$\surd$ & \cite{Sirunyan:2017nsj} \\
    \cline{2-5}
    ($5.2\!<\!\eta\!<\!6.6$)& 
    underlying event activity at $7~\tev$ &
    $\surd$ &$\surd$ & \cite{Chatrchyan:2013gfi} \\
    \cline{2-5}    
    \hline
    \hline
  \end{tabular}
  \caption{Existing LHC analyses considered for tuning for forward physics. The 3rd and 4th columns show whether the analysis is available in \textsc{HepData} and \textsc{Rivet}, respectively. } 
\label{table:tuninganalyses}
\end{table}

In addition to data from other experiments, it might also possible to use data from FASER or \FASERnu in the future. The pilot emulsion detectors used in 2018 have measured the muon flux in TI12/TI18 and found good agreement with simulations. A dedicated analysis is ongoing and the potential of this data to constrain forward particle production is being investigated. In addition, the FASER spectrometer, which will start operating in 2021, will not only be able to measure the muon flux more precisely, but will also be able to measure the muon energy spectrum. Finally, \FASERnu will measure the neutrino spectrum over a large energy range from about $100~\gev$ up to a few $\tev$. Since the cross section is known at low energies $E<360~\gev$, the low-energy part of the measured neutrino spectrum could be used to constrain or calibrate the neutrino flux.

%%******************************************
\subsection{Propagation through the LHC} 
\label{sec:bdsim}
%%******************************************

The main production mode of neutrinos arriving at \FASERnu is via the decay of light hadrons, such as pions and kaons. These particles are long-lived and decay downstream from the ATLAS IP, which requires us to model their propagation and absorption in the LHC beam pipe. Importantly, placed around the beamline are  both quadrupole and dipole magnets which deflect charged hadrons, often resulting in non-trivial trajectories. Additionally, secondary collisions of particles with the LHC infrastructure and subsequent showers can also contribute to neutrino production. Therefore, a dedicated simulation of particle propagation that accurately describes the forward infrastructure is needed to obtain reliable predictions for the neutrino flux. 

The propagation of an event from creation in a proton-proton collision at the ATLAS IP up to the FASER detector can be studied using the Beam Delivery Simulation (BDSIM)~\cite{bdsim}. BDSIM tracks particles as they travel across the accelerator lattice and simulates their possible interactions with matter along their trajectory. BDSIM operates using an underlying \textsc{Geant4} model of the LHC beamline components, as well as the associated surrounding materials including the tunnel shielding and soil material. 

\begin{figure}[tbh]
\includegraphics[width=1\textwidth]{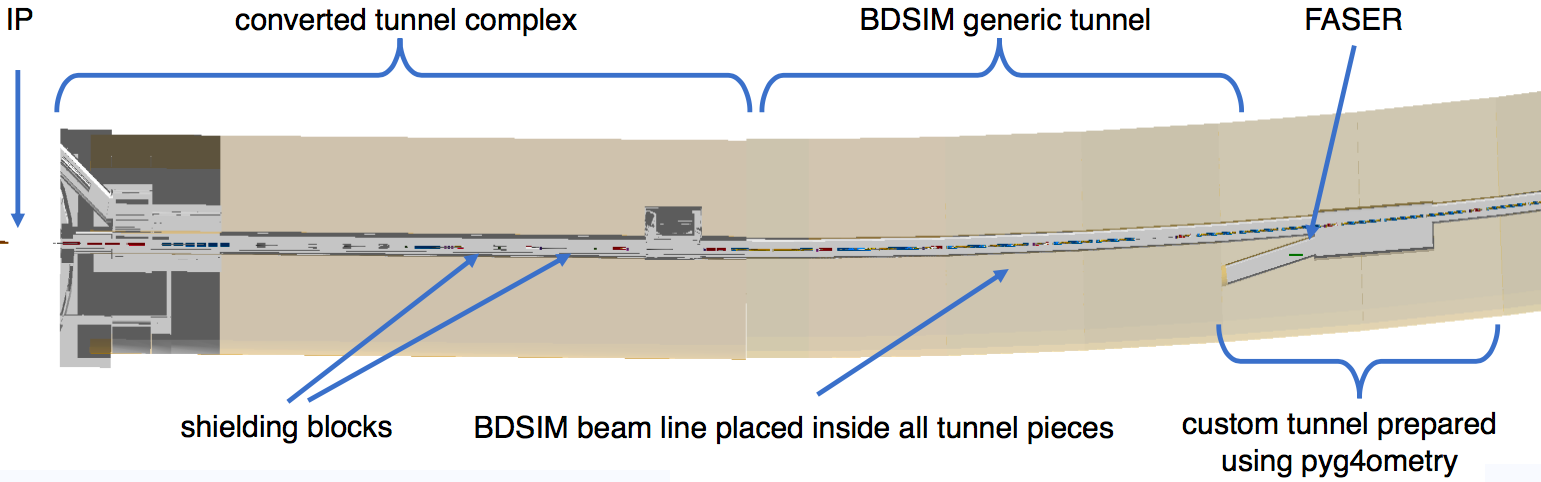}
\caption{Top view of the tunnel section between the ATLAS IP and the FASER location.}
\label{fig:geo}
\end{figure}

A BDSIM model for FASER, shown in \figref{geo}, has been implemented using a combination of existing and custom-created geometries.  The optics for accelerator components (aperture and magnet strengths) were taken from the MADX model of 2018 collision optics ($\beta^*=30\,$cm)~\cite{madx}. The section between the ATLAS IP and the end of the straight section (0 to $260~\m$ from the ATLAS IP) is based on an existing FLUKA model~\cite{fluka}.  The model for the part of the tunnel located between $260~\m$ and $450~\m$ is based on a generic BDSIM model, which uses the LHC tunnel dimensions and materials. The last part of the tunnel, including the UJ12 cavern and the TI12 tunnel, which houses the FASER detector, was created using \textsc{pyg4ometry}, a python package generating GDML files for use in BDSIM. Additional elements important for this model were created in \textsc{pyg4ometry}: the TAN (target absorber neutral), an absorber stopping neutral particles created at the ATLAS IP, and shielding blocks located in the tunnel at various locations around the beam pipe designed to decrease the radiation in the tunnel (see the left panel of \figref{custom}). The TAN is especially important for FASER simulations, as it produces secondary particles to which FASER is sensitive.

\begin{figure}[h]
	\includegraphics[height=0.2\textwidth]{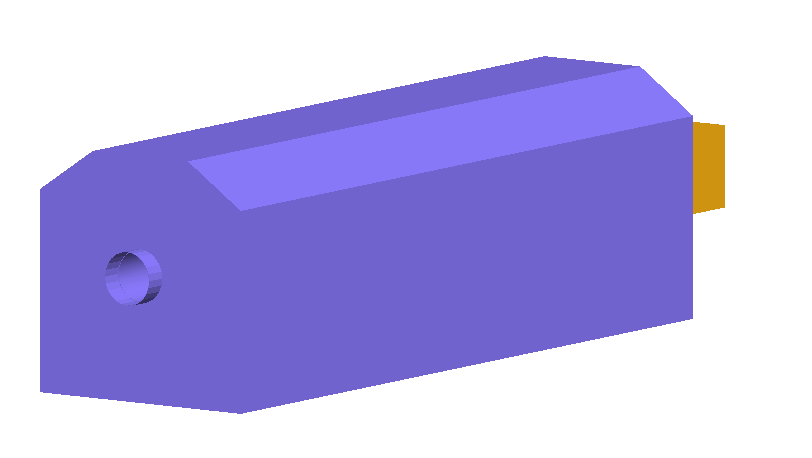}
		\includegraphics[height=0.2\textwidth]{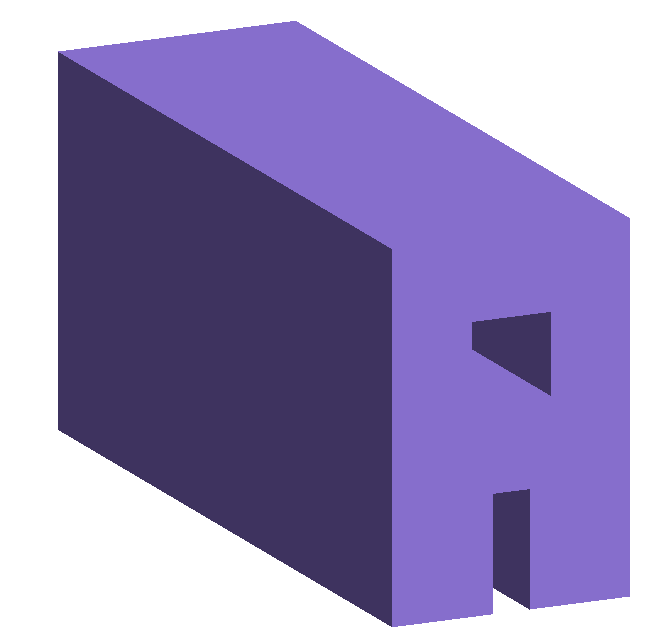}
		\includegraphics[height=0.2\textwidth]{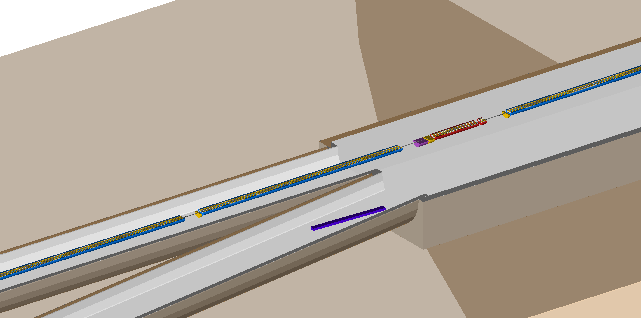}
		\caption{\textsc{Pyg4ometry} custom geometry of the TAN (left), the beam pipe shielding block (center), and the tunnel geometry at the FASER location (right).}
	\label{fig:custom}
\end{figure}

BDSIM simulates particles inside the model using particle accelerator tracking routines and the FTFP$\_$BERT \textsc{Geant4} reference physics list~\cite{Geant4PhysicsLists}. The input to BDSIM are particles produced in the ATLAS collisions simulated by the CRMC event generator \cite{crmc}, which simulates proton-proton collisions at $\sqrt{s}=14~\tev$. Simulated particle hits are collected on the so-called FASER Interface Plane, which is located right in front of the FASER detector (at $475~\m$ from the ATLAS IP, centered around the line of sight). The simulation data includes the type, position, momentum, and energy of particles traversing the FASER Interface Plane. Additionally, BDSIM allows us to analyze the origin and trajectory of particles. 

The goal of this BDSIM simulation is to estimate the flux and energy of particles reaching the FASER Interface Plane to determine the expected event rate for the detector. In addition, the simulation will make it possible to (i) study the effect of beam configurations, for example, the beam crossing angle, (ii) understand the importance of secondary neutrino production mechanisms, for example in the TAN, (iii) understand additional sources of uncertainties associated with the simulation, for example, the propagation of hadrons through the forward magnets, and (iv) study the muon flux at FASER, including their production mechanisms and propagation through the LHC infrastructure, as well as the feasibility of using muon measurements as input to constrain forward particle production. The simulation results will be compared to data taken during Run 2 at the planned location of the FASER detector.

%%******************************************
\subsection{High Energy Neutrino Interactions}
\label{sec:nuc}
%%******************************************

Once the neutrino flux is known, additional simulations will be performed to obtain the expected number of neutrino interactions and event kinematics. It is well known that large-nuclei nuclear effects will modify the interaction cross section. When measuring the neutrino cross section, these features would be considered part of the measurement. In contrast, when constraining the forward production cross section assuming SM interaction cross sections, nuclear effects are expected to be an important theory uncertainty.  Additionally, hadronization and final state interactions (FSI) will determine the kinematics of the final state. A realistic simulation of these effects is important to avoid inducing a simulation bias, for example, in the energy estimate. In the following, we discuss these effects and how their uncertainties can be estimated. 

%%------------------------------------------
\subsubsection{Interaction Cross Section and Nuclear Effects}
%%------------------------------------------

As discussed above, the typical energy of neutrinos that interact with \FASERnu is above $100~\gev$. In this regime, neutrino interactions can be described by deep inelastic scattering (DIS)~\cite{Formaggio:2013kya,Gandhi:1995tf}. In Ref.~\cite{Abreu:2019yak}, we have estimated the interaction cross section by considering the tungsten nucleus with mass number $A=184$ as a collection of $Z=74$ protons and $N=110$ neutrons. However, for neutrino scattering on heavy nuclei, nuclear effects such as shadowing, anti-shadowing, and the EMC effect become important.\footnote{The \textsc{Genie}~\cite{Andreopoulos:2009rq, Andreopoulos:2015wxa} events generated for this study include such nuclear effects, using ``effective leading order''~\textsc{GRV98} nPDFs, which are tuned to experimental data~\cite{Bodek:2002ps}. As such, it is difficult to meaningfully discuss the systematic uncertainty on the \textsc{Genie} simulation in this context, and a move to incorporate more modern nPDF sets is an essential, but longer term, development.} Therefore the PDFs of nucleons bound within nuclei have to be modified with respect to their free-nucleon counterparts, which is described by nuclear parton distribution functions (nPDFs). The PDF $f_i^{{p/A}}(x,Q)$ of a proton bound in a nucleus with mass number $A$ is defined relative to the free parton PDF $f^{(p)}_i(x,Q)$ as
\be
  f^{(p/A)}_i(x,Q) =  R_i^A (x,Q) \,  f^{(p)}_i(x,Q) \ ,
\ee
where $R_i^A (x,Q)$ is the scale-dependent nuclear modification. From this, we can construct the PDF in a heavy nucleus $f^{(A)}_{i}(x,Q)$ with mass number $A$ as 
\be
 f^{(A)}_{i}(x,Q) = Z \, f^{(p/A)}_{i}(x,Q) +  N \, f^{(n/A)}_{i}(x,Q) \ .
\ee
Based on a variety of experimental inputs, several sets of nPDFs have been developed within the last years: \textsc{EPPS16}~\cite{Eskola:2016oht}, \textsc{nCTEQ15}~\cite{Kovarik:2015cma,Kusina:2015vfa}, and \textsc{nNNPDF1.0}~\cite{AbdulKhalek:2019mzd}. In particular, these nPDFs include error sets that allow one to estimate the uncertainty associated with the description of the nuclear effects.

\begin{figure}[tp]
\centering
\includegraphics[width=0.35\textwidth]{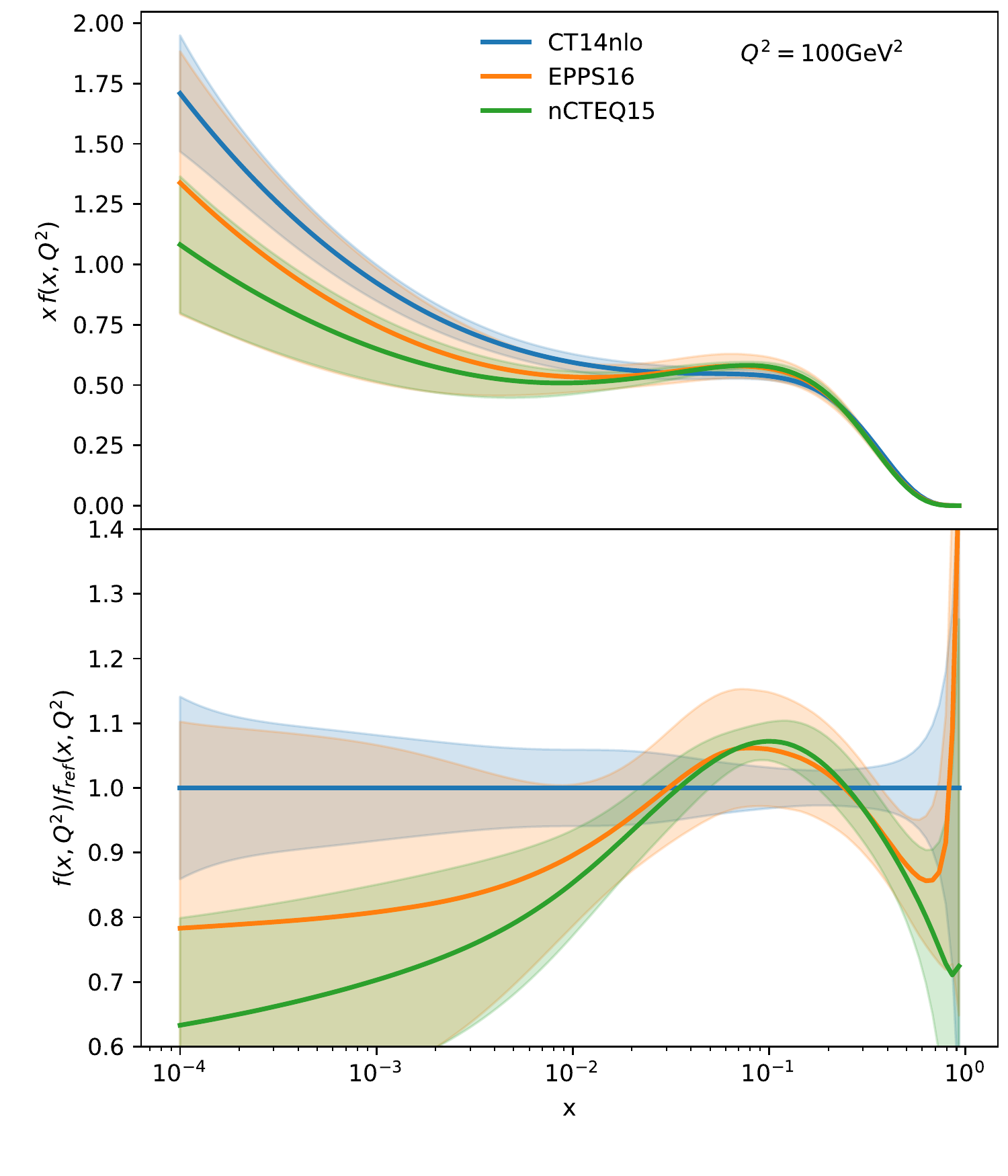} 
\hspace{1cm}
\includegraphics[width=0.35\textwidth]{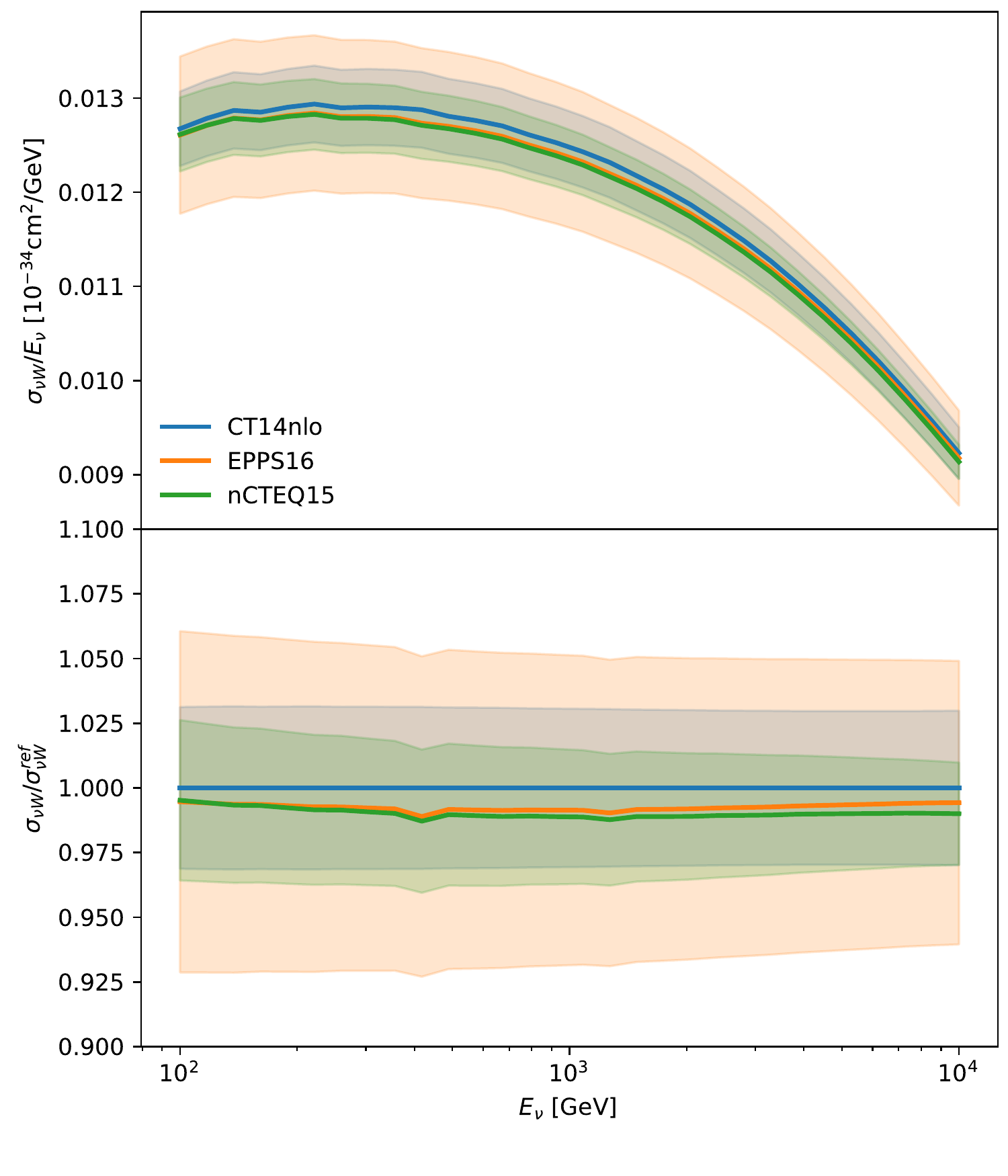} 
\caption{
\textbf{Left}: Down quark PDFs in a tungsten nucleus $ f^{(W)}_{i}(x,Q)$ for $Q^2 = 100~\gev$, using \textsc{CT14nlo} without nuclear effects (blue), \textsc{EPPS16} (orange), and \textsc{nCTEQ15} (green). 
\textbf{Right}: Interaction cross section of a muon neutrino with a tungsten nucleus, normalized by the incoming neutrino energy, using \textsc{CT14nlo} without nuclear effects (blue), \textsc{EPPS16} (orange), and \textsc{nCTEQ15} (green). 
} 
\label{fig:flux-nuclear}
\end{figure}

In the left panel of \figref{flux-nuclear}, we show the down quark PDF in a tungsten nucleus $f^{(W)}_d(x,Q)$ for $Q^2=100~\gev$ without nuclear effects (\textsc{CT14nlo} in blue) and with nuclear effects (\textsc{EPPS16} in orange and \textsc{nCTEQ15} in green). The shaded bands show nPDF uncertainties.  The lower panel shows the same PDF predictions and uncertainties relative to the central prediction of \textsc{CT14nlo}. We can clearly see that the nuclear effects cause shadowing at low $x\lesssim 0.01$, anti-shadowing at $x\sim0.1$, and the EMC effect at $x\sim0.4$. Overall, nuclear effects change the down quark PDF in tungsten by $\mathcal{O}(20\%)$. 

In the right panel of \figref{flux-nuclear}, we show the total neutrino-tungsten interaction cross section as a function of the incoming neutrino energy. We see that the nuclear effects, integrated over phase space, only change the total interaction cross section by about 1\%. The uncertainties of the cross section prediction are about 3\% for \textsc{nCTEQ15} and about 6\% for \textsc{EPPS16}.

%%------------------------------------------
\subsubsection{Hadronization and Final State Interactions}
%%------------------------------------------

A significant source of modeling uncertainty is due to the simulation of hadronization. The \textsc{Genie} simulation uses a custom model for hadronization~\cite{Yang:2009zx}, which uses \textsc{Pythia~6}~\cite{Sjostrand:2006za} for invariant masses $W^{2} \geq 3$ GeV$^{2}$, and so is most relevant here. It should be noted that \textsc{Pythia~6} has long been superceded for most collider applications, and reflects the more common use of the \textsc{Genie} software for much lower-energy physics, where DIS plays a subdominant role. The \textsc{Pythia} model in \textsc{Genie} has been tuned to neutrino--hydrogen and neutrino--deuterium scattering data~\cite{ALLEN1981385}, as is shown in Ref.~\cite{Andreopoulos:2015wxa}. Alternative tunings of the model to a wider array of datasets have been investigated by other authors~\cite{Katori:2014fxa}, but are not adopted in the default \textsc{Genie} model. Work to update the hadronization model to use \textsc{Pythia~8} is ongoing in the \textsc{Genie} Collaboration, which will benefit the work presented here, but further work will be required to assess the uncertainties of the hadronization model.

Due to the large density of nuclear matter and the large size of the tungsten nucleus, any hadron produced inside a tungsten nucleus is likely to re-interact before leaving the nucleus~\cite{Qian:2009aa}. Although such FSI do not affect the lepton, they will modify the kinematics of the final state (hadron multiplicities and energies), and disallowed final states can change the interaction rate, although the latter is unlikely at the energies relevant here. The effect on the hadronic system has, therefore, to be taken into account when modeling the neutrino interaction, which could lead to additional uncertainties, for example, when estimating the neutrino energy. 

\textsc{Genie} has two custom FSI models, referred to as the ``hA'' and ``hN'' models~\cite{INTRANUKE}. For the simulations used in Ref.~\cite{Abreu:2019yak}, the default ``hA'' model was used, which approximates the effect of FSI with an effective single interaction step, calculated separately for each particle produced at the vertex. The interaction probabilities are based on $\pi^{\pm}$--$^{56}$Fe and proton--$^{56}$Fe data (see Ref.~\cite{INTRANUKE} for details), and then extrapolating to other targets assuming $A^{\frac{2}{3}}$ scaling. The alternative ``hN'' model is a cascade model, where each particle is separately propagated from the interaction point to the edge of the nucleus, and daughter particles from re-interactions are added into the cascade. \textsc{Genie} has built-in uncertainties to rescale the total interaction probabilities for different FSI processes (e.g., inelastic or elastic collisions), but none to modify the outgoing particle kinematics of any type of interaction, so they are unlikely to be sufficient to reflect the actual uncertainty due to FSI. Estimating the size of biases to energy estimation with and without FSI is a possible route to assess the possible impact of FSI uncertainties on the analyses discussed here. 

A further step forward would be to investigate alternative models of FSI and check the effect they have on the energy estimation. Most other neutrino simulation packages use a classical cascade model similar to the \textsc{Genie} ``hN'' model~\cite{Hayato:2009zz, Juszczak:2009qa}. A significantly more sophisticated model is available from the \textsc{GiBUU} package~\cite{Buss:2011mx}, which does not factorise the particle propagation in the same way. When propagating particles out of the nucleus, at each step,  every particle feels the potential of the nucleus, and each other. This lack of factorization makes \textsc{GiBUU} more realistic, but much more computationally intensive. Although \textsc{GiBUU} is significantly more sophisticated, it is still semi-classical in that it uses free particle cross sections and modifies them, rather than full calculations for bound particles.

%%%%%%%%%%%%%%%%%%%%%%%%%%%%%%%%%%%%%%%%%%%%%%%%%%%%%%
%%%%%%%%%%%%%%%%%%%%%%%%%%%%%%%%%%%%%%%%%%%%%%%%%%%%%%
\section{Tungsten/Emulsion Detector}
\label{sec:tungsten_emulsion_detector}
%%%%%%%%%%%%%%%%%%%%%%%%%%%%%%%%%%%%%%%%%%%%%%%%%%%%%%
%%%%%%%%%%%%%%%%%%%%%%%%%%%%%%%%%%%%%%%%%%%%%%%%%%%%%%

%%******************************************
\subsection{Conceptual Detector Design}
%%******************************************

The \FASERnu neutrino detector will be placed in front of the FASER main detector, on the collision axis to maximize the number of neutrino interactions. \Figref{conceptual_design} shows a view of the neutrino detector module. The detector is made of a repeated structure of emulsion films~\cite{emulsion} interleaved with 1-mm-thick tungsten plates. The emulsion film is composed of two emulsion layers, each 70 $\mu$m thick, that are poured onto both sides of a 200-$\mu\text{m}$-thick plastic base; the film has an area of $25\,\cm \times 25\,\cm$. The whole detector contains a total of 1000 emulsion films, with a total tungsten mass of 1.2 tonnes. The total length of the detector is 1.35 m, corresponding to 285 radiation lengths $X_0$ and 10.1 hadronic interaction lengths $\lambda_\text{int}$. 
\Figref{full_chain} shows the lifecycle of emulsion films for the \FASERnu detector. Further details are discussed in the following subsections.

\begin{figure}[htpb]
\centering
\includegraphics[width=0.8\textwidth]{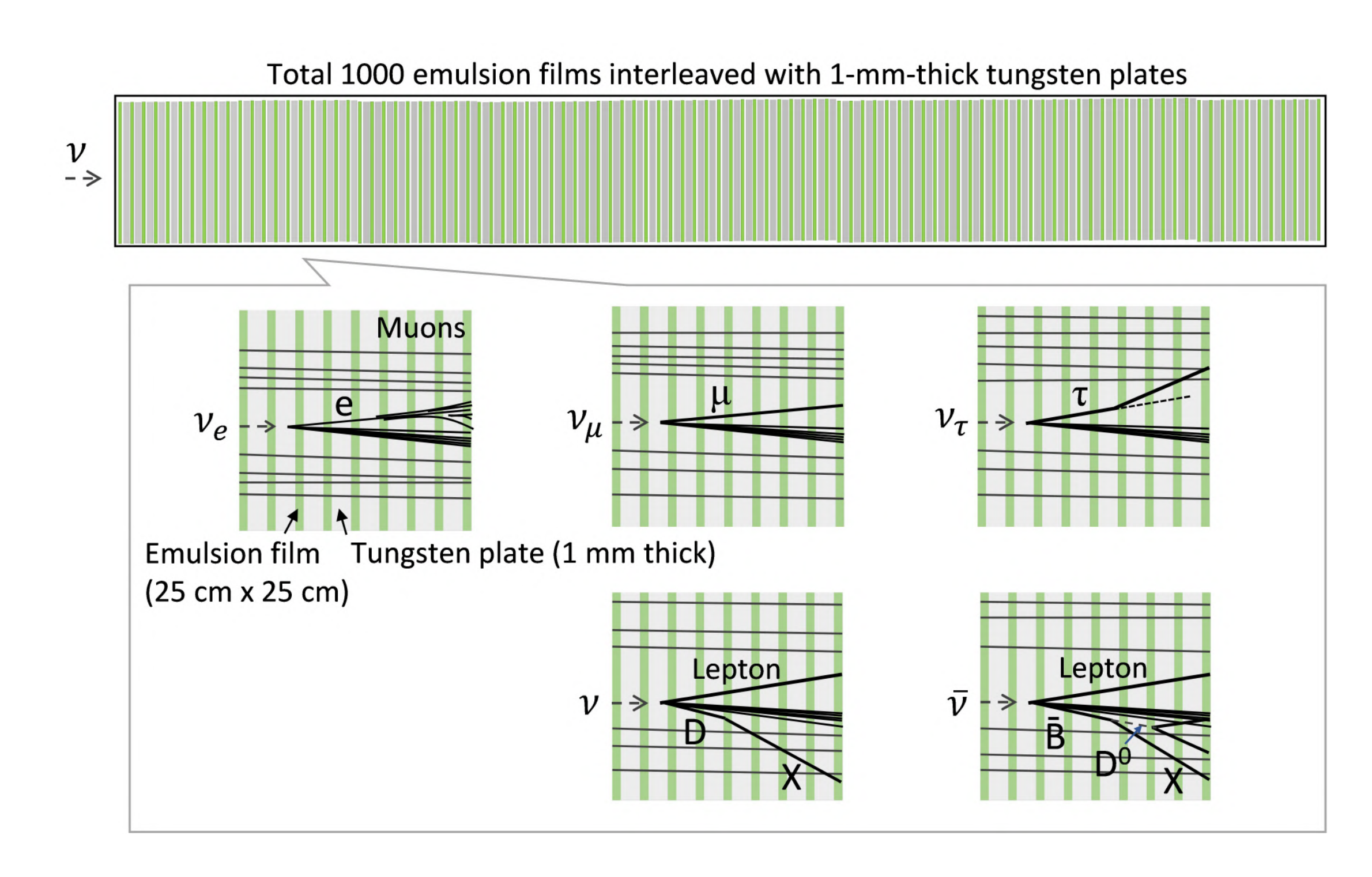}
\caption{Conceptual design of the detector structure with the topology of various neutrino event signals that can be reconstructed in the detector.
}
\label{fig:conceptual_design}
\end{figure}

\begin{figure}[htpb]
\centering
\includegraphics[width=0.85\textwidth]{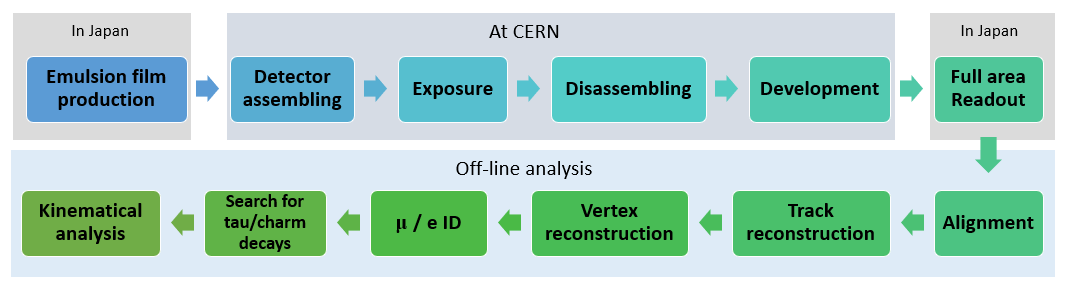}
\caption{The lifecycle of emulsion films for the \FASERnu detector, from the production of emulsion layers in Japan, to assembly, exposure, disassembly, and development at CERN, and finally to the readout in Japan and the off-line analysis of the data.
}
\label{fig:full_chain}
\end{figure}

%%------------------------------------------
\subsubsection{Emulsion Films}
%%------------------------------------------

\Figref{emulsion_film} (left) shows a photo of a standard emulsion film used in the OPERA experiment and its cross-sectional view. The film comprises two emulsion layers that were poured onto both sides of a 200 $\mu$m thick plastic base. For \FASERnu, films with a surface area of 25 cm $\times$ 25 cm will be produced. 

The emulsion sensitive units consist of silver bromide crystals, which are semiconductors with a band gap of 2.5 eV, dispersed in a gelatine substrate. The diameter of the crystals is 0.2 $\mu$m, as shown in the right-center panel of \figref{emulsion_film}.  When a charged particle passes through the emulsion, the ionization is recorded quasi-permanently, and it can then be amplified and fixed by specific chemical processes. A minimum ionising particle track is shown in the right panel of \figref{emulsion_film}.  An emulsion detector with 200~nm crystals has a position resolution of 50~nm, as shown in \figref{intrinsic_resolution}. The one-dimensional intrinsic angular resolution of a double-sided emulsion film with 200-nm-diameter crystals and a base thickness of 200 $\mu$m is therefore 0.35 mrad.

The emulsion gel and film production will be performed at an existing facility in Japan, which is currently being upgraded~\cite{pouring_system, pouring_test} (\figref{film_production}). The new facility will be ready for mass production in mid-2020.

\begin{figure}[htbp]
\begin{center}
\vspace{5mm}
\includegraphics[height=3.2cm]{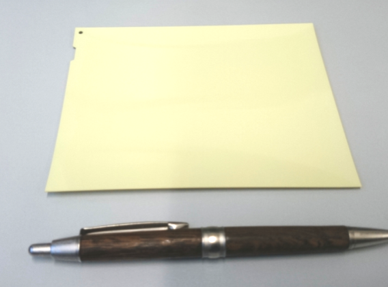}
\includegraphics[height=3.2cm]{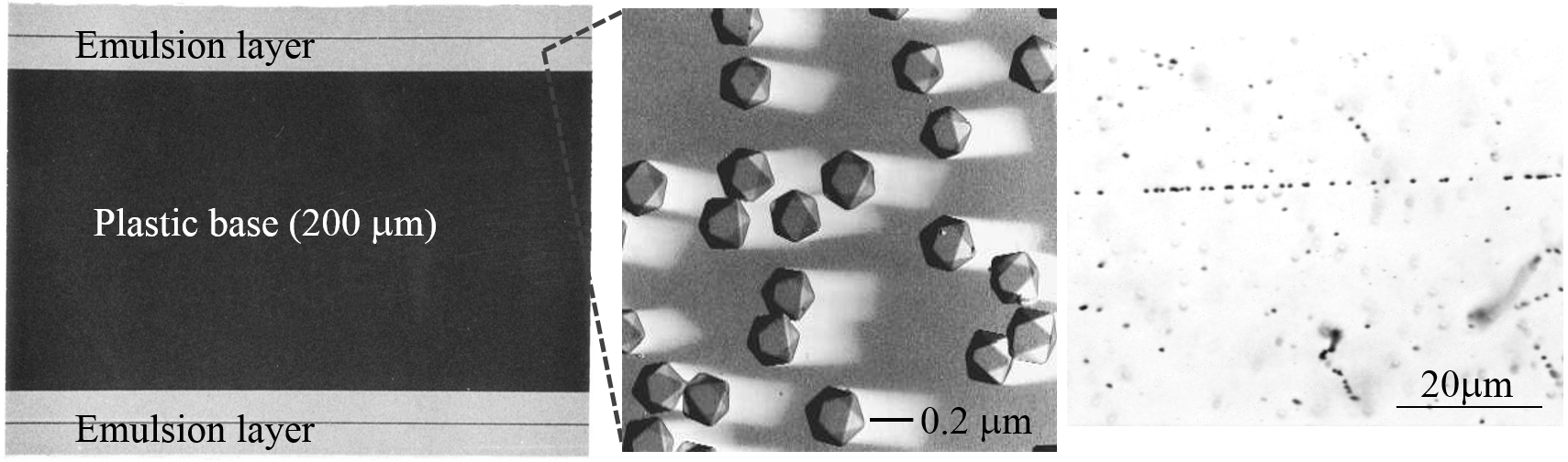}
\caption{Photo of an emulsion film (left), its cross-sectional view (left center), electron microscope image of the silver halide crystals (right center), and a minimum ionising particle track from a 10 GeV/c $\pi$ beam (right).}
\label{fig:emulsion_film}
\end{center}
\end{figure}

\begin{figure}[htbp]
\centering
\vspace{-7mm}
\includegraphics[width=0.35\textwidth]{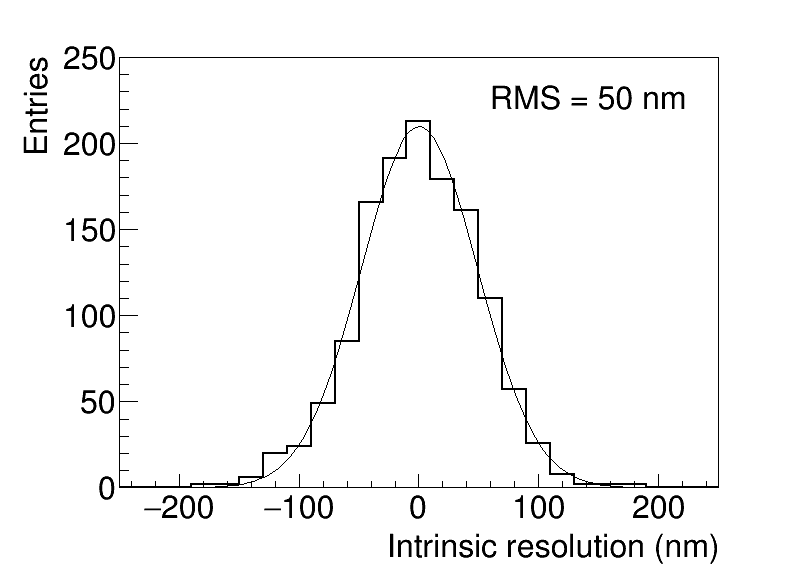}
\caption{Distribution of the distances between grains and straight-line fits to the tracks of minimum ionizing particles, showing the emulsion intrinsic spatial resolution~\cite{Amsler:2012wn}.
}
\label{fig:intrinsic_resolution}
\end{figure}

\begin{figure}[htbp]
\centering
\includegraphics[height=5cm]{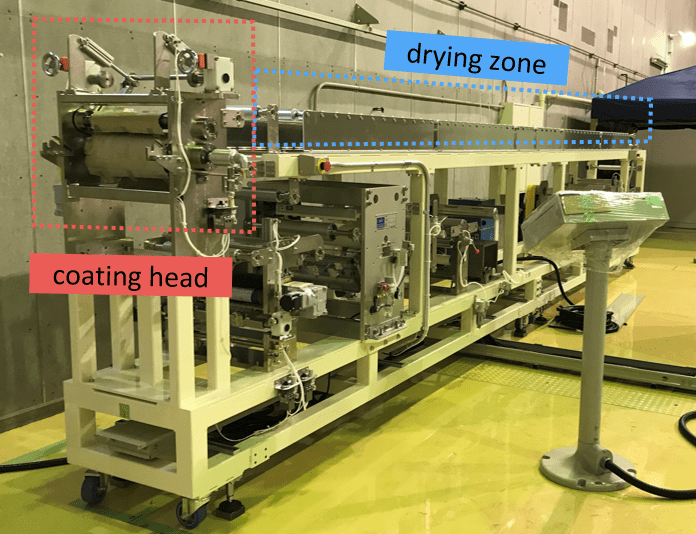}
\includegraphics[height=5cm]{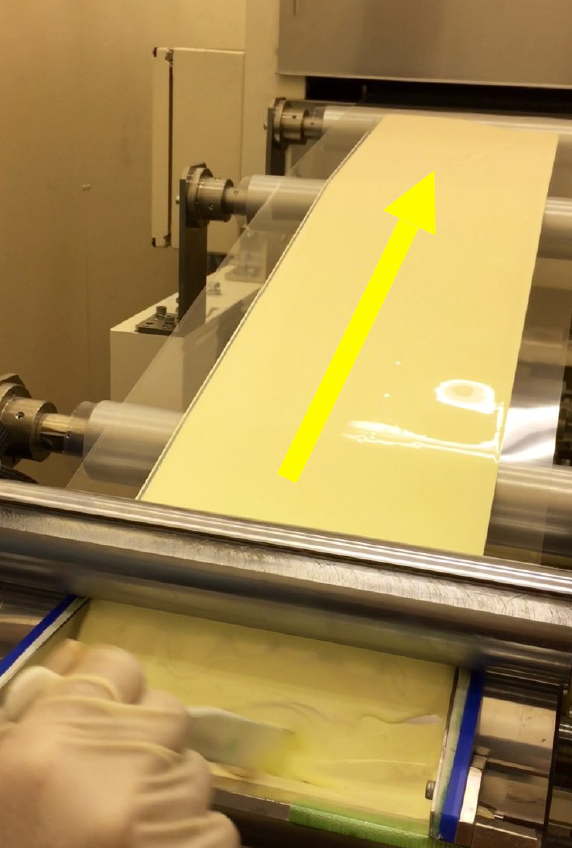}
\caption{The new system for emulsion film production at Nagoya University (left) and a test of emulsion gel pouring (right)~\cite{pouring_system, pouring_test}.}
\label{fig:film_production}
\end{figure}

%%------------------------------------------
\subsubsection{Tungsten Target}
%%------------------------------------------

Tungsten was chosen from the possible target materials shown in Table~\ref{table:target_material} for the following reasons:

\begin{itemize}
\item Its high density will allow for a higher interaction rate, keeping the detector small. Space for the detector along the beam collision axis is limited by the neutrino trench, and it's important to make the detector size small, which also makes the emulsion cost low. 

\item Its short radiation length guarantees a higher performance, both in EM shower reconstruction, keeping shower tracks to a small radius, and in momentum measurement, by means of multiple Coulomb scattering. 
\item Low radioactivity. 
\end{itemize}

\begin{table}[htbp]
\centering
\begin{tabular}{|c|c|c|c|c|c|}
\hline\hline
Material & Atomic & Density    & Interaction length & Radiation length & Thermal expansion \\
\        & number & [g/cm$^3$] & [cm]               & [mm] & $\alpha$ [\si{\times 10^{-6}K^{-1}}]\\ \hline \hline
Iron     & 26 & 7.87 & 16.8 & 17.6 & 11.8\\ \hline 
Tungsten & 74 & 19.30 & 9.9 & 3.5 & 4.5\\ \hline 
Lead     & 82 & 11.35 & 17.6 & 5.6 & 29\\ \hline \hline
\end{tabular}
\caption{Properties of possible target materials.
}
\label{table:target_material}
\end{table}

%%******************************************
\subsection{Detector Structure Implementation}
%%******************************************

The \FASERnu detector consists of 1000 layers of emulsion films interleaved with tungsten plates. The transverse dimensions are  25~cm $\times$ 25~cm. The most relevant issue in designing the detector structure is how to keep the emulsion films aligned. The position alignment has to be kept within sub-micrometer accuracy for the entire period of data taking (a few months) so that momenta can be measured by the MCS coordinate method described in \secref{energyresolution}. 

In the DONUT experiment, for example, the so-called ``Slip'' problem had to be addressed \cite{Kodama:2002dk}, as the relative position between emulsion films changed during the experiment, principally as a consequence of changes in the temperature of the environment. The thermal expansion coefficient is very different between emulsion films ($\alpha\sim 10^{-4}$/K) and metallic plates ($\alpha\sim 10^{-5}$/K), and therefore small temperature changes cause mechanical stress. This can result in a change in alignment during the measurement (multiple alignment peaks). This is depicted in \figref{slip}.  This problem can be avoided exerting a proper mechanical pressure on the emulsion films and target plates in such a way that the soft emulsion films follow the thermal expansion of metallic plates.  The thermal expansion coefficient of tungsten is very small, $\alpha=4.5\times 10^{-6}$/K. During the pilot run in 2018, the temperature was monitored in the TI12 and TI18 tunnels, as shown in \tableref{environment-temp} and its variation was found to be very small, namely about 0.1 \si{\degree C}. The linear thermal expansion of 25 cm tungsten is expected to be
\[
\sigma_x = 0.25 [\textnormal{m}] \cdot \alpha [\textnormal{K}^{-1}]\cdot \sigma_T [\textnormal{K}] = 0.25\cdot 4.5\times 10^{-6} \cdot 0.1 = 0.11 \ [\mu \textnormal{m}]
\]
This value of 0.11 $\mu$m is small, and the actual absolute ``relative'' movement between the emulsion films would be smaller.

\begin{figure}[tbp]
    \centering
    \includegraphics[width=0.8\hsize]{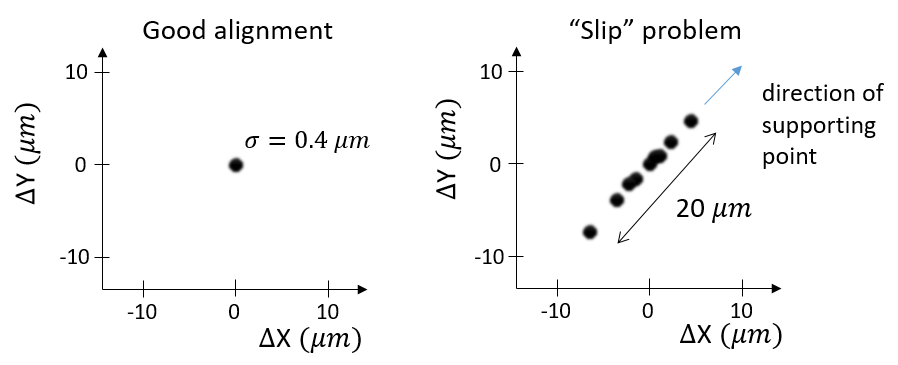}
    \caption{Schematic of the alignment problem reported in the DONUT experiment. The residuals of track position between two consecutive emulsion films are given in the $(\Delta x, \Delta y)$ plane. If the emulsion films are not pressed enough against the metallic plate, they may become misaligned.}
    \label{fig:slip}
\end{figure}

To exert sufficient pressure, a multi-step pressing mechanism will be implemented for \FASERnu, as shown in \figref{detector_structure}. First, 20 emulsion films and 20 tungsten plates will be vacuum-packed to create a modular structure in the detector. The pressure on each module is then given by the atmospheric pressure. Successively, all the 50 modules will be installed in a mechanical structure, which presses all the modules to one side. A multi-dot, deformable glue will be applied between the modules to compensate for the non-uniformity of the emulsion film thicknesses.

\begin{figure}
    \centering
    \includegraphics[width=0.9\hsize]{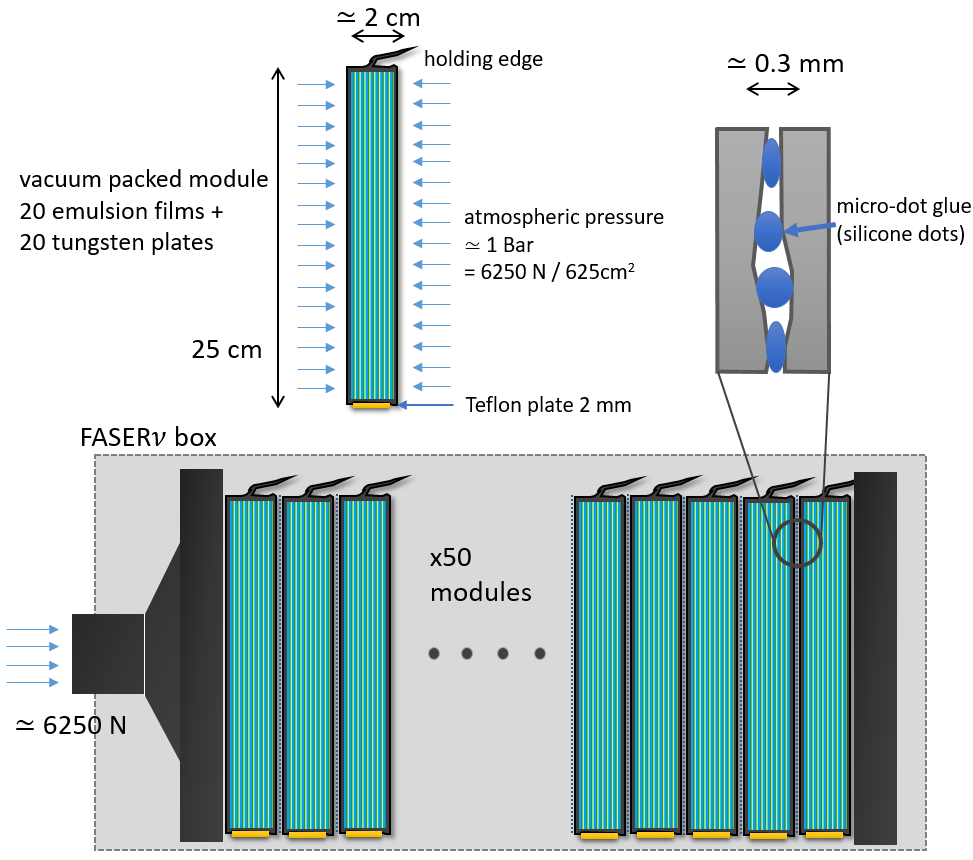}
    \caption{Module structure of the \FASERnu detector.}
    \label{fig:detector_structure}
\end{figure}

A vacuum-packed module prototype has been built with stainless-steel plates, as shown in \figref{vacuum-pack}.  An aluminum-laminated sheet was used for packing; it is custom-made by Meiwa Pax~Co., Ltd.~and can keep the vacuum for years. For the sake of safety, since each module weighs about 24 kg, a sucker cup lifter will be used.

\begin{figure}
    \centering
    \includegraphics[height=5.5cm]{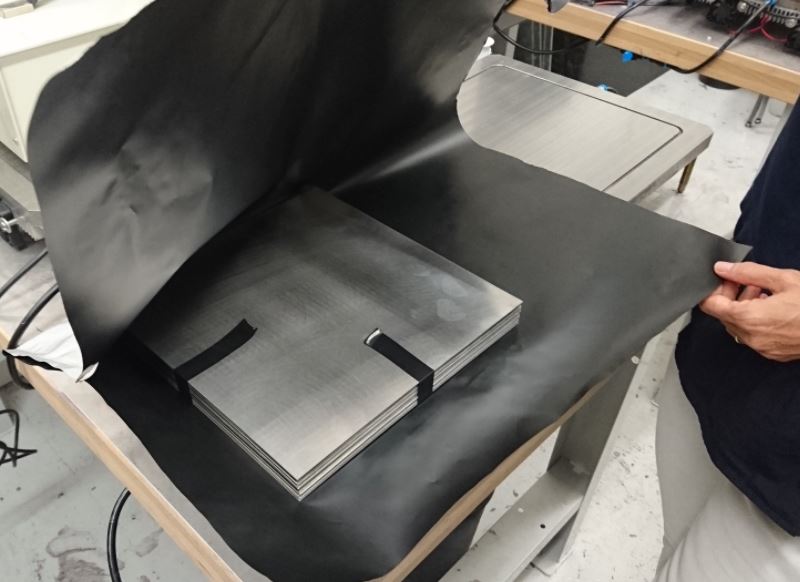}
    \includegraphics[height=5.5cm]{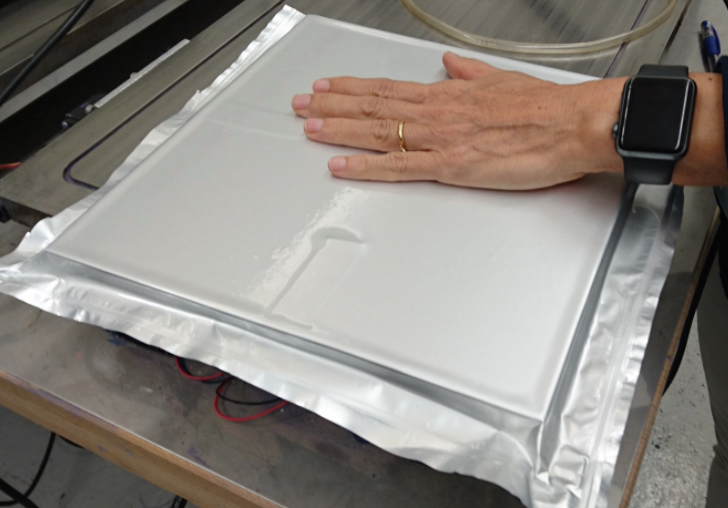}
    \includegraphics[height=5.5cm]{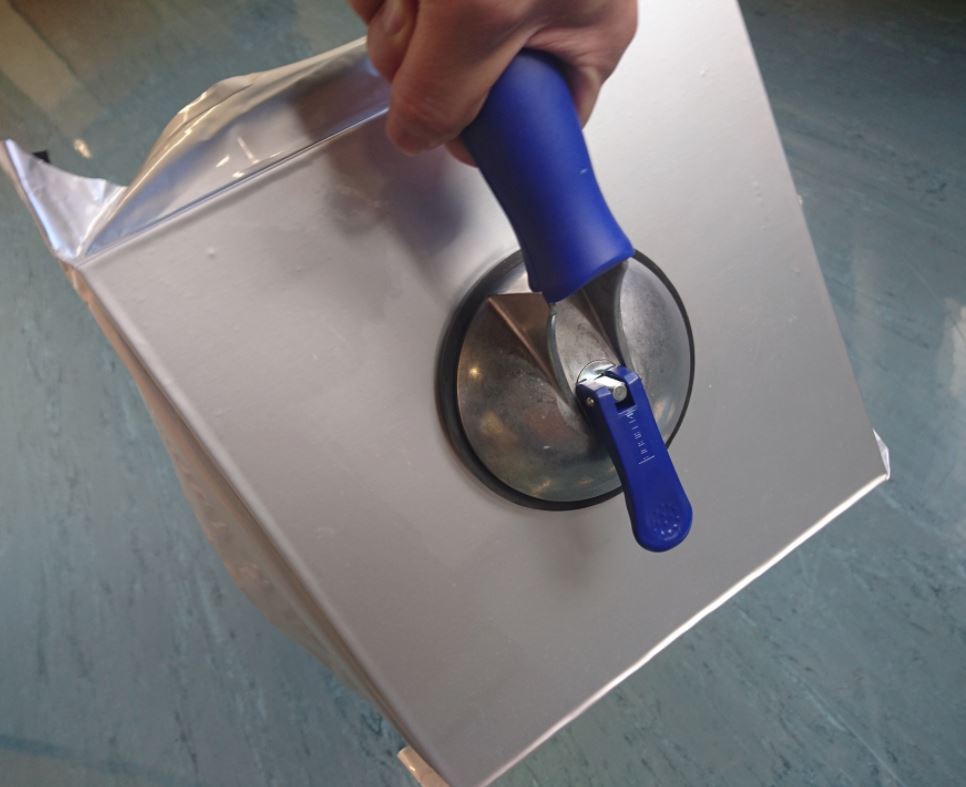}
    \caption{Assembly of a vacuum-packed module with stainless-steel plates (top), and a module being lifted by a suction cup handler (bottom). }
    \label{fig:vacuum-pack}
\end{figure}

The mechanical support, including a presser, has been designed and is shown in \figref{detector_structure2}. Given the shape of the trench, the width of \FASERnu is limited to 30~cm. Therefore the main structure, which must support the detector's entire 1.2~tonne weight, is planned on the top and bottom of the detector. It has adjustable legs to optimize the height with respect to the beam crossing angle by approximately 7 cm. The 50 vacuum-packed modules will be housed in the structure, and a force of 6250~N will be imposed by the presser (green part) located upstream.

\begin{figure}
    \centering
    \includegraphics[width=\hsize]{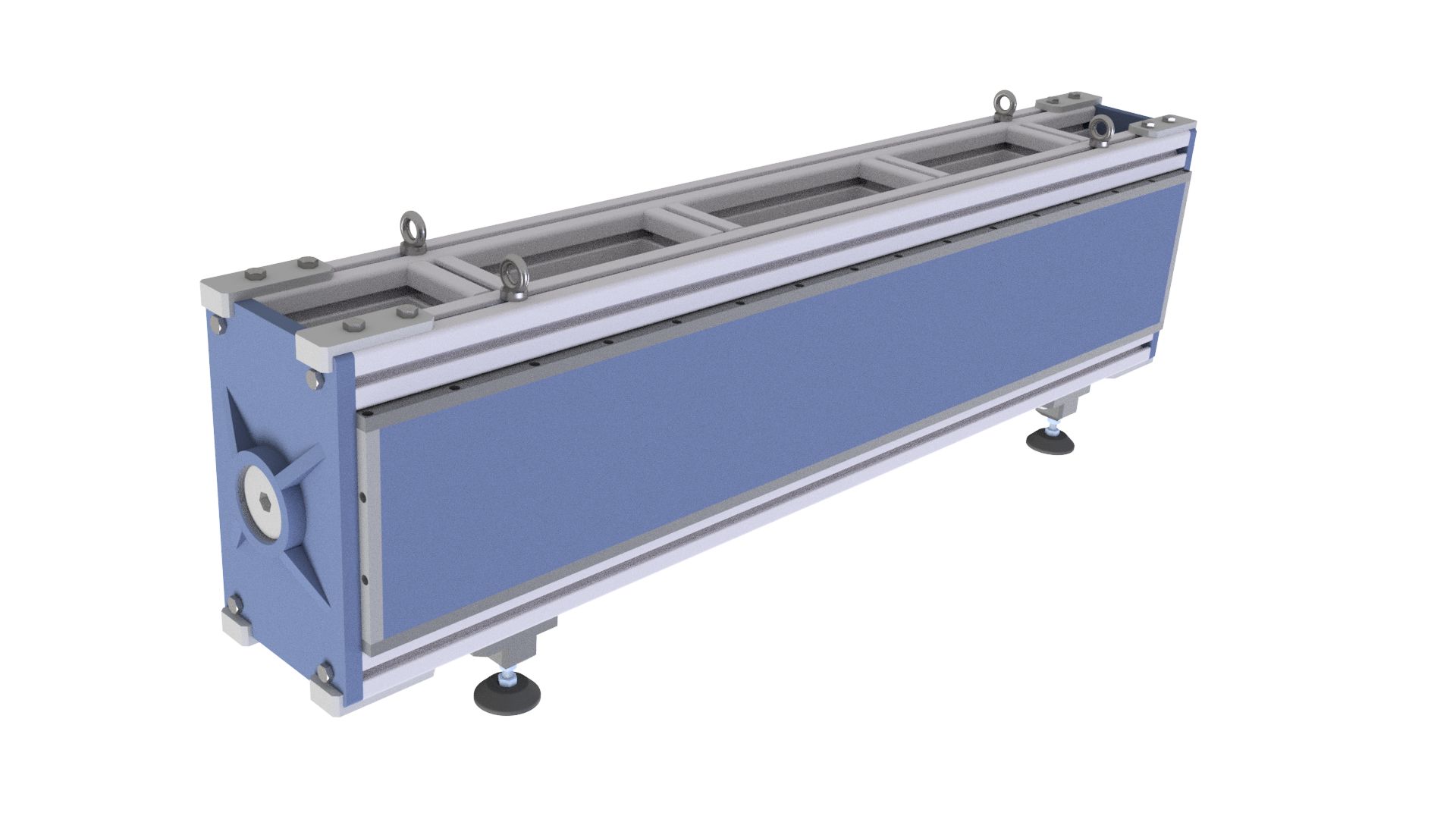}
    \includegraphics[width=0.45\hsize]{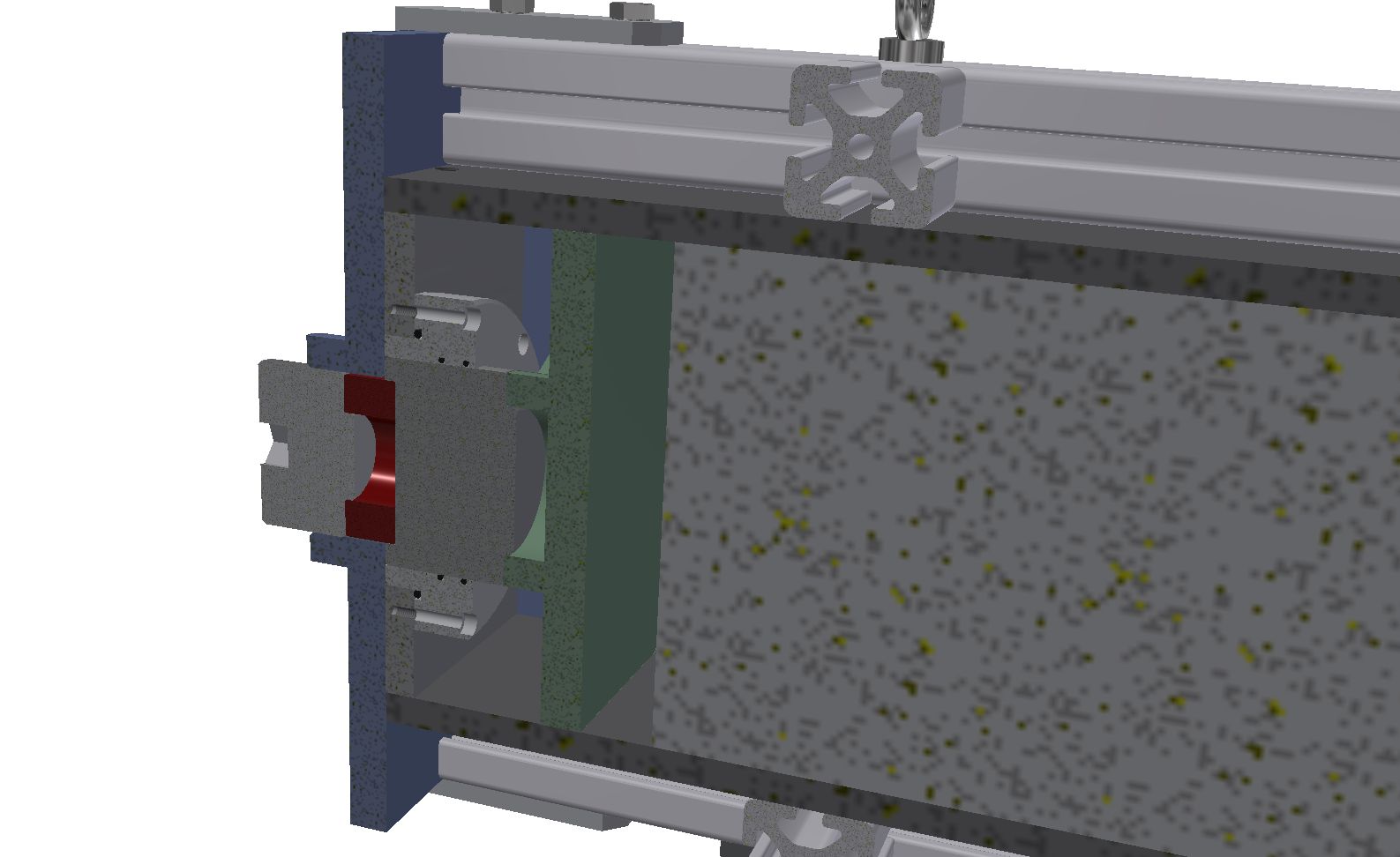}
    \includegraphics[width=0.45\hsize]{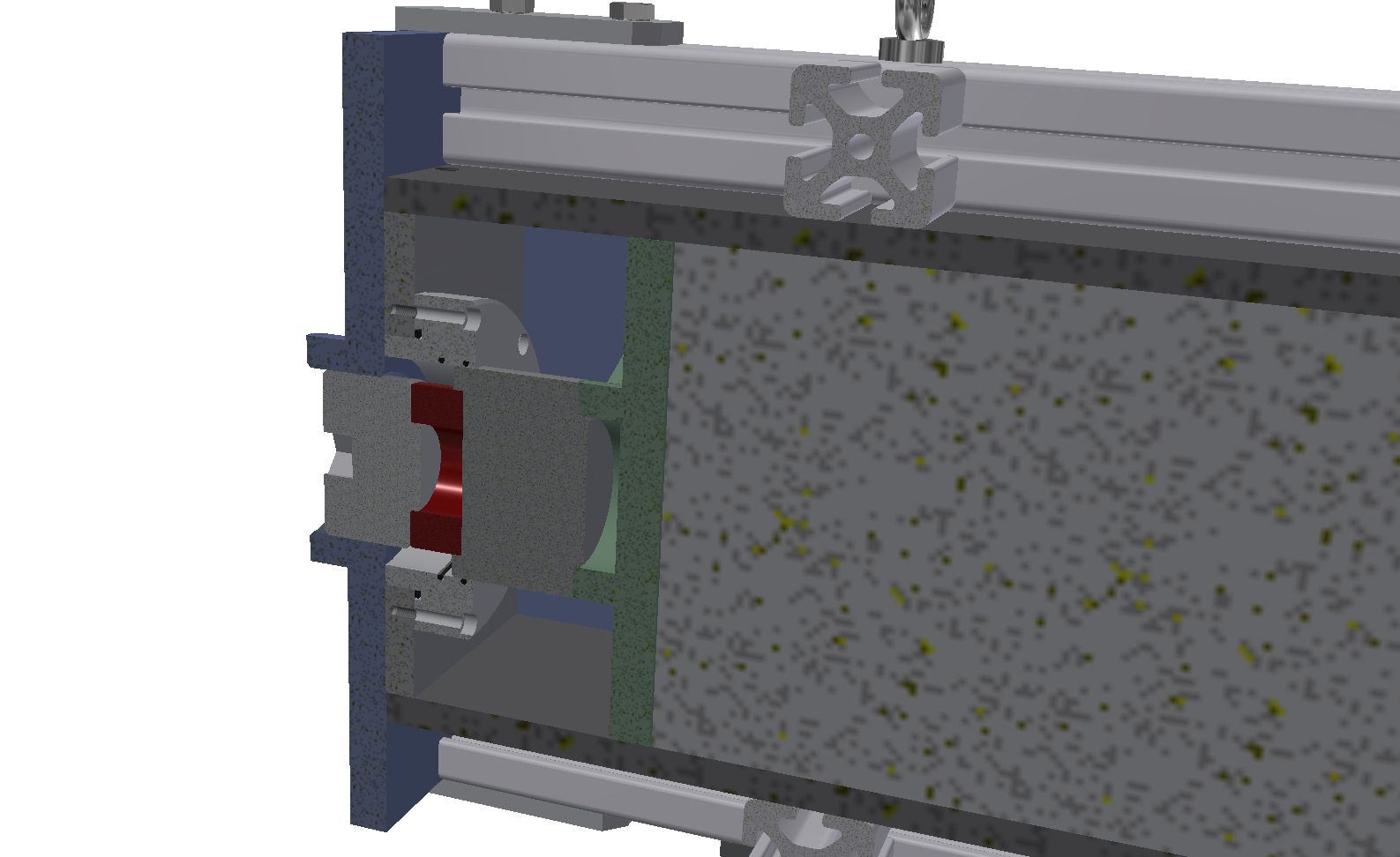}
    \caption{Detector support structure and pressing mechanism (neutrinos come from the left). Due to the narrow space, the weight of the detector is supported by the structures at the top and bottom. 50 vacuum-packed modules are inserted and a force of 6250 N is imposed by the presser (green part) located upstream. }
    \label{fig:detector_structure2}
\end{figure}

Charged particle tracks recorded in the emulsion detectors tend to fade over time. This ``fading'' effect is accelerated at high temperature. Although the temperature at the TI12 tunnel is not very high (18 \si{\degree C}), the fading might not be negligible. We are considering an option to actively cool the \FASERnu detector to 10 \si{\degree C} with a proper cooling system. It needs to be carefully designed because temperature instabilities, which have a direct impact on alignment stability, can be more problematic than fading. Fluctuations of the environmental temperature caused by the electronics of the FASER main detector could also have an impact. If the rms variation is more than 0.5 \si{\degree C}, it would be worth investigating the possibility of active cooling. We will run in 2021 without active cooling, and we will implement it later if it is found to be necessary.

%%******************************************
\subsection{Assembly}
\label{sec:assembly}
%%******************************************

The \FASERnu detector will be assembled at CERN, in the dark room of Building 169. The dark room was set up for the CHORUS experiment and has been used by several experiments, including OPERA (CNGS1), AEgIS (AD6), SHiP (P350), DsTau (NA65), and for beam tests employing emulsion detectors. A picture of the dark room is shown in the left panel of \figref{darkroom}. It would be ideal to use this facility for detector assembly and the chemical processing of emulsions for \FASERnu.

\begin{figure}
    \centering
    \includegraphics[height=6.1cm]{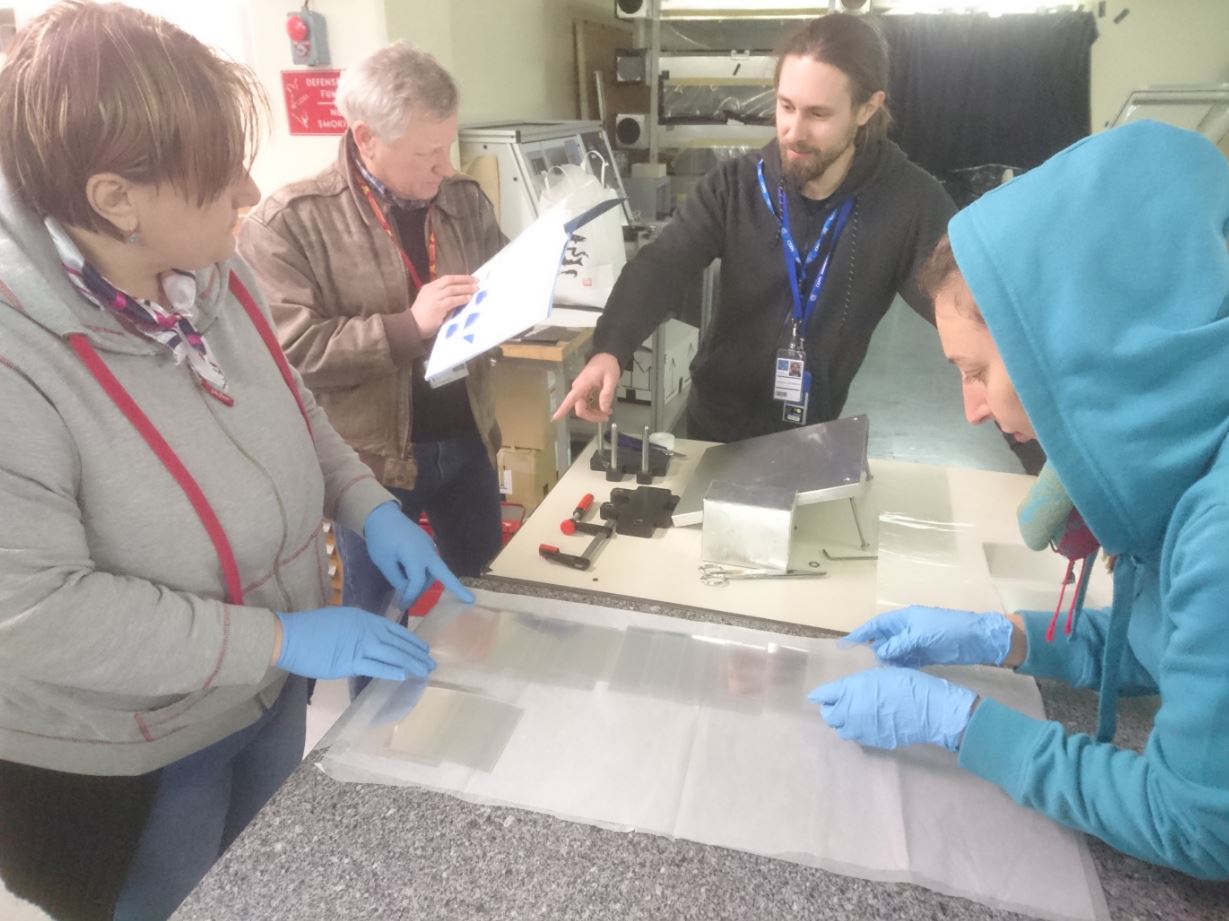}
    \includegraphics[height=6.1cm]{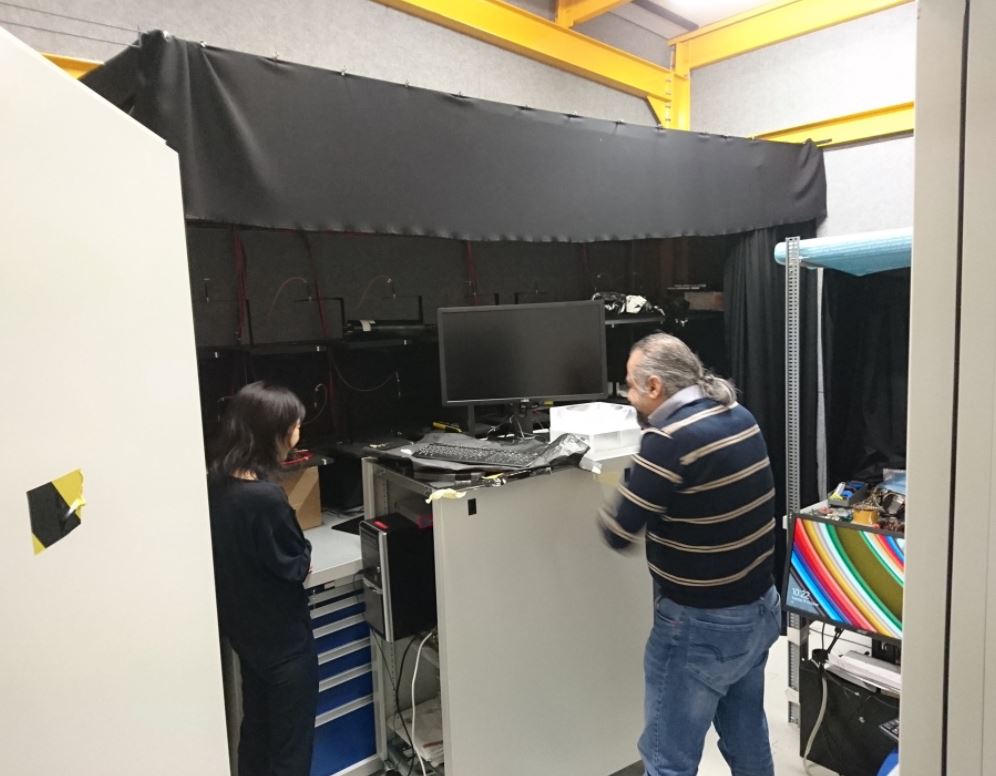}
    \caption{Dark room facilities at CERN. \textbf{Left}: The common facility dedicated to emulsion activities in Building 169. \textbf{Right}: The dark room in IdeaSquare. This room is currently used for photo sensor testing, but could be used for assembling and disassembling the \FASERnu detector.}
    \label{fig:darkroom}
\end{figure}

In the event of activity overlap, the use of this dark room has to be coordinated among experiments. As a backup solution, we are investigating the possibility to upgrade the dark room facility available at the IdeaSquare building (shown in the right panel of \figref{darkroom}) for assembling and disassembling the \FASERnu detector. The chemical treatment of emulsion films will not be feasible in the IdeaSquare facility.

Since heavy modules have to be positioned with high precision, we will assemble the entire detector at a surface facility. The 1.3~tonne detector will then be transported to the experimental site in one piece. This minimizes the amount of underground work under restricted conditions, namely, with personal protective equipment and the tight timeline of Technical Stops. It also minimizes the number of objects crossing over the LHC beamline, reducing the risk of accidents.

%%******************************************
\subsection{Transport}
\label{sec:transport}
%%******************************************

The \FASERnu detector will be brought down to the LHC tunnel using the PM15 elevator at Point 1, which can withstand a load of 3 tonnes.  It will then be transported along the LHC beamline on an electric cart. To prevent any possible damage to the LHC magnets or interconnects, the electric vehicles will be equipped with collision detectors and their speed will be limited to 3 km/h.  The \FASERnu detector will then be carried over the LHC and QRL cryo-line in UJ12 using the crane already employed for the main FASER detector installation, as shown in \figref{UJ12}. A protection device has been installed under this crane, with dimensions similar to those of \FASERnu and a 1.5 tonne load capability, which will allow operations even when the LHC is cold.\footnote{For heavy load transportation along the LHC tunnel and over the LHC when the machine is cold, a special permit is required from the CERN-TE department safety officer. According to him, it will be granted, as long as the above described procedures are followed.}

\begin{figure}
    \centering
    \includegraphics[height=5.5cm]{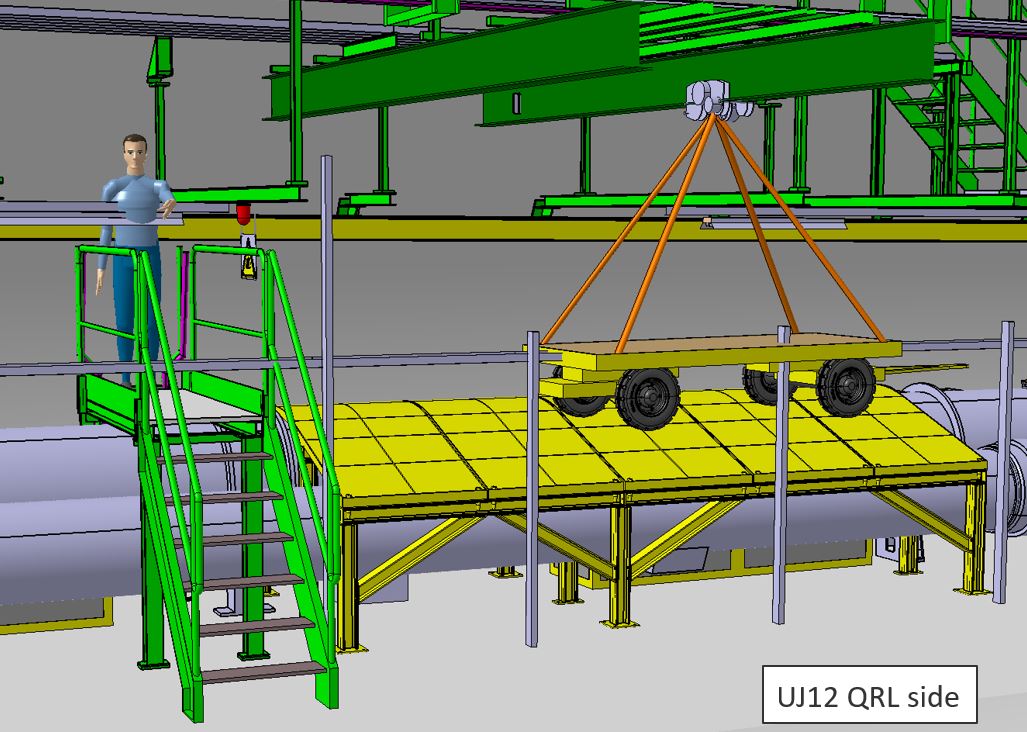}
    \includegraphics[height=5.5cm]{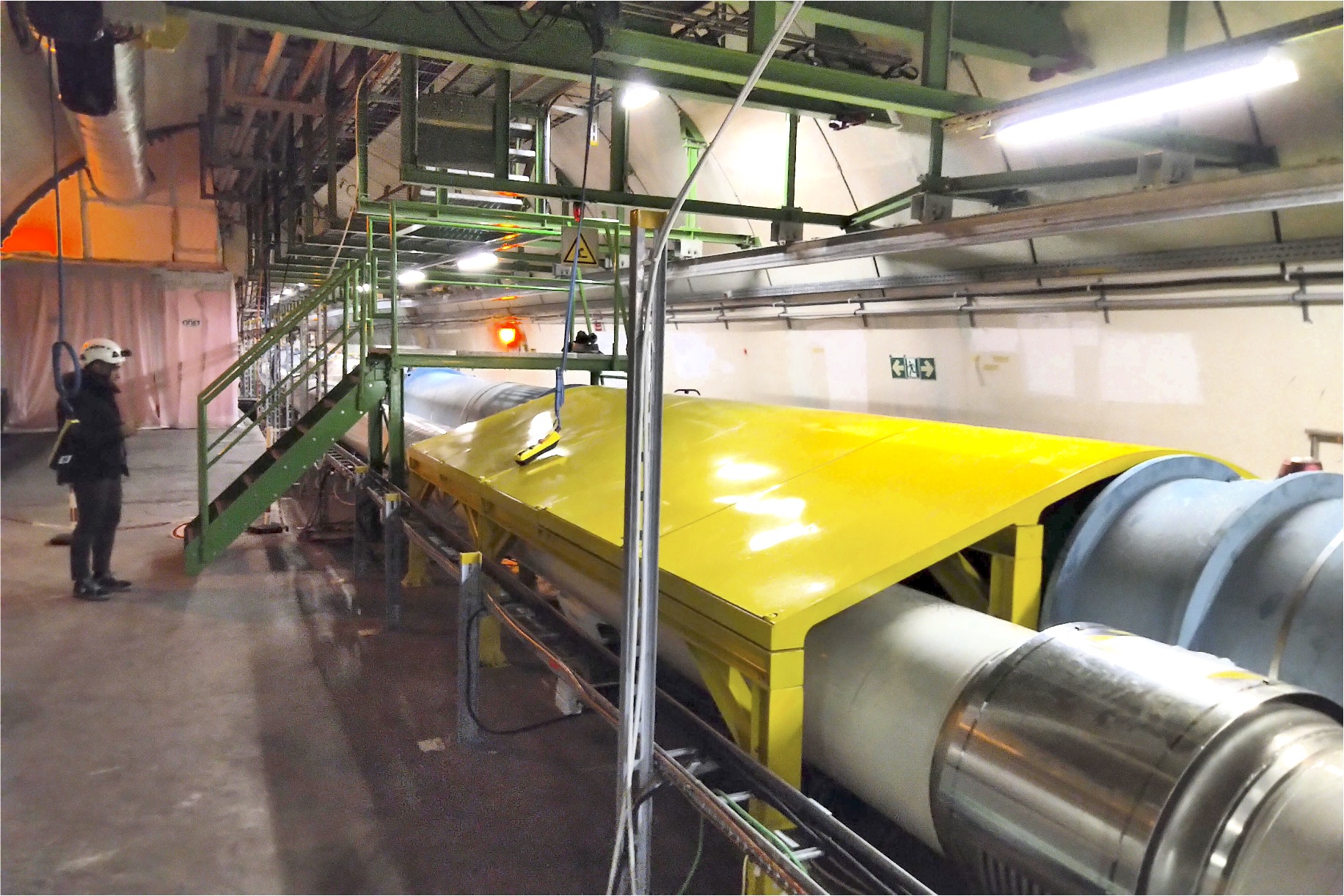}
\caption{The schematics of the transport setup at UJ12 (left) and a current photo if the site (right). The yellow QRL protection will allow us to transport \FASERnu even when the LHC is cold.}
    \label{fig:UJ12}
\end{figure}

The detector will be installed into the neutrino trench in front of the FASER detector by using the crane that will be installed in TI12 for FASER installation (scheduled to be installed in March 2020). Since the crane has to reach the neutrino trench, an additional rail will be needed in TI12, as shown in \figref{transport}.

Because of the \FASERnu operations, the crane in UJ12  will be used more frequently than originally planned. Consequently, the transport group proposes to install a crane transport platform. The additional cost for the transport infrastructure on top of that already approved to be installed for FASER is about 20 kCHF.

\begin{figure}
    \centering
    \includegraphics[width=0.8\textwidth]{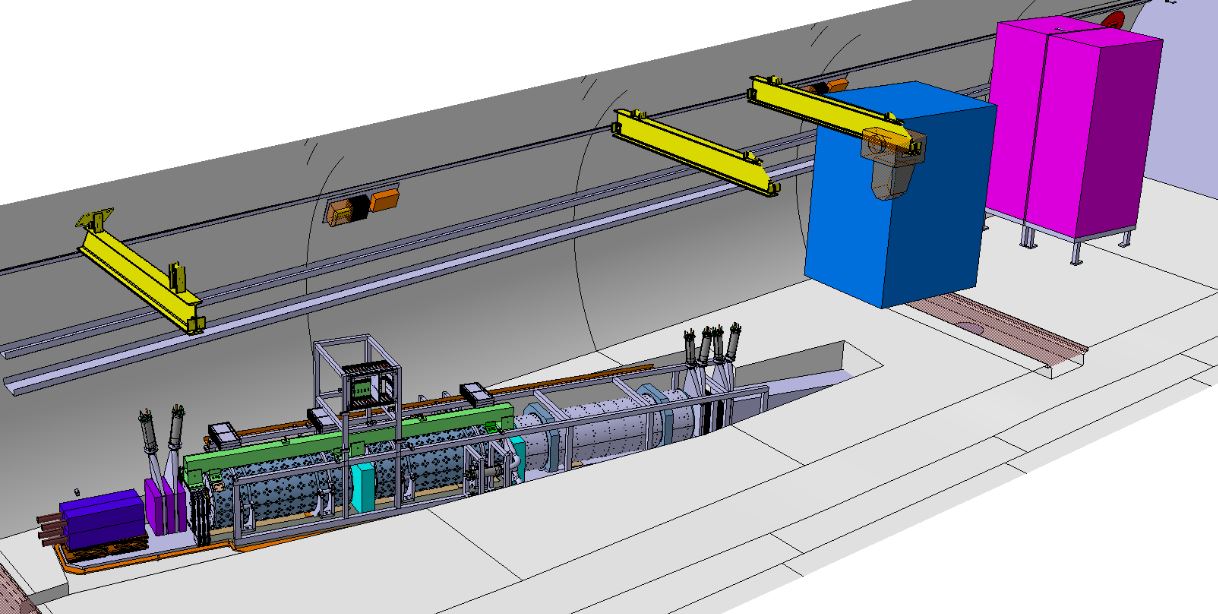}
    \caption{The rails (yellow beams) to be installed at the FASER site. The third rail (on the far right) is newly added to install \FASERnu.}
    \label{fig:transport}
\end{figure}

The Handling Engineering team (EN-HE) at CERN will be in charge of detector transportation. According to them, the transport works (installation and removal) would take $\sim \frac{1}{2}$ a day and be feasible during the Technical Stops. 

%%******************************************
\subsection{Environmental Monitoring}
%%******************************************

The monitoring of environmental conditions is essential to diagnose possible alignment problems and detector sensitivity loss. Four temperature sensors will be installed upstream and downstream and at the center-bottom and center-top of the \FASERnu detector. In addition, a humidity sensor will  be installed at the center of \FASERnu. The sensors will be connected to a TIM (Tracker Interlock Monitoring) module, and all information will be sent to the FASER DCS (Detector Control System). 

%%******************************************
\subsection{Radiation Protection}
\label{sec:rp}
%%******************************************

After the removal of the \FASERnu detector, the radiation level of \FASERnu will be measured. The radiation is expected to be very small at the FASER site. This was already checked in 2018 through \textit{in situ} measurements; no activation of the pilot emulsion detectors was observed at the time of their removal from the experimental area. 

We are in contact with the Radiation Protection (RP) team at CERN. The RP team has agreed to perform the RP scan of the detector during the Technical Stops. The scan itself would take less than 1 hour. The RP team also suggested to perform the installation/removal not on the first day of the Technical Stop, but in the following days. This is because some of the beamline components, especially the TAN, have a high activation immediately after high luminosity running. One should avoid to pass by such radioactive elements. 

In  case the detector is activated, it must stay in the buffer zone until it cools down. In order to minimize an accumulation of cosmic events, the detector should be directed towards the Jura mountains, which cover about 120 mrad from horizon at Point 1. The cosmic events and neutrino events can then be separated by the angle.

As a general feature of  heavy elements, Tungsten, can be activated relatively easily. The long term installation in TI12 over several replacements of emulsion films may cause activation of tungsten plates. The RP team is currently performing a dedicated simulation to study this effect.

%%******************************************
\subsection{Exchange of Films}
%%******************************************

The \FASERnu detector will be replaced during planned Technical Stops. The expected beam operation is described in \secref{environment}. 
In 2021 we expect 10-20~fb$^{-1}$, while, in 2022 and 2023, we may get up to $\sim$100~fb$^{-1}$ per year. In addition, we might get $\sim$100~fb$^{-1}$ in 2024 if LHC Run 3 is extended by one year. We plan to install seven sets of emulsion layers during LHC Run 3: one in 2021, three in 2022, and three in 2023. The emulsion films will be  produced a few months before each installation. The emulsion film chemical development will be performed in the dark room at CERN soon after their extraction. The exchange procedure steps are: 
(1) construction of the new emulsion modules using the unused set of tungsten plates;
(2) extraction of the exposed emulsion modules from FASER and installation of the new modules; 
(3) disassembling of the emulsion films and their chemical development in the dark room at CERN. 

It would take 3--4 days to prepare the new modules. Two to three people will work for 8--10 hours/day in the dark room, which could be divided into two shifts. The work can be done with one expert and one non-expert, and there is sufficient expertise in the Collaboration for this to be carried out. In addition, two people will be needed to conduct the extraction and installation operations. 

%%******************************************
\subsection{Chemical Development}
%%******************************************

The recorded signal in the silver bromide crystals (latent image) will be amplified by chemical development. The procedure and the chemical solutions are described in Table~\ref{tab:solutions}. 
In the developer solution, filaments of metallic silver start to grow from the latent image spec and become visible as dots under optical microscopes. The amplification gain is about $\mathcal{O}(10^8)$, and it depends on the temperature and duration of the chemical treatment.  After the development of an entire detector, which corresponds to an emulsion area of 63~\si{m^2}, about 2 tonnes of chemical waste need to be disposed (see Table~\ref{tab:solutions}). 

The chemical development will be carried out in the dark room at CERN, which will be equipped so that 200 films/day could be processed. One week will be necessary to develop 1000 films. As discussed in \secref{assembly}, the dark room of Building 169 is well-suited also for chemical development. If this facility is unavailable, we will establish another dark room at CERN. The existing emulsion development facility of the University of Bern could also be used.

\begin{table}[]
    \centering
    \small
    \begin{tabular}{|l|p{2.2cm}|p{4.8cm}|p{4.8cm}|p{1.4cm}|}
         \hline
         \hline
         Solution & \raggedright Time and Temperature & Function  & Chemical & Amount /63 \si{m^2}\\
         \hline
         \hline
         Developer & \raggedright 20 min at \si{20 \pm 0.1 \degree C}& Chemical amplification of signal with a gain of $\mathcal{O}(10^8)$& \raggedright OPERA Dev (Fujifilm), RD-90s starter (Fujifilm)& 400 L\\
         \hline
        Stopper & 10 min & Stop chemical amplification & Acetic acid& 200 L\\
         \hline
         Fixer & 90 min & Resolve unused silver bromide crystals& UR-F1 (Fujifilm)& 1150 L\\
         \hline
         Wash & $>$300 min & Wash out all chemicals & Running water & \\
         \hline
         Thickener & 20 min & \raggedright Control emulsion layer thickness & \raggedright Glycerine, Drywell (Fujifilm) & 50 L\\
         \hline
         Drying & $\sim$ 1 day & Dry films for shipment & Air at R.H.=50--60\% & \\
         \hline
         \hline
    \end{tabular}
    \caption{Solutions required for the emulsion chemical development of each \FASERnu detector, which has an emulsion film area of 63 \si{m^2}.}
    \label{tab:solutions}
\end{table}

%%%%%%%%%%%%%%%%%%%%%%%%%%%%%%%%%%%%%%%%%%%%%%%%%%%%%%
\subsection{Safety}
\label{sec:safety}
%%%%%%%%%%%%%%%%%%%%%%%%%%%%%%%%%%%%%%%%%%%%%%%%%%%%%%

The \FASERnu detector does not require electricity consumption. Furthermore it is surrounded by an aluminum support structure, and so it is not flammable.  
The relevant safety issue for \FASERnu would be the handling of heavy objects and its transportation. The basic unit of the \FASERnu detector is the vacuum-packed module, which weighs about 24~kg and will be assembled and partly handled in the dark, where special care is required. During assembly, we will handle already-furbished tungsten plates.  No powder will be present, so no special requirement or precaution will be needed for using tungsten. People should wear gloves and safety shoes. To help safe handling, a support device with vacuum suckers will be prepared.
When the assembled module is handled, the light level will be maximized, allowing for the dark room operation. Installation of the emulsion module into the support frame can be done in the light.

Additional safety issues related to detector transportation and radio protection have been discussed above in \secsref{transport}{rp}, respectively. 

%%%%%%%%%%%%%%%%%%%%%%%%%%%%%%%%%%%%%%%%%%%%%%%%%%%%%%
%%%%%%%%%%%%%%%%%%%%%%%%%%%%%%%%%%%%%%%%%%%%%%%%%%%%%%
\section{Interface Detector}
\label{sec:interface_detector}
%%%%%%%%%%%%%%%%%%%%%%%%%%%%%%%%%%%%%%%%%%%%%%%%%%%%%%
%%%%%%%%%%%%%%%%%%%%%%%%%%%%%%%%%%%%%%%%%%%%%%%%%%%%%%

The \FASERnu emulsion detector is a stand-alone detector component. As discussed in Ref.~\cite{Abreu:2019yak}, this emulsion detector will allow us to detect neutrinos, to estimate their energies, and to separate them from background. 

At the same time, there are benefits to coupling \FASERnu to the FASER spectrometer, which is located immediately downstream of \FASERnu. This can be done by means of an interface detector, an additional silicon tracker layer that interfaces \FASERnu with the main detector, as shown in \figref{detector_upgrade}. Such a hybrid emulsion/electronic detector has been successfully demonstrated in several experiments, such as E653~\cite{Kodama:1993wg}, WA75~\cite{Albanese:1985wk}, CHORUS~\cite{Eskut:2007rn}, DONuT~\cite{Kodama:2007aa}, and OPERA~\cite{Acquafredda:2009zz}. 
If an event in the emulsion detector and the interface detector can be matched, a combined analysis will be possible, giving the following improvements:

\begin{figure}[htpb]
\centering
\includegraphics[width=0.8\textwidth]{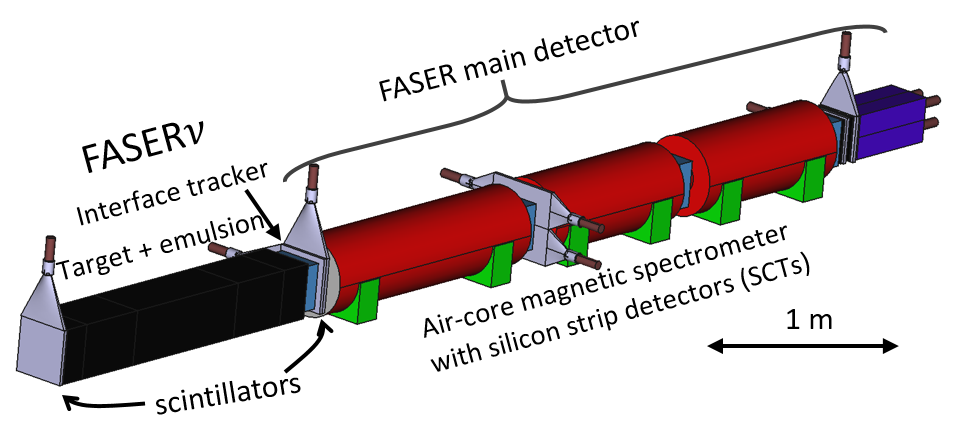}
\caption{Upgrade of the \FASERnu detector to include an interface detector, which couples \FASERnu to the FASER main detector. The components of the overall detector include the \FASERnu emulsion detector (black), scintillators (grey), the interface detector and additional tracking layers (blue), magnets (red), and the electromagnetic calorimeter (purple). 
}
\label{fig:detector_upgrade}
\end{figure}

\begin{description}[leftmargin=0.16in]

\item [Charge Identification] The FASER spectrometer can distinguish muons from anti-muons, and thus provide the charge information needed to distinguish muon neutrino and anti-muon neutrino interactions. The lepton charge identification would allow us to measure neutrino and anti-neutrino cross sections separately, making the measurements significantly more incisive. Charge identification for electrons and taus would be difficult because of their paths; electrons make electromagnetic showers and taus mostly decay into hadrons.
\item [Improvement of energy resolution] The FASER spectrometer can provide an additional measurement of the charged particle momenta and thus improve the energy resolution. 

\item [Background rejection] By only using the emulsion detector it is often impossible to relate spatially disconnected segments of an event, for example a muon and an associated muon-initiated neutral hadron interaction (background). The time information provided by the interface detector would allow us to correlate the different parts of the same event and reject muon-initiated backgrounds. Moreover, a scintillator positioned upstream of the emulsion detector provides an additional opportunity to identify an incoming muon, improving the discrimination between the muon-induced background and the neutrino interaction events.  
\end{description}

A typical event in \FASERnu+FASER is shown in \figref{interface}. The  tracks detected in \FASERnu will appear as hits in the interface detector, and matches of the hit patterns can be analyzed. In each set of \FASERnu data ($\sim 30~\ifb$), about $4 \times 10^8$ muons are expected (see \secref{bg}), most of which are single-track events. On the other hand, we expect $\mathcal{O}(10^4)$ neutrino and hadron events. If we require 1 mm positional matching in both dimensions of the  detector ($25\,\cm \times 25\,\cm$) and an angular matching of 10 mrad in the 50-mrad angular spread of background particles, the probability of fake matching is $(1/250)^2 (10/50)^2 = 6\times 10^{-7}$. By requiring that two or more tracks match, the correspondence between events in \FASERnu and the FASER main detector will be uniquely identified.

The interface detector should have a spatial resolution of the order of 0.1 mm.  Our current baseline is to copy the tracker station of the FASER spectrometer, which is made of ATLAS silicon strip detectors (SCT)~\cite{ATLAS:1997ag, ATLAS:1997af}. The SCT modules consist of two layers of strip detectors with a pitch of 80 $\mu$m, that are tilted by 40 mrad. The effective resolution is 23 $\mu$m in one direction, and 580 $\mu$m in the other. Each tracker station has three planes of 8 SCT modules. Each plane has an area of 24 cm $\times$ 24 cm. 

\begin{figure}
    \centering
    \includegraphics[width=0.8\textwidth]{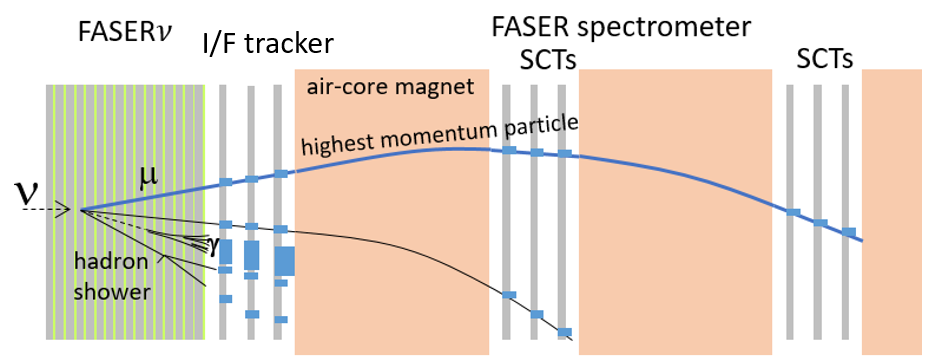}
    \caption{Schematic of the FASER/\FASERnu global reconstruction with an additional interface tracker.}
    \label{fig:interface}
\end{figure}

In October 2019, we requested that the ATLAS SCT Collaboration allow us to use an additional 40 spare SCT modules, sufficient to construct 5 tracking planes, for the interface detector. (This request is in addition to the 80 spare modules already granted to us to construct the FASER main detector tracking stations.)  The SCT Collaboration kindly approved this request in October 2019. The quality assessment of these SCT modules will be performed by the end of 2019.

To avoid any interference between construction of \FASERnu and the construction and commissioning of the FASER main detector, we plan to install the interface detector after the 2021 run, during the year-end Technical Stop. This would also allow us to carefully design the interface detector.

As an example, a simulated $\nu_\mu$ CC neutrino interaction event with $E_\nu=1~\tev$ is shown in the top panel of \figref{g4_muon_neut}. Due to the long lateral length of \FASERnu of $10~\lambda_{int}$, a large fraction of hadrons would interact before exiting \FASERnu. Therefore it is reasonable to use the last track segments in \FASERnu to match the hits in the interface detector. These \FASERnu tracks will be compared with all events in the interface detector which have multiple tracks. Such multi-track events would also frequently be generated by primary high energy muons due to the electromagnetic showers resulting from knock-on electrons. An example event is shown in the bottom panel of \figref{g4_muon_neut} for a $1~\tev$ muon. However, such muon-initiated background events can be removed using an additional veto scintillator which will be installed at the upstream part of \FASERnu to tag the incoming muons. 

\begin{figure}
 \centering
 \includegraphics[width=16cm]{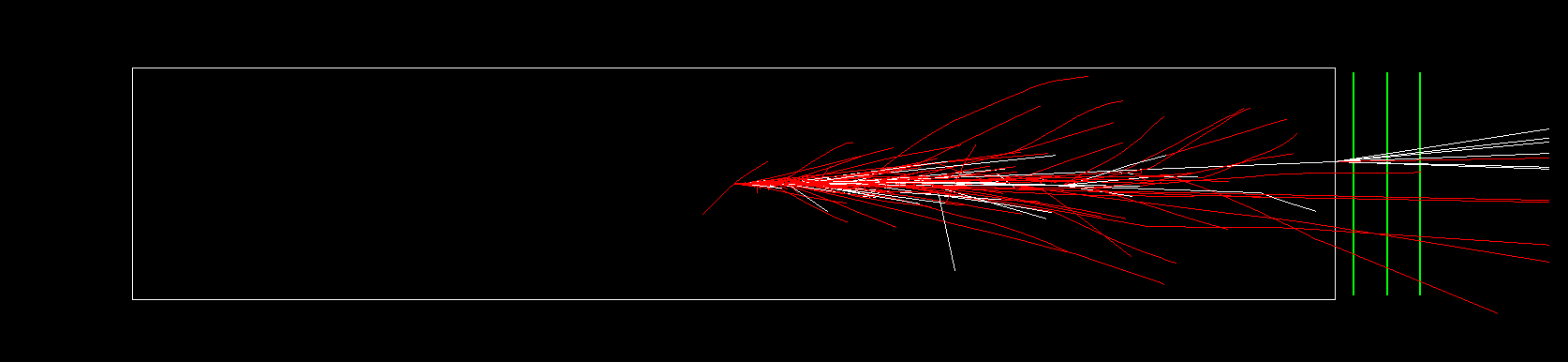}
 \includegraphics[width=16cm]{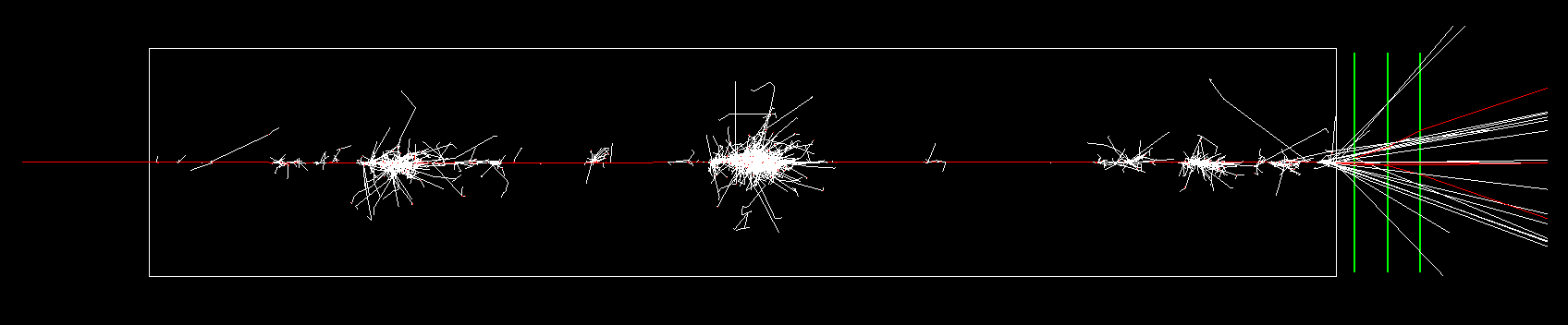}
 \caption{ Illustration of the primary and secondary particle trajectories of a 1 TeV $\nu_\mu$ (top) and a 1 TeV $\mu^-$ (bottom) through the emulsion and interface detectors. The white box indicates the \FASERnu emulsion detector, and the green vertical lines indicate the silicon planes of the interface detector. The particles enter from the left. The primary interaction of the incident neutrino with the tungsten nucleus is set at the center of \FASERnu. The charged (neutral) particles are indicated by red (white) lines. In the neutrino event, only particles with momenta larger than 1 GeV are shown, corresponding to the track selection in the emulsion detector. In the muon event, lower-energy particles ($E\gtrsim 0.1$ MeV) are also shown, corresponding to the threshold of the SCT. }
 \label{fig:g4_muon_neut}
\end{figure}

Event matching of upstream neutrino events is expected to be less efficient due to the long lateral thickness, which means charged hadrons produced in the neutrino decay can interact before reaching the interface detector. Additionally, the active transverse area of the FASER spectrometer is a circle of radius 10 cm which is smaller than the emulsion detector (25 cm $\times$ 25 cm).  The fiducial volume of \FASERnu where matching to the FASER spectrometer is possible will therefore be reduced compared to the full \FASERnu volume. Nevertheless, the limited efficiency/acceptance can be compensated by the abundant statistics of $\nu_\mu$ and $\bar{\nu}_\mu$ events. Therefore the goal to separately measure the neutrino and anti-neutrino cross sections can be achieved. Further detailed studies of the design and performance of global reconstruction are under way.

%%%%%%%%%%%%%%%%%%%%%%%%%%%%%%%%%%%%%%%%%%%%%%%%%%%%%%
%%%%%%%%%%%%%%%%%%%%%%%%%%%%%%%%%%%%%%%%%%%%%%%%%%%%%%
\section{Offline Analysis}
\label{sec:offline_analysis}
%%%%%%%%%%%%%%%%%%%%%%%%%%%%%%%%%%%%%%%%%%%%%%%%%%%%%%
%%%%%%%%%%%%%%%%%%%%%%%%%%%%%%%%%%%%%%%%%%%%%%%%%%%%%%

%%******************************************
\subsection{Detector Simulation Framework}
%%******************************************

We are preparing a complete simulation to study neutrino interactions, as well as background muon interactions, with the \FASERnu detector. \textsc{Geant4}~\cite{Agostinelli:2002hh} is used for the emulsion and interface detectors. The emulsion detector is composed of 1000 layers, with each layer consisting of a 1 mm thick tungsten plate and an emulsion film, as described in \secref{tungsten_emulsion_detector}. 

The interface detector is made of three silicon planes with transverse dimensions of 24~cm$\times$24~cm. Each plane is composed of eight  6~cm$\times$12~cm modules, and each module consists of a 380 $\mu$m carbon baseboard (thermal pyrolytic graphite with a density of 2.2~g/cm$^3$), and two silicon strip layers glued to each side of the baseboard. The silicon layer has a thickness of 285 $\mu$m, and the glue (epoxy with a density of 0.95~g/cm$^3$) has a thickness of 25 $\mu$m on each side of the baseboard. The separation between two adjacent planes is set to 3.5 cm, and the separation between the first interface plane and the emulsion detector is 2.0 cm. The glue material is mainly made of the elements H, C, O, and N. 

The neutrino interactions will be generated, for example, by the \textsc{Genie} and Pythia neutrino interaction generators. The generated primary particles  from the neutrino-nucleus interaction are further passed to \textsc{Geant4} for subsequent simulations. Figure \ref{fig:g4event} shows such a neutrino interaction event visualization. The initial neutrino $\nu_\mu$  energy is 1 TeV, and the first primary interaction is generated by \textsc{Genie}. This  simulation framework will be used for future studies.

\begin{figure}
    \centering
    \includegraphics[height=5cm]{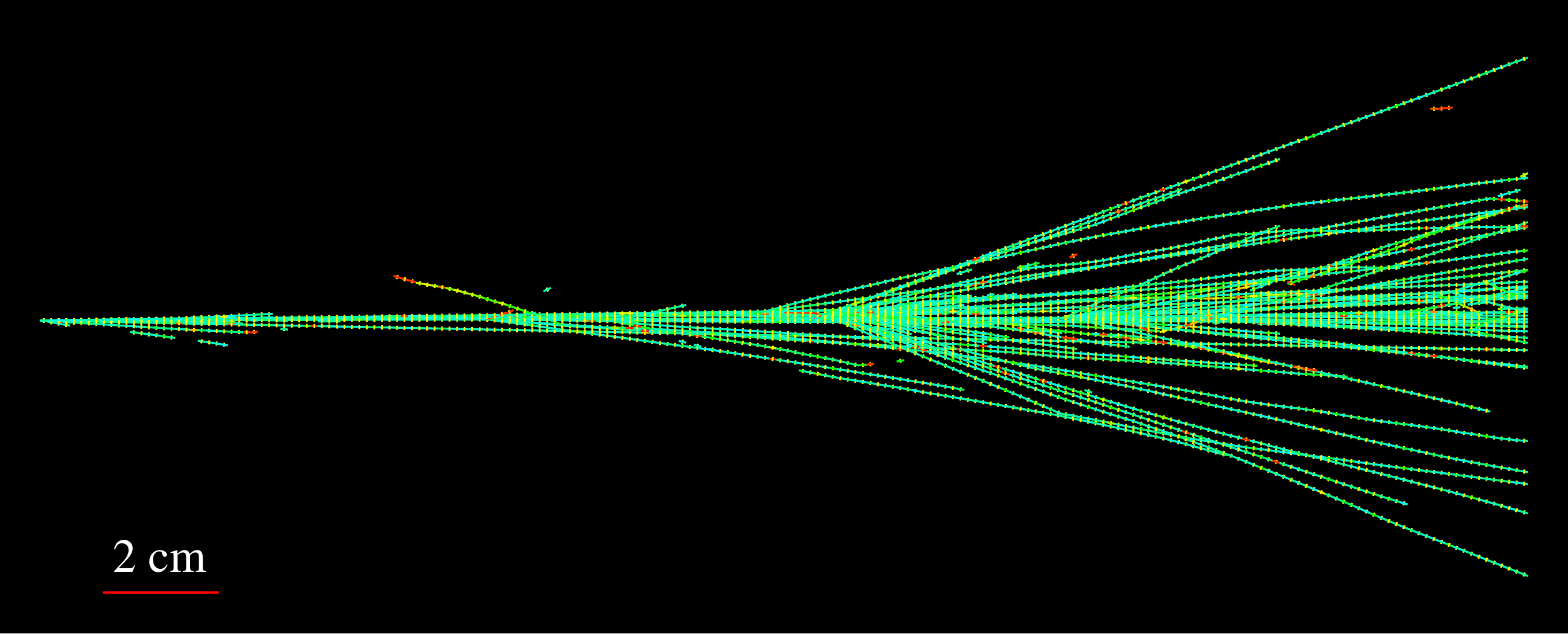}  
    \includegraphics[height=5cm]{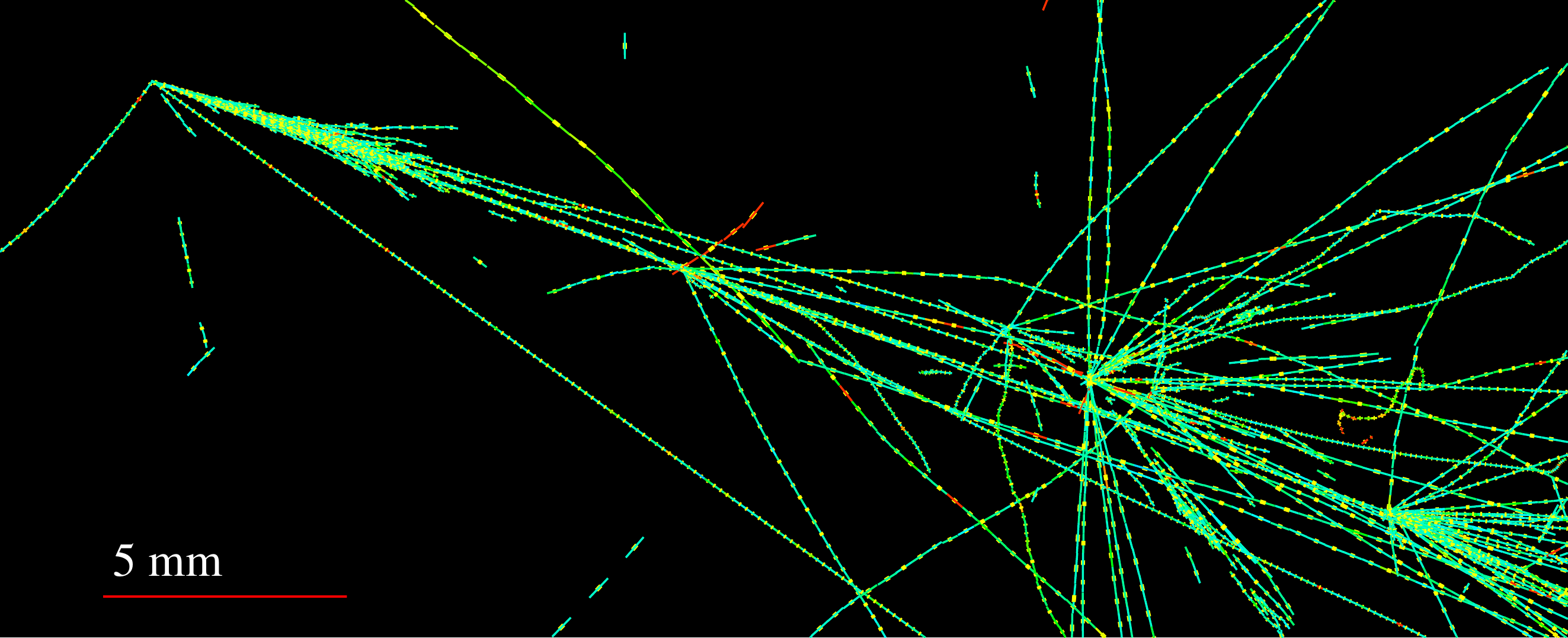}
    \caption{1 TeV $\nu_\mu$ CC interactions generated by \textsc{Genie} and propagated in \FASERnu by \textsc{Geant4}. The hits on emulsion films are shown along the depth of 200 tungsten plates, corresponding to $57\, X_0$ and $2\,\lambda_{\text{int}}$. Cuts on momentum ($P>0.3$ GeV) and angle ($\tan\theta<0.5$, where $\theta$ is the angle with respect to the neutrino direction) are applied. The color shows $dE/dx$, where green indicates minimum-ionizing particles and red represents heavily-ionizing particles. \textbf{Top}: side view. \textbf{Bottom}: tilted view.
     }
    \label{fig:g4event}
\end{figure}

%%******************************************
\subsection{Emulsion Readout}
%%******************************************

The emulsion readout system takes a sequence of tomographic images by changing the focal plane through each emulsion layer, as shown in \figref{readout}. The digitized images are then analyzed to recognize sequences of aligned grains as a track segment (microtracks). In \FASERnu, the neutrino event analysis will be based on readout of the full emulsion detectors by the Hyper Track Selector (HTS) system~\cite{Yoshimoto:2017ufm} in Japan, which is the fastest readout system at present. The HTS system includes a dedicated lens, camera, XYZ-axis stage, and computer cluster for image processing; see \figref{hts}. It takes 22 tomographic images, and 16 successive images in the emulsion layer are used for track recognition.

\begin{figure}[htbp]
\begin{center}
\vspace{5mm}
\includegraphics[height=5.5cm]{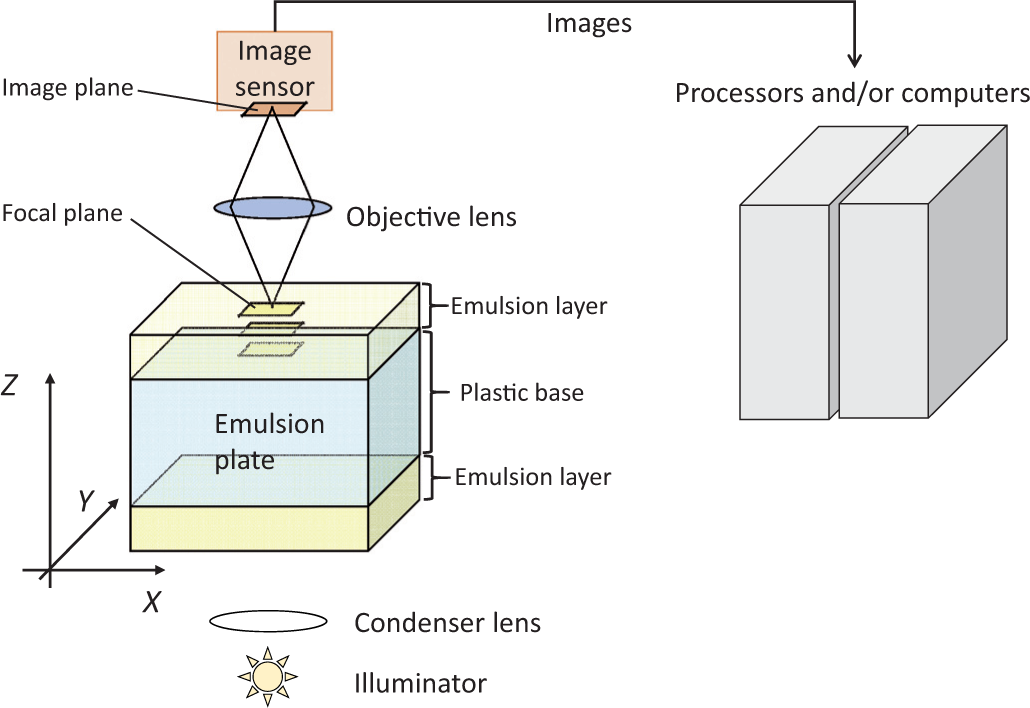}
\caption{Schematic view of the emulsion readout system~\cite{Yoshimoto:2017ufm}.
}
\label{fig:readout}
\end{center}
\end{figure}

\begin{figure}[htbp]
\begin{center}
\includegraphics[height=7cm]{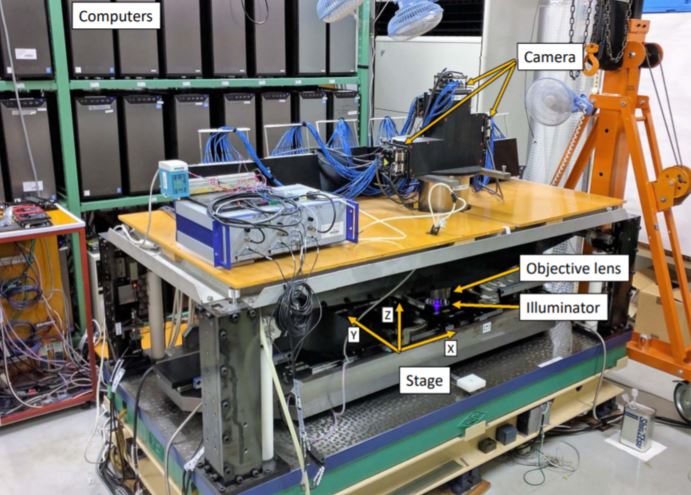}
\caption{The fast emulsion readout system (HTS) in Japan~\cite{Yoshimoto:2017ufm}.
}
\label{fig:hts}
\end{center}
\end{figure}

Conventional systems use a field of view of 0.12 mm $\times$ 0.12 mm or 0.3 mm $\times$ 0.4 mm. HTS makes use of a custom-made objective lens with a much larger field of view of 5.1 mm $\times$ 5.1 mm and a magnification of 12.1. 
The optical path is divided into six, and correspondingly the image is projected on to six mosaic camera modules. Each mosaic camera module consists of 12 2.2-Mpixel image sensors. In total, 72 image sensors work in parallel to build the large field of view. The raw image data throughput from 72 image sensors amounts to 48 GBytes/s, which is then processed in real time by 36 tracking computers with two GPUs each.  The readout speed of the HTS system is 0.45~m$^2$/hour/layer, which is a big leap from the previous generations, as shown in Table~\ref{table:scanning_speed}. 
By the time of the runs in 2021/2022, an upgraded HTS system (HTS2, which will be about 5 times faster) will also be available. The baseline plan for \FASERnu is to use HTS, since its performance, such as the readout speed and resolution, is already proven. The total emulsion film surface to be analyzed in \FASERnu is 189 m$^2$/year implying a readout time of 840 hours/year. Assuming some hours of machine time each day, it will be possible to finish reading out the data taken in each year within a year.

\begin{table}[htbp]
\centering
\begin{tabular}{|l|c|c|}
\hline
\hline
\                        & Field of view [mm$^2$]      & Readout speed [cm$^2$/hour/layer] \\
\hline 
SUTS (used in OPERA)     & 0.04                & 72            \\ \hline 
HTS \ (running)          & 25                  & 4500          \\ \hline 
HTS2 (under development) & 50                  & 25000         \\ \hline
\hline
\end{tabular}
\caption{Comparison of recent emulsion scanning systems and their performance properties.
}
\label{table:scanning_speed}
\end{table}

The information stored by the HTS system is the track segments recorded in the top and bottom layers of films, or ``microtracks.''  By connecting the microtracks on both layers, ``basetracks'' are formed, as shown in \figref{basetrack}, which are the basic unit used in later processing. Each basetrack provides 3D coordinates $\vec{X}=(x,y,z)$, 3D vector information $\vec{V}=(\tan\theta_x,\tan\theta_y,1)$, and $dE/dx$ parameters, where the vector information has an angular resolutions of 2 mrad. 

\begin{figure}[tbp]
\begin{center}
\includegraphics[width=0.55\textwidth]{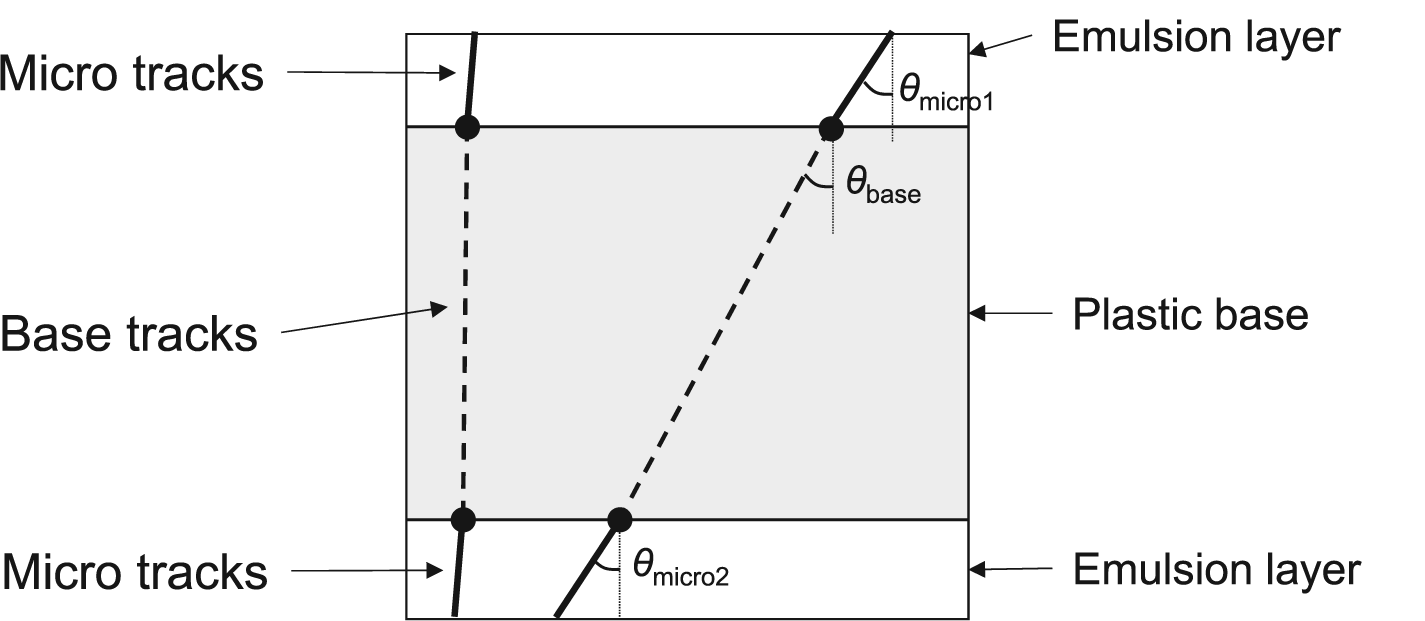}
\caption{Definitions of microtracks and basetracks.
}
\label{fig:basetrack}
\end{center}
\end{figure}

%%******************************************
\subsection{Data Reconstruction}
%%******************************************

\FASERnu expects high-energy particles ($\sim$ TeV) at high density ($\mathcal{O}(10^5)\mu/\cm^2$). The reconstruction of these particles requires software dedicated to such a high-energy and high-density environment, for which the experience accumulated for the DsTau experiment (NA65) can be directly used~\cite{Ariga:2018gan,Aoki:2019jry}. As an example, the films may be aligned by using high-energy muon tracks. The data processing is divided into sub volumes, for example, 2 cm $\times$ 2 cm $\times$ 30 emulsion films. The alignment precision has a dependence on the processing area size, which is due to a non-linear distortion of the plastic base. With this method, an alignment accuracy of 0.4 $\mu$m has been reached in the work of the DsTau Collaboration~\cite{Aoki:2019jry}, as shown in \figref{alignment}. Depending on the purpose, the processing unit will be optimized.

\begin{figure}[tbp]
    \centering
    \includegraphics[height=4.5cm]{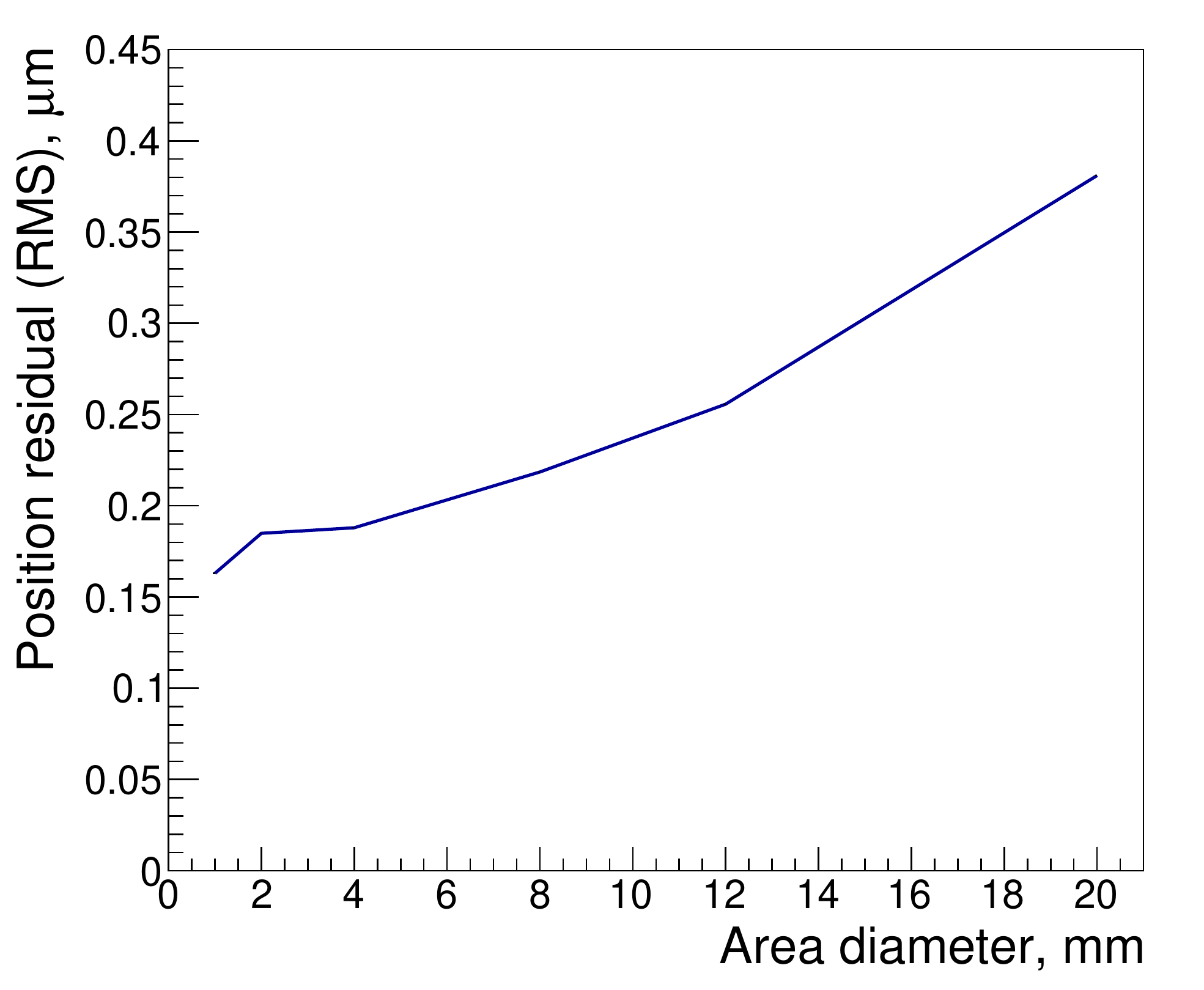}
    \includegraphics[height=4.5cm]{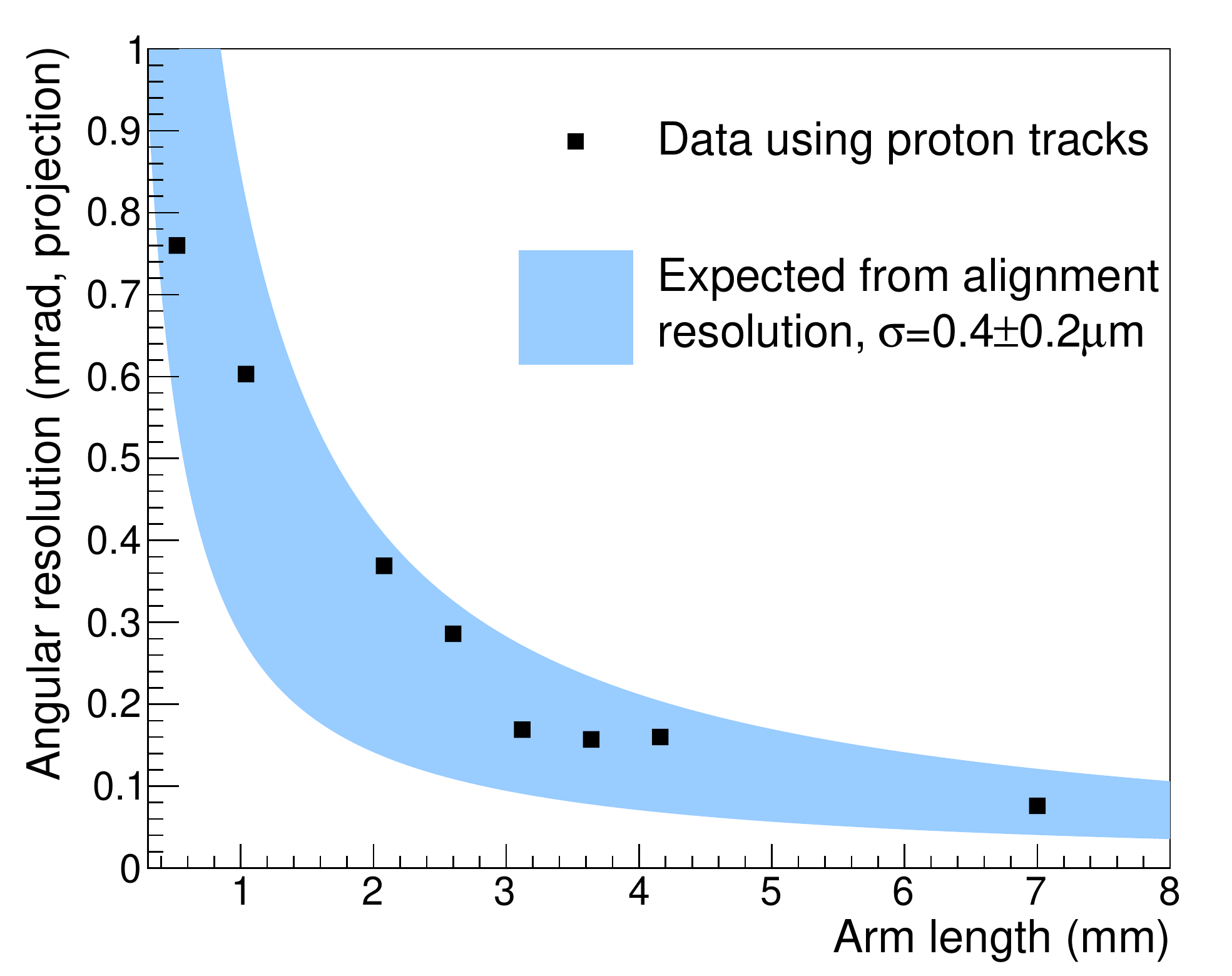}
    \caption{\textbf{Left}: Alignment resolution obtained in data as a function of the diameter of the data processing unit. The position displacement of basetracks from the reconstructed track (a straight line) is evaluated. \textbf{Right}: Angular resolution of tracks obtained by the HTS system as a function of reconstructed track length. The $x$-axis is the length used for the angular measurement. These results are from the work of the DsTau Collaboration in Ref.~\cite{Aoki:2019jry}.}
    \label{fig:alignment}
\end{figure}

The basic concept of track reconstruction is based on the correspondence of basetracks on different films in position and angular space. The widely-used algorithm uses a correspondence test of two consecutive basetracks. However, 
the track density in \FASERnu (10$^5$--10$^6$ /cm$^2$ in a small angular space) is relatively high. The conventional reconstruction tools for OPERA, which had a track density of 10$^2$--10$^3$ /cm$^2$ in a large angular space, are not appropriate. This is especially true if two or more tracks with similar direction (within a few mrad) become close to each other (in a few micron), because the algorithm may not resolve the correct paths.

A new tracking algorithm has recently been developed by the DsTau Collaboration to reconstruct tracks in environments with high track density and narrow angular spreads. Some detail of the algorithm is described in Ref.~\cite{Aoki:2019jry}. The reconstruction tools will be  implemented in the \FASERnu analysis.

After the processing to reconstruct tracks in the full area of the emulsion films, a systematic analysis will be performed to locate neutrino interactions. The analysis for the neutrino interactions is described in Ref.~\cite{Abreu:2019yak}. 

%%******************************************
\subsection{Neutrino Energy Reconstruction}
\label{sec:energyresolution}
%%******************************************

An important part of the analysis of \FASERnu data is the reconstruction of the neutrino energy. Information from the track reconstruction and the identification of the neutrino vertex will be used to estimate the neutrino energy. This requires studying charged particles and the electromagnetic component associated with both the leptonic and hadronic recoil products of neutrino scatterings. The latter are important because of the possible large momentum transfer in DIS. 

Because of its high spatial resolution, \FASERnu will be able to precisely determine the final state topology and estimate kinematic quantities. In particular one can employ (i) topological variables, such as track multiplicity and the slopes of tracks, (ii) track momenta measured with the use of the multiple Coulomb scattering (MCS) method~\cite{Kodama:2002dk}, and (iii) the energy measurement of the EM component performed by the analysis of shower development in the emulsion detector. A more detailed discussion of the relevant variables and methods has been given in Ref.~\cite{Abreu:2019yak}.

We have considered two strategies for energy reconstruction, based on only the visible energy measurements and also a multivariate analysis. In fact, even a simple sum of the visible energy of charged particles and EM showers already gives a relatively good estimate of the neutrino energy, as illustrated in the left panel of \figref{ann}. For energies above $1~\tev$, however, this is limited by the precision of the momentum measurement in the MCS method. 

Further improvement can be achieved by the use of a combination of the aforementioned topological and kinematical variables serving as an input to an artificial neural network (ANN). We present example results of such an analysis in the central panel of \figref{ann} for $\nu_\mu$ CC events. These have been obtained by employing the ANN algorithm implemented using the MLP package in the CERN ROOT framework~\cite{Antcheva:2009zz}. The algorithm was trained on data simulated in \textsc{Genie} and by taking into account realistic EM energy and charged particle momentum reconstruction resolutions. Here we haven't done this study with a realistic density of background tracks. The EM shower reconstruction would be affected, therefore we accounted for a relatively large uncertainty on the shower energy measurement of $50\%$. Thanks to the use of the ANN algorithm, the expected resolution of the neutrino energy measurement is about $30\%$ for a wide range of energies between $100~\gev$ and several $\tev$, as shown in the right panel of \figref{ann}. 

\begin{figure}[t]
\centering
\includegraphics[width=\textwidth]{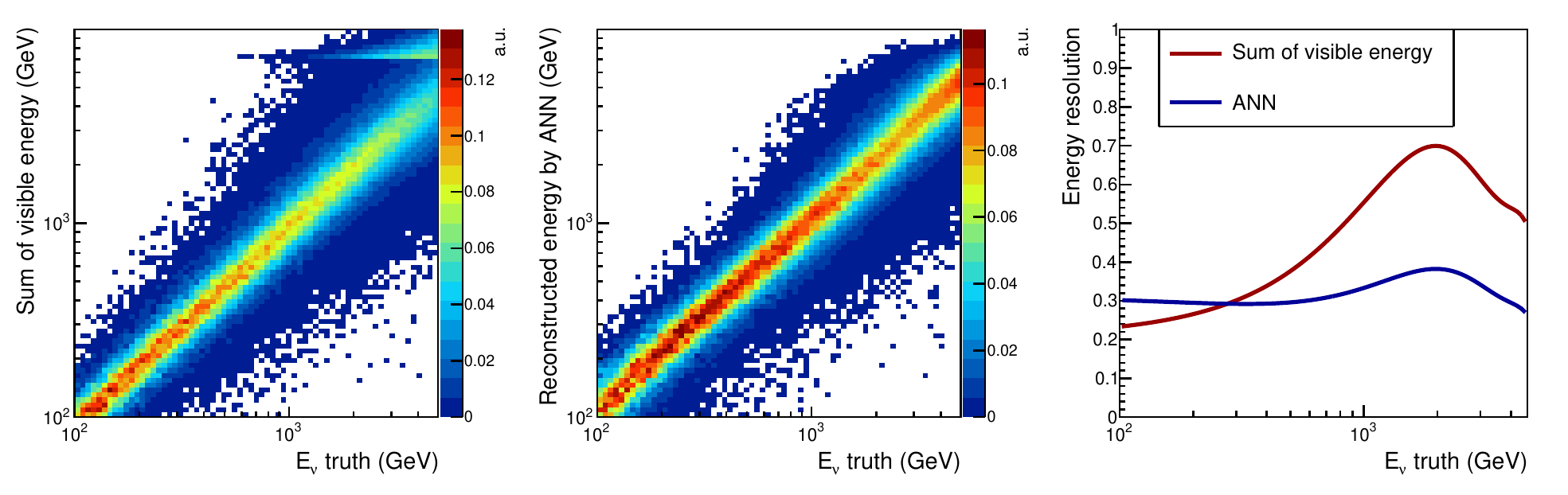}
\caption{\textbf{Left}: Neutrino energy and sum of visible energy (momentum of charged particles and energy of electromagnetic showers) for $\nu_\mu$ CC samples with at least five charged tracks $n_{\text{tr}} \geq 5$, with smearing (MC). 
\textbf{Center}: Neutrino energy reconstruction based on the ANN. \textbf{Right}: $\Delta E_{\nu}^{\text{ANN}} / E_{\nu}^{\text{true}}$ for the same sample. An energy resolution of 30\% (rms) was obtained for the energies of interest. From Ref.~\cite{Abreu:2019yak}.
}
\label{fig:ann}
\end{figure}

Prior to the physics run, thorough tests will be performed to cross-validate the ANN algorithm with respect to the simple estimate of visible energy at $E_\nu<1~\tev$, as well as to assess the impact of possible inaccuracies in \textsc{Genie} modeling of high-energy neutrino interactions in tungsten, as discussed in \secref{nuc}. 

%%%%%%%%%%%%%%%%%%%%%%%%%%%%%%%%%%%%%%%%%%%%%%%%%%%%%%
%%%%%%%%%%%%%%%%%%%%%%%%%%%%%%%%%%%%%%%%%%%%%%%%%%%%%%
\section{Coordination, Schedule, and Cost}
\label{sec:cost_and_schedule}
%%%%%%%%%%%%%%%%%%%%%%%%%%%%%%%%%%%%%%%%%%%%%%%%%%%%%%
%%%%%%%%%%%%%%%%%%%%%%%%%%%%%%%%%%%%%%%%%%%%%%%%%%%%%%

%%******************************************
\subsection{Coordination and Schedule}
%%******************************************

\FASERnu is an additional component of the FASER detector, and its construction and operation will be organized by the FASER Collaboration.  Given the already-tight schedule of the FASER main detector construction, the FASER Collaboration will put priority on the completion the main detector. The part of \FASERnu that is independent of the resources required for the main detector, namely the tungsten/emulsion detector, will be ready in time for data-taking in 2021.  On the other hand, the interface detector would require a non-negligible effort from the spectrometer team. 
Simulation studies are ongoing to evaluate the effectiveness of the matching between \FASERnu events and the main spectrometer using the interface detector. Based on these results, the implementation of the interface detector could take place after the end of the 2021 run possibly in time for 2022 data-taking. The lack of an interface detector in 2021 will have a minor impact on physics performance, as we expect only a small fraction of the total luminosity to be gathered in 2021 (see \tableref{operation}).

The global schedule of \FASERnu is given in \figref{schedule}. The emulsion films will be produced in Japan a few months before installation, which is optimal to avoid unnecessary cosmic background, but still provides enough time to assess the quality of the emulsion films. The films will then be shipped to CERN by air. The \FASERnu detector will be assembled right before installation to minimize background events from cosmic neutral hadron interactions.  The emulsion films will be replaced in every Technical Stop. For 2021, since the expected integrated luminosity is relatively small, the detector will remain in the TI12 location longer than in later years. However, various diagnostic tests will be conducted by sampling parts of the detector, which will allow timely feedback to optimize subsequent runs. The chemical processing of emulsion films will be done after each data taking, and the developed films will then be sent to the scanning facility at the remote site. The readout of emulsion films will take about 4 months (depending on the machine time and shift organization for exchanging films) for each chunk of data. As a result, the data collection time and the readout time for each set of emulsion films are roughly comparable. 

The reconstruction tools will be shared with the DsTau Collaboration, which largely saves the effort of software development. The FASER Collaboration is currently working on the analysis of the 2018 pilot runs, which provides a good training sample to establish the data processing pipeline. A large part of the software will be ready by the 2021 run.

In parallel to the planning, construction, and operation of the experiment, we will improve our understanding of neutrino fluxes and the corresponding uncertainties. As outlined in \secref{flux}, these efforts include (i) the creation of a dedicated forward physics tune, based on available external/internal data to quantify the uncertainties from hadronic interaction models, (ii) the full implementation of beam transport by BDSIM, and (iii) the analysis of nuclear effects and their impact on the neutrino energy estimation. These efforts will involve \FASERnu associate members as well as external groups, such as the \textsc{Professor} team, which is already involved in the preparation of this Technical Proposal.

\begin{figure}[tbp]
    \centering
    \includegraphics[width=\textwidth]{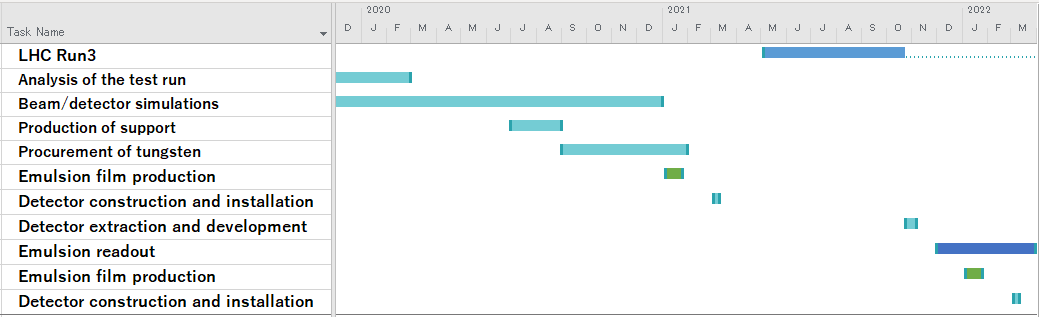}
    \includegraphics[width=\textwidth]{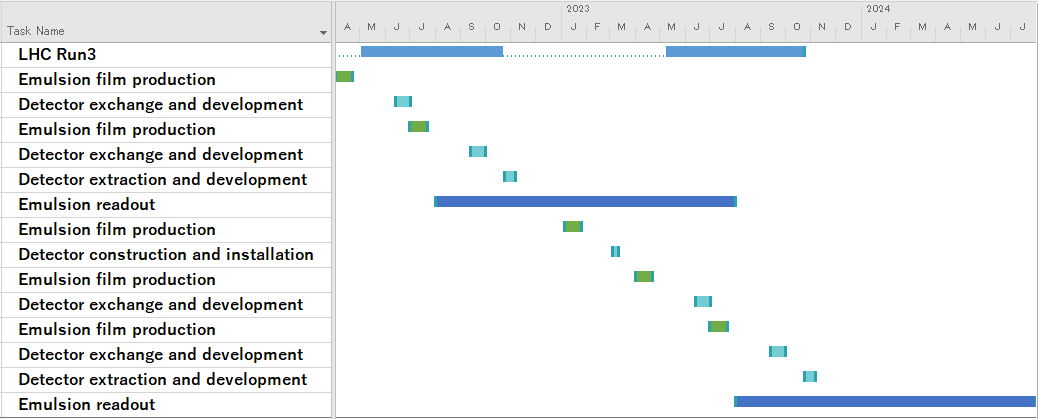}
    \caption{Timeline of the project. 
    }
    \label{fig:schedule}
\end{figure}

\if0
\begin{table}[htbp]
\centering
\begin{tabular}{|l|r|r|}
    \hline
    Tasks                            & Period (days) & Workload (FTE) \\
    \hline
    Emulsion film production         & 105 & 2.00 \\
    \hline
    Detector construction            &  21 & 2.00 \\
    \hline
    Detector installation/extraction &  14 & 1.00 \\
    \hline
    Emulsion development             &  98 & 1.60 \\
    \hline
    Emulsion readout (readout)       & 560 & 0.35 \\
    \hline
    Emulsion readout (film exchange) & 560 & 0.15 \\
    \hline
\end{tabular}
\caption{Needed time and workload for each task.}
\label{table:workload}
\end{table}
\fi

%%******************************************
\subsection{Cost Estimate}
%%******************************************

The cost estimates for the tungsten/emulsion detector and for the interface detector are summarized in Tables~\ref{table:tun_emul_detector_cost} and \ref{table:interface_detector_cost}, respectively. The largest budget items are the emulsion gel and tungsten plates. For the tungsten plates, we have included two sets, so that a replacement detector can be prepared ahead of time, and the \FASERnu detector can be replaced quickly during the short Technical Stops. If LHC Run 3 is extended by one year, an additional 135 kCHF for emulsion gel and about 23 kCHF for other consumables will be required. 

The CERN host lab costs are summarized in Table \ref{table:host_lab_cost}. It includes infrastructure work. These costs are relatively small, given the infrastructure work that is already underway to prepare TI12 for FASER, but the cost of transportation infrastructure includes 20 kCHF for the additional transport infrastructure discussed in \secref{transport}.

\begin{table}[htbp]
\centering
\begin{tabular}{|l|r|}
\hline
{\bf Item}   & {\bf Cost [kCHF]} \\
\hline
Emulsion gel for 440 m$^2$ & 315 \\
Emulsion film production cost for 440 m$^2$ & 32 \\
Tungsten plates, 1200 kg (first set) & 173 \\
Tungsten plates, 1200 kg (second set) & 173 \\
Packing materials & 5 \\
Support structure & 12 \\
Chemicals for emulsion development & 20 \\
Tools for emulsion development & 5 \\
Racks for emulsion film storage & 5 \\
Computing server & 10 \\
\hline
\textbf{Total} & \textbf{750} \\
\hline 
[Emulsion gel for 2024 running] & [135] \\
{[}Additional consumables for 2024 running{]} & [23] \\
\hline
\textbf{[Total including 2024 running]} & \textbf{[908]} \\
\hline 
\end{tabular}
\caption{Cost estimate for the tungsten/emulsion detector.
}
\label{table:tun_emul_detector_cost}
\end{table}

\begin{table}[htbp]
\centering
\begin{tabular}{|l|r|}
\hline
{\bf Item}   &{\bf Cost [kCHF]} \\
\hline
SCT modules           &  0 \\
Tracker mechanics     & 20 \\
Chiller               &  0 \\
Connection to EN-CV   &  2 \\
Tracker Readout Board &  5 \\
Powersupplies         & 31 \\
Cables                & 11 \\
Flex cable            & 10 \\
Patch panel           & 10 \\
Tracker Interlock and Monitoring Board &  3 \\
\hline
\textbf{Total} & \textbf{92} \\
\hline
\end{tabular}
\caption{Cost estimate for the interface detector. 
}
\label{table:interface_detector_cost}
\end{table}

\begin{table}[htbp]
\centering
\begin{tabular}{|l|r|}
\hline
{\bf Item}   &{\bf Cost [kCHF]} \\
\hline
Transport infrastructure  &  20 \\
Chemical disposal         &  20 \\
Computing storage (EOS disk space) &  30 \\
\hline
\textbf{Total} & \textbf{70} \\
\hline
\end{tabular}
\caption{The host lab costs. 
}
\label{table:host_lab_cost}
\end{table}

\newpage
%%%%%%%%%%%%%%%%%%%%%%%%%%%%%%%%%%%%%%%%%%%%%%%%%%%%%%
%%%%%%%%%%%%%%%%%%%%%%%%%%%%%%%%%%%%%%%%%%%%%%%%%%%%%%
\section*{Acknowledgments}
%%%%%%%%%%%%%%%%%%%%%%%%%%%%%%%%%%%%%%%%%%%%%%%%%%%%%%
%%%%%%%%%%%%%%%%%%%%%%%%%%%%%%%%%%%%%%%%%%%%%%%%%%%%%%

The FASER Collaboration gratefully acknowledges invaluable assistance from the CERN Physics Beyond Colliders study group; the LHC Tunnel Region Experiment (TREX) working group; and the following CERN teams:  survey, safety, radioprotection, transport, civil engineering, beam instrumentation and FLUKA simulations. We are grateful to the DsTau Collaboration for providing their spare emulsion films for the {\em in situ} measurements in 2018 and Yosuke Suzuki for helping with the film production; as well as to Stefan Hoeche, Yoshitaka Itow, Masahiro Komatsu, Michelangelo Mangano, Steven Mrenna, Tanguy Pierog, Hiroki Rokujo, Hank Sobel, and Ralf Ulrich for useful discussions. This work was supported in part by grants from the Heising-Simons Foundation (Grant Nos.~2018-1135 and 2019-1179) and the Simons Foundation (Grant No.~623683). AA is supported in part by the Albert Einstein Center for Fundamental Physics, University of Bern. The work of TA is supported by JSPS KAKENHI Grant Number JP 19H01909 and a research grant from the Mitsubishi Foundation. PBD acknowledges the support of the U.S.~Department of Energy under Grant Contract desc0012704. JLF~is supported in part by U.S.~National Science Foundation Grant Nos.~PHY-1620638 and PHY-1915005 and Simons Investigator Award \#376204. IG~is supported in part by U.S.~Department of Energy Grant DOE-SC0010008. FK~is supported in part by DE-AC02-76SF00515 and in part by the Gordon and Betty Moore Foundation through Grant GBMF6210. ST~is supported in part by the Lancaster-Manchester-Sheffield Consortium for Fundamental Physics under STFC grant ST/P000800/1. This work is supported in part by the Swiss National Science Foundation. 

\bibliography{references}

\providecommand{\href}[2]{#2}\begingroup\raggedright\begin{thebibliography}{10}

\bibitem{Abreu:2019yak}
{\bf FASER} Collaboration, H.~Abreu {\em et al.}, ``{Detecting and Studying
  High-Energy Collider Neutrinos with FASER at the LHC},''
\href{http://arxiv.org/abs/1908.02310}{{\tt arXiv:1908.02310 [hep-ex]}}.
%%CITATION = ARXIV:1908.02310;%%.

\bibitem{Feng:2017uoz}
J.~L. Feng, I.~Galon, F.~Kling, and S.~Trojanowski, ``{ForwArd Search
  ExpeRiment at the LHC},''
  \href{http://dx.doi.org/10.1103/PhysRevD.97.035001}{{\em Phys. Rev.} {\bf
  D97} (2018) no.~3, 035001},
\href{http://arxiv.org/abs/1708.09389}{{\tt arXiv:1708.09389 [hep-ph]}}.
%%CITATION = ARXIV:1708.09389;%%.

\bibitem{Ariga:2018zuc}
{\bf FASER} Collaboration, A.~Ariga {\em et al.}, ``{Letter of Intent for
  FASER: ForwArd Search ExpeRiment at the LHC},''
  \href{http://arxiv.org/abs/1811.10243}{{\tt arXiv:1811.10243
  [physics.ins-det]}}. \url{https://cds.cern.ch/record/2642351}.
Submitted to the CERN LHCC on 18 July 2018.
%%CITATION = ARXIV:1811.10243;%%.

\bibitem{Ariga:2018pin}
{\bf FASER} Collaboration, A.~Ariga {\em et al.}, ``{Technical Proposal for
  FASER: ForwArd Search ExpeRiment at the LHC},''
  \href{http://arxiv.org/abs/1812.09139}{{\tt arXiv:1812.09139
  [physics.ins-det]}}. \url{http://cds.cern.ch/record/2651328}.
Submitted to the CERN LHCC on 7 November 2018.
%%CITATION = ARXIV:1812.09139;%%.

\bibitem{Ariga:2018uku}
{\bf FASER} Collaboration, A.~Ariga {\em et al.}, ``{FASER's physics reach for
  long-lived particles},''
  \href{http://dx.doi.org/10.1103/PhysRevD.99.095011}{{\em Phys. Rev.} {\bf
  D99} (2019) no.~9, 095011},
\href{http://arxiv.org/abs/1811.12522}{{\tt arXiv:1811.12522 [hep-ph]}}.
%%CITATION = ARXIV:1811.12522;%%.

\bibitem{emulsion}
A.~Ariga, T.~Ariga, G.~De~Lellis, A.~Ereditato, and K.~Niwa, {\em Particle
  Physics Reference Library - Vol. 2. Detectors for Particles and Radiation},
  ch.~Nuclear Emulsions.
\newblock Springer, in press. ISBN: 978-3-030-35317-9.

\bibitem{Beni:2019pyp}
N.~Beni {\em et al.}, ``{XSEN: a $\nu$N Cross Section Measurement using High
  Energy Neutrinos from pp collisions at the LHC},''
\href{http://arxiv.org/abs/1910.11340}{{\tt arXiv:1910.11340
  [physics.ins-det]}}.
%%CITATION = ARXIV:1910.11340;%%.

\bibitem{Baltay:1988au}
C.~Baltay {\em et al.}, ``{$\nu_\mu - \nu_e$ Universality in Charged Current
  Neutrino Interactions},''
\href{http://dx.doi.org/10.1103/PhysRevD.41.2653}{{\em Phys. Rev.} {\bf D41}
  (1990)  2653}.
%%CITATION = PHRVA,D41,2653;%%.

\bibitem{Kodama:2007aa}
{\bf DONuT} Collaboration, K.~Kodama {\em et al.}, ``{Final tau-neutrino
  results from the DONuT experiment},''
  \href{http://dx.doi.org/10.1103/PhysRevD.78.052002}{{\em Phys. Rev.} {\bf
  D78} (2008)  052002},
\href{http://arxiv.org/abs/0711.0728}{{\tt arXiv:0711.0728 [hep-ex]}}.
%%CITATION = ARXIV:0711.0728;%%.

\bibitem{Tanabashi:2018oca}
{\bf Particle Data Group} Collaboration, M.~Tanabashi {\em et al.}, ``{Review
  of Particle Physics},''
\href{http://dx.doi.org/10.1103/PhysRevD.98.030001}{{\em Phys. Rev.} {\bf D98}
  (2018) no.~3, 030001}.
%%CITATION = PHRVA,D98,030001;%%.

\bibitem{Aartsen:2017kpd}
{\bf IceCube} Collaboration, M.~G. Aartsen {\em et al.}, ``{Measurement of the
  multi-TeV neutrino cross section with IceCube using Earth absorption},''
  \href{http://dx.doi.org/10.1038/nature24459}{{\em Nature} {\bf 551} (2017)
  596--600},
\href{http://arxiv.org/abs/1711.08119}{{\tt arXiv:1711.08119 [hep-ex]}}.
%%CITATION = ARXIV:1711.08119;%%.

\bibitem{Bustamante:2017xuy}
M.~Bustamante and A.~Connolly, ``{Extracting the Energy-Dependent
  Neutrino-Nucleon Cross Section above 10 TeV Using IceCube Showers},''
  \href{http://dx.doi.org/10.1103/PhysRevLett.122.041101}{{\em Phys. Rev.
  Lett.} {\bf 122} (2019) no.~4, 041101},
\href{http://arxiv.org/abs/1711.11043}{{\tt arXiv:1711.11043 [astro-ph.HE]}}.
%%CITATION = ARXIV:1711.11043;%%.

\bibitem{Agafonova:2018auq}
{\bf OPERA} Collaboration, N.~Agafonova {\em et al.}, ``{Final Results of the
  OPERA Experiment on $\nu_\tau$ Appearance in the CNGS Neutrino Beam},''
  \href{http://dx.doi.org/10.1103/PhysRevLett.121.139901,
  10.1103/PhysRevLett.120.211801}{{\em Phys. Rev. Lett.} {\bf 120} (2018)
  no.~21, 211801}, \href{http://arxiv.org/abs/1804.04912}{{\tt arXiv:1804.04912
  [hep-ex]}}.
[Erratum: Phys. Rev. Lett.121,no.13,139901(2018)].
%%CITATION = ARXIV:1804.04912;%%.

\bibitem{Li:2017dbe}
{\bf Super-Kamiokande} Collaboration, Z.~Li {\em et al.}, ``{Measurement of the
  tau neutrino cross section in atmospheric neutrino oscillations with
  Super-Kamiokande},'' \href{http://dx.doi.org/10.1103/PhysRevD.98.052006}{{\em
  Phys. Rev.} {\bf D98} (2018) no.~5, 052006},
\href{http://arxiv.org/abs/1711.09436}{{\tt arXiv:1711.09436 [hep-ex]}}.
%%CITATION = ARXIV:1711.09436;%%.

\bibitem{Aartsen:2019tjl}
{\bf IceCube} Collaboration, M.~G. Aartsen {\em et al.}, ``{Measurement of
  Atmospheric Tau Neutrino Appearance with IceCube DeepCore},''
  \href{http://dx.doi.org/10.1103/PhysRevD.99.032007}{{\em Phys. Rev.} {\bf
  D99} (2019) no.~3, 032007},
\href{http://arxiv.org/abs/1901.05366}{{\tt arXiv:1901.05366 [hep-ex]}}.
%%CITATION = ARXIV:1901.05366;%%.

\bibitem{Goncharov:2001qe}
{\bf NuTeV} Collaboration, M.~Goncharov {\em et al.}, ``{Precise Measurement of
  Dimuon Production Cross-Sections in $\nu_{\mu}$ Fe and $\bar{\nu}_{\mu}$ Fe
  Deep Inelastic Scattering at the Tevatron.},''
  \href{http://dx.doi.org/10.1103/PhysRevD.64.112006}{{\em Phys. Rev.} {\bf
  D64} (2001)  112006},
\href{http://arxiv.org/abs/hep-ex/0102049}{{\tt arXiv:hep-ex/0102049
  [hep-ex]}}.
%%CITATION = HEP-EX/0102049;%%.

\bibitem{Rabinowitz:1993xx}
S.~A. Rabinowitz {\em et al.}, ``{Measurement of the strange sea distribution
  using neutrino charm production},''
\href{http://dx.doi.org/10.1103/PhysRevLett.70.134}{{\em Phys. Rev. Lett.} {\bf
  70} (1993)  134--137}.
%%CITATION = PRLTA,70,134;%%.

\bibitem{KayisTopaksu:2011mx}
A.~Kayis-Topaksu {\em et al.}, ``{Measurement of charm production in neutrino
  charged-current interactions},''
  \href{http://dx.doi.org/10.1088/1367-2630/13/9/093002}{{\em New J. Phys.}
  {\bf 13} (2011)  093002},
\href{http://arxiv.org/abs/1107.0613}{{\tt arXiv:1107.0613 [hep-ex]}}.
%%CITATION = ARXIV:1107.0613;%%.

\bibitem{Aartsen:2016nxy}
{\bf IceCube} Collaboration, M.~G. Aartsen {\em et al.}, ``{The IceCube
  Neutrino Observatory: Instrumentation and Online Systems},''
  \href{http://dx.doi.org/10.1088/1748-0221/12/03/P03012}{{\em JINST} {\bf 12}
  (2017) no.~03, P03012},
\href{http://arxiv.org/abs/1612.05093}{{\tt arXiv:1612.05093 [astro-ph.IM]}}.
%%CITATION = ARXIV:1612.05093;%%.

\bibitem{Collaboration:2011nsa}
{\bf ANTARES} Collaboration, M.~Ageron {\em et al.}, ``{ANTARES: the first
  undersea neutrino telescope},''
  \href{http://dx.doi.org/10.1016/j.nima.2011.06.103}{{\em Nucl. Instrum.
  Meth.} {\bf A656} (2011)  11--38},
\href{http://arxiv.org/abs/1104.1607}{{\tt arXiv:1104.1607 [astro-ph.IM]}}.
%%CITATION = ARXIV:1104.1607;%%.

\bibitem{Avrorin:2013uyc}
{\bf BAIKAL} Collaboration, A.~D. Avrorin {\em et al.}, ``{The
  prototyping/early construction phase of the BAIKAL-GVD project},''
  \href{http://dx.doi.org/10.1016/j.nima.2013.10.064}{{\em Nucl. Instrum.
  Meth.} {\bf A742} (2014)  82--88},
\href{http://arxiv.org/abs/1308.1833}{{\tt arXiv:1308.1833 [astro-ph.IM]}}.
%%CITATION = ARXIV:1308.1833;%%.

\bibitem{Adrian-Martinez:2016fdl}
{\bf KM3Net} Collaboration, S.~Adrian-Martinez {\em et al.}, ``{Letter of
  intent for KM3NeT 2.0},''
  \href{http://dx.doi.org/10.1088/0954-3899/43/8/084001}{{\em J. Phys.} {\bf
  G43} (2016) no.~8, 084001},
\href{http://arxiv.org/abs/1601.07459}{{\tt arXiv:1601.07459 [astro-ph.IM]}}.
%%CITATION = ARXIV:1601.07459;%%.

\bibitem{Aartsen:2016xlq}
{\bf IceCube} Collaboration, M.~G. Aartsen {\em et al.}, ``{Observation and
  Characterization of a Cosmic Muon Neutrino Flux from the Northern Hemisphere
  using six years of IceCube data},''
  \href{http://dx.doi.org/10.3847/0004-637X/833/1/3}{{\em Astrophys. J.} {\bf
  833} (2016) no.~1, 3},
\href{http://arxiv.org/abs/1607.08006}{{\tt arXiv:1607.08006 [astro-ph.HE]}}.
%%CITATION = ARXIV:1607.08006;%%.

\bibitem{Ferrari:2005zk}
A.~Ferrari, P.~R. Sala, A.~Fasso, and J.~Ranft, {\em {FLUKA: A Multi-particle
  Transport Code (Program Version 2005)}}.
\newblock CERN Yellow Reports: Monographs. CERN, Geneva, 2005.
\newblock
\url{http://cds.cern.ch/record/898301}.
\newblock
%%CITATION = CERN-2005-010;%%.

\bibitem{Battistoni:2015epi}
G.~Battistoni {\em et al.}, ``{Overview of the FLUKA code},''
\href{http://dx.doi.org/10.1016/j.anucene.2014.11.007}{{\em Annals Nucl.
  Energy} {\bf 82} (2015)  10--18}.
%%CITATION = ANEND,82,10;%%.

\bibitem{FLUKAstudy}
{CERN Sources, Targets, and Interactions Group}, M.~Sabate-Gilarte, F.~Cerutti,
  and A.~Tsinganis, ``{Characterization of the radiation field for the FASER
  experiment},''.

\bibitem{TimePix}
{\bf Medipix3} Collaboration, C.~Brezina, Y.~Fu, M.~De~Gaspari, V.~Gromov,
  X.~Llopart, T.~Poikela, F.~Zappon, and A.~Kruth, ``The timepix3 chip,'' tech.
  rep., 2014.
\newblock
  \url{https://indico.cern.ch/event/267425/attachments/477859/661149/Timepix3_final.pdf}.

\bibitem{Pierog:2013ria}
T.~Pierog, I.~Karpenko, J.~M. Katzy, E.~Yatsenko, and K.~Werner, ``{EPOS LHC:
  Test of collective hadronization with data measured at the CERN Large Hadron
  Collider},'' \href{http://dx.doi.org/10.1103/PhysRevC.92.034906}{{\em Phys.
  Rev.} {\bf C92} (2015)  034906},
\href{http://arxiv.org/abs/1306.0121}{{\tt arXiv:1306.0121 [hep-ph]}}.
%%CITATION = ARXIV:1306.0121;%%.

\bibitem{Ostapchenko:2010vb}
S.~Ostapchenko, ``{Monte Carlo treatment of hadronic interactions in enhanced
  Pomeron scheme: I. QGSJET-II model},''
  \href{http://dx.doi.org/10.1103/PhysRevD.83.014018}{{\em Phys. Rev.} {\bf
  D83} (2011)  014018},
\href{http://arxiv.org/abs/1010.1869}{{\tt arXiv:1010.1869 [hep-ph]}}.
%%CITATION = ARXIV:1010.1869;%%.

\bibitem{Ahn:2009wx}
E.-J. Ahn, R.~Engel, T.~K. Gaisser, P.~Lipari, and T.~Stanev, ``{Cosmic ray
  interaction event generator SIBYLL 2.1},''
  \href{http://dx.doi.org/10.1103/PhysRevD.80.094003}{{\em Phys. Rev.} {\bf
  D80} (2009)  094003},
\href{http://arxiv.org/abs/0906.4113}{{\tt arXiv:0906.4113 [hep-ph]}}.
%%CITATION = ARXIV:0906.4113;%%.

\bibitem{Riehn:2015oba}
F.~Riehn, R.~Engel, A.~Fedynitch, T.~K. Gaisser, and T.~Stanev, ``{A new
  version of the event generator Sibyll},'' {\em PoS} {\bf ICRC2015} (2016)
  558,
\href{http://arxiv.org/abs/1510.00568}{{\tt arXiv:1510.00568 [hep-ph]}}.
%%CITATION = ARXIV:1510.00568;%%.

\bibitem{Riehn:2017mfm}
F.~Riehn, H.~P. Dembinski, R.~Engel, A.~Fedynitch, T.~K. Gaisser, and
  T.~Stanev, ``{The hadronic interaction model SIBYLL 2.3c and Feynman
  scaling},'' \href{http://dx.doi.org/10.22323/1.301.0301}{{\em PoS} {\bf
  ICRC2017} (2018)  301}, \href{http://arxiv.org/abs/1709.07227}{{\tt
  arXiv:1709.07227 [hep-ph]}}.
[35,301(2017)].
%%CITATION = ARXIV:1709.07227;%%.

\bibitem{Fedynitch:2018cbl}
A.~Fedynitch, F.~Riehn, R.~Engel, T.~K. Gaisser, and T.~Stanev, ``{The hadronic
  interaction model Sibyll-2.3c and inclusive lepton fluxes},''
  \href{http://dx.doi.org/10.1103/PhysRevD.100.103018}{{\em Phys. Rev.} {\bf
  D100} (2019) no.~10, 103018},
\href{http://arxiv.org/abs/1806.04140}{{\tt arXiv:1806.04140 [hep-ph]}}.
%%CITATION = ARXIV:1806.04140;%%.

\bibitem{Sjostrand:2006za}
T.~Sjostrand, S.~Mrenna, and P.~Z. Skands, ``{PYTHIA 6.4 Physics and Manual},''
  \href{http://dx.doi.org/10.1088/1126-6708/2006/05/026}{{\em JHEP} {\bf 05}
  (2006)  026},
\href{http://arxiv.org/abs/hep-ph/0603175}{{\tt arXiv:hep-ph/0603175
  [hep-ph]}}.
%%CITATION = HEP-PH/0603175;%%.

\bibitem{Sjostrand:2007gs}
T.~Sjostrand, S.~Mrenna, and P.~Z. Skands, ``{A Brief Introduction to PYTHIA
  8.1},'' \href{http://dx.doi.org/10.1016/j.cpc.2008.01.036}{{\em Comput. Phys.
  Commun.} {\bf 178} (2008)  852--867},
\href{http://arxiv.org/abs/0710.3820}{{\tt arXiv:0710.3820 [hep-ph]}}.
%%CITATION = ARXIV:0710.3820;%%.

\bibitem{Buckley:2010ar}
A.~Buckley, J.~Butterworth, L.~Lonnblad, D.~Grellscheid, H.~Hoeth, J.~Monk,
  H.~Schulz, and F.~Siegert, ``{Rivet user manual},''
  \href{http://dx.doi.org/10.1016/j.cpc.2013.05.021}{{\em Comput. Phys.
  Commun.} {\bf 184} (2013)  2803--2819},
\href{http://arxiv.org/abs/1003.0694}{{\tt arXiv:1003.0694 [hep-ph]}}.
%%CITATION = ARXIV:1003.0694;%%.

\bibitem{Buckley:2009bj}
A.~Buckley, H.~Hoeth, H.~Lacker, H.~Schulz, and J.~E. von Seggern,
  ``{Systematic event generator tuning for the LHC},''
  \href{http://dx.doi.org/10.1140/epjc/s10052-009-1196-7}{{\em Eur. Phys. J.}
  {\bf C65} (2010)  331--357},
\href{http://arxiv.org/abs/0907.2973}{{\tt arXiv:0907.2973 [hep-ph]}}.
%%CITATION = ARXIV:0907.2973;%%.

\bibitem{Adriani:2017jys}
{\bf LHCf} Collaboration, O.~Adriani {\em et al.}, ``{Measurement of forward
  photon production cross-section in proton–proton collisions at $\sqrt{s}$ =
  13 TeV with the LHCf detector},''
  \href{http://dx.doi.org/10.1016/j.physletb.2017.12.050}{{\em Phys. Lett.}
  {\bf B780} (2018)  233--239},
\href{http://arxiv.org/abs/1703.07678}{{\tt arXiv:1703.07678 [hep-ex]}}.
%%CITATION = ARXIV:1703.07678;%%.

\bibitem{Aspell:2012ux}
{\bf TOTEM} Collaboration, G.~Antchev {\em et al.}, ``{Measurement of the
  forward charged particle pseudorapidity density in $pp$ collisions at
  $\sqrt{s} = 7$ TeV with the TOTEM experiment},''
  \href{http://dx.doi.org/10.1209/0295-5075/98/31002}{{\em EPL} {\bf 98} (2012)
  no.~3, 31002},
\href{http://arxiv.org/abs/1205.4105}{{\tt arXiv:1205.4105 [hep-ex]}}.
%%CITATION = ARXIV:1205.4105;%%.

\bibitem{ATLAS:2017rme}
{\bf ATLAS and LHCf} Collaboration, ``{Measurement of contributions of
  diffractive processes to forward photon spectra in pp collisions at
  $\sqrt{s}$ = 13 TeV},'' Tech. Rep. ATLAS-CONF-2017-075, CERN, Geneva, Nov,
  2017.
\newblock \url{https://cds.cern.ch/record/2291387}.

\bibitem{Adriani:2018ess}
{\bf LHCf} Collaboration, O.~Adriani {\em et al.}, ``{Measurement of inclusive
  forward neutron production cross section in proton-proton collisions at $
  \sqrt{s}=13 $ TeV with the LHCf Arm2 detector},''
  \href{http://dx.doi.org/10.1007/JHEP11(2018)073}{{\em JHEP} {\bf 11} (2018)
  073},
\href{http://arxiv.org/abs/1808.09877}{{\tt arXiv:1808.09877 [hep-ex]}}.
%%CITATION = ARXIV:1808.09877;%%.

\bibitem{Adriani:2015nwa}
{\bf LHCf} Collaboration, O.~Adriani {\em et al.}, ``{Measurement of very
  forward neutron energy spectra for 7 TeV proton–proton collisions at the
  Large Hadron Collider},''
  \href{http://dx.doi.org/10.1016/j.physletb.2015.09.041}{{\em Phys. Lett.}
  {\bf B750} (2015)  360--366},
\href{http://arxiv.org/abs/1503.03505}{{\tt arXiv:1503.03505 [hep-ex]}}.
%%CITATION = ARXIV:1503.03505;%%.

\bibitem{Adriani:2015iwv}
{\bf LHCf} Collaboration, O.~Adriani {\em et al.}, ``{Measurements of
  longitudinal and transverse momentum distributions for neutral pions in the
  forward-rapidity region with the LHCf detector},''
  \href{http://dx.doi.org/10.1103/PhysRevD.94.032007}{{\em Phys. Rev.} {\bf
  D94} (2016) no.~3, 032007},
\href{http://arxiv.org/abs/1507.08764}{{\tt arXiv:1507.08764 [hep-ex]}}.
%%CITATION = ARXIV:1507.08764;%%.

\bibitem{Adriani:2012ap}
{\bf LHCf} Collaboration, O.~Adriani {\em et al.}, ``{Measurement of forward
  neutral pion transverse momentum spectra for $\sqrt{s}$ = 7TeV proton-proton
  collisions at LHC},''
  \href{http://dx.doi.org/10.1103/PhysRevD.86.092001}{{\em Phys. Rev.} {\bf
  D86} (2012)  092001},
\href{http://arxiv.org/abs/1205.4578}{{\tt arXiv:1205.4578 [hep-ex]}}.
%%CITATION = ARXIV:1205.4578;%%.

\bibitem{Antchev:2014lez}
{\bf TOTEM} Collaboration, G.~Antchev {\em et al.}, ``{Measurement of the
  forward charged particle pseudorapidity density in pp collisions at $\sqrt{s}
  = 8$ TeV using a displaced interaction point},''
  \href{http://dx.doi.org/10.1140/epjc/s10052-015-3343-7}{{\em Eur. Phys. J.}
  {\bf C75} (2015) no.~3, 126},
\href{http://arxiv.org/abs/1411.4963}{{\tt arXiv:1411.4963 [hep-ex]}}.
%%CITATION = ARXIV:1411.4963;%%.

\bibitem{Chatrchyan:2014qka}
{\bf CMS, TOTEM} Collaboration, S.~Chatrchyan {\em et al.}, ``{Measurement of
  pseudorapidity distributions of charged particles in proton-proton collisions
  at $\sqrt{s}$ = 8 TeV by the CMS and TOTEM experiments},''
  \href{http://dx.doi.org/10.1140/epjc/s10052-014-3053-6}{{\em Eur. Phys. J.}
  {\bf C74} (2014) no.~10, 3053},
\href{http://arxiv.org/abs/1405.0722}{{\tt arXiv:1405.0722 [hep-ex]}}.
%%CITATION = ARXIV:1405.0722;%%.

\bibitem{Sirunyan:2017nsj}
{\bf CMS} Collaboration, A.~M. Sirunyan {\em et al.}, ``{Measurement of the
  inclusive energy spectrum in the very forward direction in proton-proton
  collisions at $ \sqrt{s}=13 $ TeV},''
  \href{http://dx.doi.org/10.1007/JHEP08(2017)046}{{\em JHEP} {\bf 08} (2017)
  046},
\href{http://arxiv.org/abs/1701.08695}{{\tt arXiv:1701.08695 [hep-ex]}}.
%%CITATION = ARXIV:1701.08695;%%.

\bibitem{Chatrchyan:2013gfi}
{\bf CMS} Collaboration, S.~Chatrchyan {\em et al.}, ``{Study of the Underlying
  Event at Forward Rapidity in pp Collisions at $\sqrt{s}$ = 0.9, 2.76, and 7
  TeV},'' \href{http://dx.doi.org/10.1007/JHEP04(2013)072}{{\em JHEP} {\bf 04}
  (2013)  072},
\href{http://arxiv.org/abs/1302.2394}{{\tt arXiv:1302.2394 [hep-ex]}}.
%%CITATION = ARXIV:1302.2394;%%.

\bibitem{bdsim}
L.~Nevay, J.~Snuverink, A.~Abramov, L.~Deacon, H.~Garcia-Morales, S.~Gibson,
  R.~Kwee-Hinzmann, H.~Pikhartova, W.~Shields, S.~Walker, and S.~Boogert,
  ``{BDSIM: An Accelerator Tracking Code with Particle-Matter Interactions},''
  2018.

\bibitem{madx}
H.~Grote and F.~Schmidt, ``{MAD-X: An upgrade from MAD8},''
{\em Conf. Proc.} {\bf C030512} (2003)  3497.
%%CITATION = CONFP,C030512,3497;%%.

\bibitem{fluka}
T.~Böhlen, F.~Cerutti, M.~Chin, A.~Fassò, A.~Ferrari, P.~Ortega, A.~Mairani,
  P.~Sala, G.~Smirnov, and V.~Vlachoudis, ``{The FLUKA Code: Developments and
  Challenges for High Energy and Medical Applications},''
  \href{http://dx.doi.org/https://doi.org/10.1016/j.nds.2014.07.049}{{\em
  Nuclear Data Sheets} {\bf 120} (2014)  211 -- 214}.
  \url{http://www.sciencedirect.com/science/article/pii/S0090375214005018}.

\bibitem{Geant4PhysicsLists}
A.~Ribon {\em et al.}, ``Status of geant4 hadronic physics for the simulation
  of lhc experiments at the start of the lhc physics program,''
  CERN-LCGAPP-2010-02.
  \url{https://lcgapp.cern.ch/project/docs/noteStatusHadronic2010.pdf}.

\bibitem{crmc}
C.~Baus, T.~Pierog, and R.~Ulrich, ``{Cosmic Ray Monte Carlo (CRMC)},''.
  \url{https://web.ikp.kit.edu/rulrich/crmc.html}.

\bibitem{Formaggio:2013kya}
J.~A. Formaggio and G.~P. Zeller, ``{From eV to EeV: Neutrino Cross Sections
  Across Energy Scales},''
  \href{http://dx.doi.org/10.1103/RevModPhys.84.1307}{{\em Rev. Mod. Phys.}
  {\bf 84} (2012)  1307--1341},
\href{http://arxiv.org/abs/1305.7513}{{\tt arXiv:1305.7513 [hep-ex]}}.
%%CITATION = ARXIV:1305.7513;%%.

\bibitem{Gandhi:1995tf}
R.~Gandhi, C.~Quigg, M.~H. Reno, and I.~Sarcevic, ``{Ultrahigh-energy neutrino
  interactions},'' \href{http://dx.doi.org/10.1016/0927-6505(96)00008-4}{{\em
  Astropart. Phys.} {\bf 5} (1996)  81--110},
\href{http://arxiv.org/abs/hep-ph/9512364}{{\tt arXiv:hep-ph/9512364
  [hep-ph]}}.
%%CITATION = HEP-PH/9512364;%%.

\bibitem{Andreopoulos:2009rq}
C.~Andreopoulos {\em et al.}, ``{The GENIE Neutrino Monte Carlo Generator},''
  \href{http://dx.doi.org/10.1016/j.nima.2009.12.009}{{\em Nucl. Instrum.
  Meth.} {\bf A614} (2010)  87--104},
\href{http://arxiv.org/abs/0905.2517}{{\tt arXiv:0905.2517 [hep-ph]}}.
%%CITATION = ARXIV:0905.2517;%%.

\bibitem{Andreopoulos:2015wxa}
C.~Andreopoulos, C.~Barry, S.~Dytman, H.~Gallagher, T.~Golan, R.~Hatcher,
  G.~Perdue, and J.~Yarba, ``{The GENIE Neutrino Monte Carlo Generator: Physics
  and User Manual},''
\href{http://arxiv.org/abs/1510.05494}{{\tt arXiv:1510.05494 [hep-ph]}}.
%%CITATION = ARXIV:1510.05494;%%.

\bibitem{Bodek:2002ps}
A.~Bodek and U.~K. Yang, ``{Higher twist, xi(omega) scaling, and effective LO
  PDFs for lepton scattering in the few GeV region},''
  \href{http://dx.doi.org/10.1088/0954-3899/29/8/369}{{\em J. Phys.} {\bf G29}
  (2003)  1899--1906},
\href{http://arxiv.org/abs/hep-ex/0210024}{{\tt arXiv:hep-ex/0210024
  [hep-ex]}}.
%%CITATION = HEP-EX/0210024;%%.

\bibitem{Eskola:2016oht}
K.~J. Eskola, P.~Paakkinen, H.~Paukkunen, and C.~A. Salgado, ``{EPPS16: Nuclear
  parton distributions with LHC data},''
  \href{http://dx.doi.org/10.1140/epjc/s10052-017-4725-9}{{\em Eur. Phys. J.}
  {\bf C77} (2017) no.~3, 163},
\href{http://arxiv.org/abs/1612.05741}{{\tt arXiv:1612.05741 [hep-ph]}}.
%%CITATION = ARXIV:1612.05741;%%.

\bibitem{Kovarik:2015cma}
K.~Kovarik {\em et al.}, ``{nCTEQ15 - Global analysis of nuclear parton
  distributions with uncertainties in the CTEQ framework},''
  \href{http://dx.doi.org/10.1103/PhysRevD.93.085037}{{\em Phys. Rev.} {\bf
  D93} (2016) no.~8, 085037},
\href{http://arxiv.org/abs/1509.00792}{{\tt arXiv:1509.00792 [hep-ph]}}.
%%CITATION = ARXIV:1509.00792;%%.

\bibitem{Kusina:2015vfa}
A.~Kusina {\em et al.}, ``{nCTEQ15 - Global analysis of nuclear parton
  distributions with uncertainties},''
  \href{http://dx.doi.org/10.22323/1.247.0041}{{\em PoS} {\bf DIS2015} (2015)
  041},
\href{http://arxiv.org/abs/1509.01801}{{\tt arXiv:1509.01801 [hep-ph]}}.
%%CITATION = ARXIV:1509.01801;%%.

\bibitem{AbdulKhalek:2019mzd}
{\bf NNPDF} Collaboration, R.~Abdul~Khalek, J.~J. Ethier, and J.~Rojo,
  ``{Nuclear parton distributions from lepton-nucleus scattering and the impact
  of an electron-ion collider},''
  \href{http://dx.doi.org/10.1140/epjc/s10052-019-6983-1}{{\em Eur. Phys. J.}
  {\bf C79} (2019) no.~6, 471},
\href{http://arxiv.org/abs/1904.00018}{{\tt arXiv:1904.00018 [hep-ph]}}.
%%CITATION = ARXIV:1904.00018;%%.

\bibitem{Yang:2009zx}
T.~Yang, C.~Andreopoulos, H.~Gallagher, K.~Hoffmann, and P.~Kehayias, ``{A
  Hadronization Model for Few-GeV Neutrino Interactions},''
  \href{http://dx.doi.org/10.1140/epjc/s10052-009-1094-z}{{\em Eur. Phys. J.}
  {\bf C63} (2009)  1--10},
\href{http://arxiv.org/abs/0904.4043}{{\tt arXiv:0904.4043 [hep-ph]}}.
%%CITATION = ARXIV:0904.4043;%%.

\bibitem{ALLEN1981385}
P.~Allen {\em et al.}, ``Multiplicity distributions in neutrino-hydrogen
  interactions,''
  \href{http://dx.doi.org/https://doi.org/10.1016/0550-3213(81)90532-0}{{\em
  Nuclear Physics B} {\bf 181} (1981) no.~3, 385 -- 402}.
  \url{http://www.sciencedirect.com/science/article/pii/0550321381905320}.

\bibitem{Katori:2014fxa}
T.~Katori and S.~Mandalia, ``{PYTHIA hadronization process tuning in the GENIE
  neutrino interaction generator},''
  \href{http://dx.doi.org/10.1088/0954-3899/42/11/115004}{{\em J. Phys.} {\bf
  G42} (2015) no.~11, 115004},
\href{http://arxiv.org/abs/1412.4301}{{\tt arXiv:1412.4301 [hep-ex]}}.
%%CITATION = ARXIV:1412.4301;%%.

\bibitem{Qian:2009aa}
X.~Qian {\em et al.}, ``{Experimental Study of the A(e,e'$\pi^+$) Reaction on
  $^1$H, $^2$H, $^{12}$C, $^{27}$Al, $^{63}$Cu and $^{197}$Au},''
  \href{http://dx.doi.org/10.1103/PhysRevC.81.055209}{{\em Phys. Rev.} {\bf
  C81} (2010)  055209},
\href{http://arxiv.org/abs/0908.1616}{{\tt arXiv:0908.1616 [nucl-ex]}}.
%%CITATION = ARXIV:0908.1616;%%.

\bibitem{INTRANUKE}
S.~Dytman, ``{Final state interactions in neutrino-nucleus experiments},''
{\em Acta Phys. Polon.} {\bf B40} (2009)  2445--2460.
%%CITATION = APPOA,B40,2445;%%.

\bibitem{Hayato:2009zz}
Y.~Hayato, ``{A neutrino interaction simulation program library NEUT},''
{\em Acta Phys. Polon.} {\bf B40} (2009)  2477--2489.
%%CITATION = APPOA,B40,2477;%%.

\bibitem{Juszczak:2009qa}
C.~Juszczak, ``{Running NuWro},'' {\em Acta Phys. Polon.} {\bf B40} (2009)
  2507--2512,
\href{http://arxiv.org/abs/0909.1492}{{\tt arXiv:0909.1492 [hep-ex]}}.
%%CITATION = ARXIV:0909.1492;%%.

\bibitem{Buss:2011mx}
O.~Buss, T.~Gaitanos, K.~Gallmeister, H.~van Hees, M.~Kaskulov, O.~Lalakulich,
  A.~B. Larionov, T.~Leitner, J.~Weil, and U.~Mosel, ``{Transport-theoretical
  Description of Nuclear Reactions},''
  \href{http://dx.doi.org/10.1016/j.physrep.2011.12.001}{{\em Phys. Rept.} {\bf
  512} (2012)  1--124},
\href{http://arxiv.org/abs/1106.1344}{{\tt arXiv:1106.1344 [hep-ph]}}.
%%CITATION = ARXIV:1106.1344;%%.

\bibitem{pouring_system}
H.~Rokujo, ``Nuclear emulsion production facility.''. Exploration of Particle
  Physics and Cosmology with Neutrinos Workshop2019,
  \url{https://indico.cern.ch/event/815137/contributions/3444731}.

\bibitem{pouring_test}
K.~Sugimura, H.~Rokujo, N.~Naganawa, and M.~Nakamura, ``Construction of
  emulsion film pouring system in nagoya university.''. Fall Meeting of
  Federation of Imaging Society 2018.

\bibitem{Amsler:2012wn}
C.~Amsler {\em et al.}, ``{A new application of emulsions to measure the
  gravitational force on antihydrogen},''
  \href{http://dx.doi.org/10.1088/1748-0221/8/02/P02015}{{\em JINST} {\bf 8}
  (2013)  P02015},
\href{http://arxiv.org/abs/1211.1370}{{\tt arXiv:1211.1370 [physics.ins-det]}}.
%%CITATION = ARXIV:1211.1370;%%.

\bibitem{Kodama:2002dk}
K.~Kodama {\em et al.}, ``{Detection and analysis of tau neutrino interactions
  in DONUT emulsion target},''
\href{http://dx.doi.org/10.1016/S0168-9002(02)01555-3}{{\em Nucl. Instrum.
  Meth.} {\bf A493} (2002)  45--66}.
%%CITATION = NUIMA,A493,45;%%.

\bibitem{Kodama:1993wg}
{\bf Fermilab E653} Collaboration, K.~Kodama {\em et al.}, ``{Measurement of
  beauty hadron pair production in 600-GeV/c pi- emulsion interactions.},''
\href{http://dx.doi.org/10.1016/0370-2693(93)91446-T}{{\em Phys. Lett.} {\bf
  B303} (1993)  359--367}.
%%CITATION = PHLTA,B303,359;%%.

\bibitem{Albanese:1985wk}
J.~P. Albanese {\em et al.}, ``{Direct Observation of the Decay of Beauty
  Particles Into Charm Particles},''
\href{http://dx.doi.org/10.1016/0370-2693(85)91389-9}{{\em Phys. Lett.} {\bf
  158B} (1985)  186--192}.
%%CITATION = PHLTA,158B,186;%%.

\bibitem{Eskut:2007rn}
{\bf CHORUS} Collaboration, E.~Eskut {\em et al.}, ``{Final results on
  $\nu_{\mu} \to \nu_{\tau}$ oscillation from the CHORUS experiment},''
  \href{http://dx.doi.org/10.1016/j.nuclphysb.2007.10.023}{{\em Nucl. Phys.}
  {\bf B793} (2008)  326--343},
\href{http://arxiv.org/abs/0710.3361}{{\tt arXiv:0710.3361 [hep-ex]}}.
%%CITATION = ARXIV:0710.3361;%%.

\bibitem{Acquafredda:2009zz}
R.~Acquafredda {\em et al.}, ``{The OPERA experiment in the CERN to Gran Sasso
  neutrino beam},''
\href{http://dx.doi.org/10.1088/1748-0221/4/04/P04018}{{\em JINST} {\bf 4}
  (2009)  P04018}.
%%CITATION = JINST,4,P04018;%%.

\bibitem{ATLAS:1997ag}
{\bf ATLAS} Collaboration, {\em {ATLAS inner detector: Technical Design Report,
  1}}.
\newblock Technical Design Report ATLAS. CERN, Geneva, 1997.
\newblock
\url{https://cds.cern.ch/record/331063}.
\newblock
%%CITATION = CERN-LHCC-97-16;%%.

\bibitem{ATLAS:1997af}
{\bf ATLAS} Collaboration, S.~Haywood, L.~Rossi, R.~Nickerson, and
  A.~Romaniouk, {\em {ATLAS inner detector: Technical Design Report, 2}}.
\newblock Technical Design Report ATLAS. CERN, Geneva, 1997.
\newblock
\url{https://cds.cern.ch/record/331064}.
\newblock
%%CITATION = CERN-LHCC-97-17;%%.

\bibitem{Agostinelli:2002hh}
{\bf GEANT4} Collaboration, S.~Agostinelli {\em et al.}, ``{GEANT4: A
  Simulation toolkit},''
\href{http://dx.doi.org/10.1016/S0168-9002(03)01368-8}{{\em Nucl. Instrum.
  Meth.} {\bf A506} (2003)  250--303}.
%%CITATION = NUIMA,A506,250;%%.

\bibitem{Yoshimoto:2017ufm}
M.~Yoshimoto, T.~Nakano, R.~Komatani, and H.~Kawahara, ``{Hyper-track selector
  nuclear emulsion readout system aimed at scanning an area of one thousand
  square meters},'' \href{http://dx.doi.org/10.1093/ptep/ptx131}{{\em PTEP}
  {\bf 2017} (2017) no.~10, 103H01},
\href{http://arxiv.org/abs/1704.06814}{{\tt arXiv:1704.06814
  [physics.ins-det]}}.
%%CITATION = ARXIV:1704.06814;%%.

\bibitem{Ariga:2018gan}
{\bf DsTau} Collaboration, T.~Ariga, ``{Study of tau-neutrino production at the
  CERN SPS},''
\href{http://dx.doi.org/10.22323/1.340.0240}{{\em PoS} {\bf ICHEP2018} (2019)
  240}.
%%CITATION = POSCI,ICHEP2018,240;%%.

\bibitem{Aoki:2019jry}
{\bf DsTau} Collaboration, S.~Aoki {\em et al.}, ``{DsTau: Study of tau
  neutrino production with 400 GeV protons from the CERN-SPS},''
  \href{http://dx.doi.org/10.1007/JHEP01(2020)033}{{\em JHEP} {\bf 01} (2020)
  33},
\href{http://arxiv.org/abs/1906.03487}{{\tt arXiv:1906.03487 [hep-ex]}}.
%%CITATION = ARXIV:1906.03487;%%.

\bibitem{Antcheva:2009zz}
I.~Antcheva {\em et al.}, ``{ROOT: A C++ framework for petabyte data storage,
  statistical analysis and visualization},''
  \href{http://dx.doi.org/10.1016/j.cpc.2009.08.005}{{\em Comput. Phys.
  Commun.} {\bf 180} (2009)  2499--2512},
\href{http://arxiv.org/abs/1508.07749}{{\tt arXiv:1508.07749
  [physics.data-an]}}.
%%CITATION = ARXIV:1508.07749;%%.

\end{thebibliography}\endgroup

\end{document}